\documentclass[11pt]{amsart}
\usepackage{graphicx}
\usepackage{amscd}
\usepackage{amsbsy}
\usepackage{amsmath}
\usepackage{amstext}
\usepackage{amsfonts}
\usepackage{amssymb}
\newtheorem{theorem}{Theorem}
\theoremstyle{plain}

\newtheorem{corollary}{Corollary}

\newtheorem{definition}{Definition}
\newtheorem{example}{Example}

\newtheorem{lemma}{Lemma}

\newtheorem{proposition}{Proposition}
\newtheorem{remark}{Remark}

\numberwithin{equation}{section}

\makeindex

\begin{document}
\title[A Rossetta Stone for Quantum Mechanics]{A Rosetta Stone for Quantum Mechanics with\\an Introduction to Quantum Computation\\Version 1.5}
\author{Samuel J. Lomonaco, Jr.}
\address{Dept. of Comp. Sci. \& Elect. Engr.\\
University of Maryland Baltimore County\\
1000 Hilltop Circle\\
Baltimore, MD 21250}
\email{E-Mail: Lomonaco@UMBC.EDU}
\urladdr{WebPage: http://www.csee.umbc.edu/\symbol{126}lomonaco}
\thanks{This work was partially supported by ARO Grant \#P-38804-PH-QC and the L-O-O-P
Fund. The author gratefully acknowledges the hospitality of the University of
Cambridge Isaac Newton Institute for Mathematical Sciences, Cambridge,
England, where some of this work was completed. \ \ I would also like to thank
the other AMS\ Short Course lecturers, Howard Brandt, Dan Gottesman, Lou
Kauffman, Alexei Kitaev, Peter Shor, Umesh Vazirani and the many Short Course
participants for their support. \ (Copyright 2000.) \ }
\date{June 20, 2000}
\maketitle

\begin{abstract}
The purpose of these lecture notes is to provide readers, who have some
mathematical background but little or no exposure to quantum mechanics and
quantum computation, with enough material to begin reading the research
literature in quantum computation and quantum information theory. \ This paper
is a written version of the first of eight one hour lectures given in the
American Mathematical Society (AMS)\ Short Course on Quantum Computation held
in conjunction with the Annual Meeting of the AMS in Washington, DC, USA in
January 2000, and will be published in the AMS\ PSAPM volume entitled
``Quantum Computation.''. \ 

Part 1 of the paper is a preamble introducing the reader to the concept of the qubit,

Part 2 gives an introduction to quantum mechanics covering such topics as
Dirac notation, quantum measurement, Heisenberg uncertainty, Schr\"{o}dinger's
equation, density operators, partial trace, multipartite quantum systems, the
Heisenberg versus the Schr\"{o}dinger picture, quantum entanglement, EPR
paradox, quantum entropy.

Part 3 gives a brief introduction to quantum computation, covering such topics
as elementary quantum computing devices, wiring diagrams, the no-cloning
theorem, quantum teleportation, Shor's algorithm, Grover's algorithm.

Many examples are given to illustrate underlying principles. \ A table of
contents as well as an index are provided for readers who wish to ``pick and
choose.'' \ Since this paper is intended for a diverse audience, it is written
in an informal style at varying levels of difficulty and sophistication, from
the very elementary to the more advanced.
\end{abstract}\tableofcontents

\part{{\protect\Large Preamble}\bigskip}

\bigskip

\section{\bigskip\textbf{Introduction}}

\quad\bigskip

\medskip

These lecture notes were written for the American Mathematical Society (AMS)
Short Course on Quantum Computation held 17-18 January 2000 in conjunction
with the Annual Meeting of the AMS in Washington, DC in January 2000. \ The
notes are intended for readers with some mathematical background but with
little or no exposure to quantum mechanics. \ The purpose of these notes is to
provide such readers with enough material in quantum mechanics and quantum
computation to begin reading the vast literature on quantum computation,
quantum cryptography, and quantum information theory.

\bigskip

The paper was written in an informal style. \ Whenever possible, each new
topic was begun with the introduction of the underlying motivating intuitions,
and then followed by an explanation of the accompanying mathematical finery.
\ Hopefully, once having grasped the basic intuitions, the reader will find
that the remaining material easily follows.\bigskip

Since this paper is intended for a diverse audience, it was written at varying
levels of difficulty and sophistication, from the very elementary to the more
advanced. \ A large number of examples have been included. \ An index and
table of contents are provided for those readers who prefer to ``pick and
choose.'' \ Hopefully, this paper will provide something of interest for everyone.

\bigskip

Because of space limitations, these notes are, of necessity, far from a
complete overview of quantum mechanics. \ For example, only finite dimensional
Hilbert spaces are considered, thereby avoiding the many pathologies that
always arise when dealing with infinite dimensional objects. \ Many important
experiments that are traditionally part of the standard fare in quantum
mechanics texts (such as for example, the Stern-Gerlach experiment, Young's
two slit experiment, the Aspect experiment) have not been mentioned in this
paper. \ We leave it to the reader to decide if these notes have achieved
their objective.

\bigskip

The final version of this paper together with all the other lecture notes of
the AMS\ Short Course on Quantum Computation will be published as a book in
the AMS\ PSAPM Series entitled ``Quantum Computation.''

\bigskip

\section{\textbf{The classical world}}

\bigskip

\subsection{Introducing the Shannon bit.}

\qquad\bigskip

Since one of the objectives of this paper is to discuss quantum information,
we begin with a brief discussion of classical information. \ \bigskip

The Shannon bit is so well known in our age of information that it needs
little, if any, introduction. \ As we all know, the Shannon bit is like a very
decisive individual. \ It is either 0 or 1, but by no means both at the same
time. \ The Shannon bit has become so much a part of our every day lives that
we take many of its properties for granted. \ For example, we take for granted
that Shannon bits can be copied.

\subsection{\textbf{\bigskip}Polarized light: Part I. The classical perspective}

\qquad\bigskip

Throughout this paper the quantum polarization states of light will be used to
provide concrete illustrations of underlying quantum mechanical principles.
\ So we also begin with a brief discussion of polarized light from the
classical perspective.

\index{Polarized Light}

Light waves in the vacuum are transverse electromagnetic (EM) waves with both
electric and magnetic field vectors perpendicular to the direction of
propagation and also to each other. (See figure 1.) \medskip%

\begin{center}
\includegraphics[
height=1.7513in,
width=3.6556in
]%
{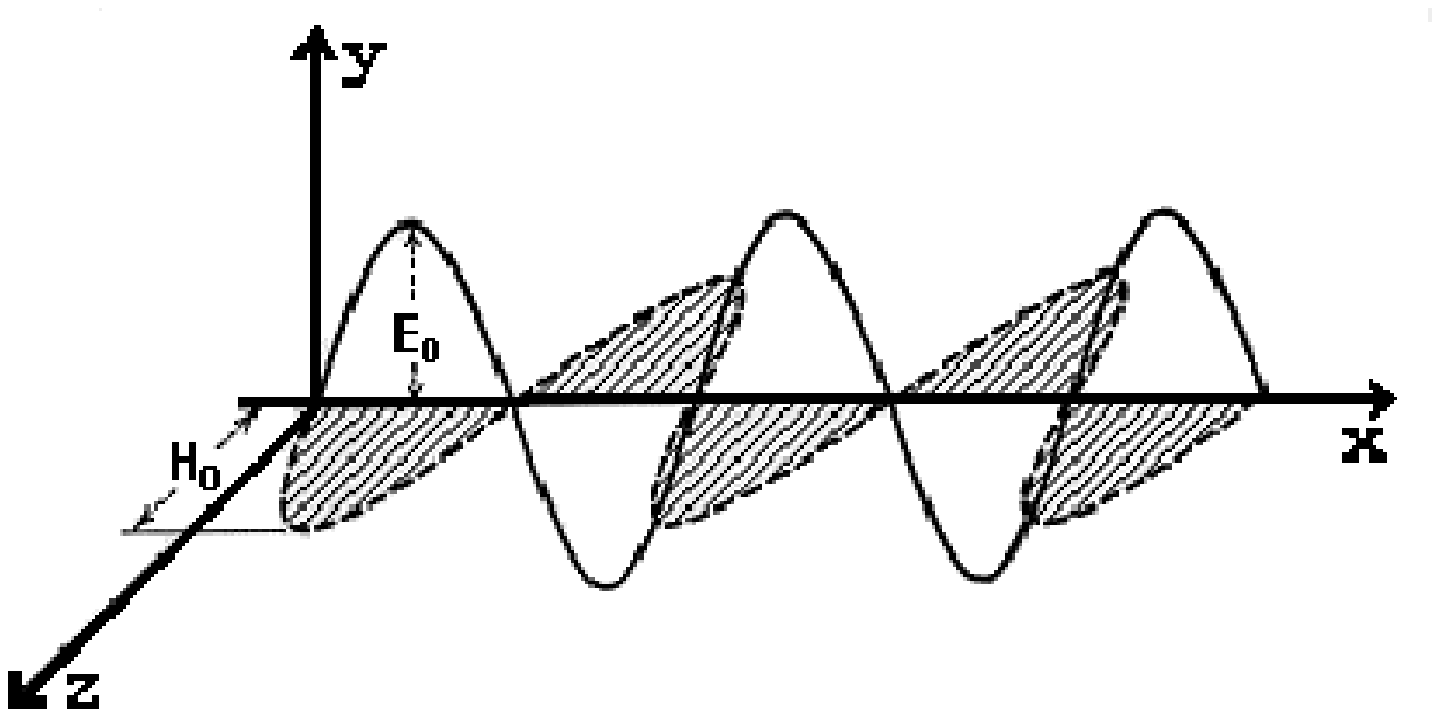}%
\\
Figure 1. A linearly polarized electromagnetic wave.
\end{center}

\noindent If the electric field vector is always parallel to a fixed line,
then the EM wave is said to be \textbf{linearly polarized}. If the electric
field vector rotates about the direction of propagation forming a
right-(left-)handed screw, it is said to be \textbf{right} (\textbf{left})
\textbf{elliptically} \textbf{polarized}. If the rotating electric field
vector inscribes a circle, the EM wave is said to be right-or
left-\textbf{circularly polarized}.

\index{Polarized Light}

\section{\textbf{The quantum world}}

\bigskip

\subsection{Introducing the qubit -- But what is a qubit?}

\qquad\bigskip

Many of us may not be as familiar with the quantum bit of information, called
a \textbf{qubit}. \ Unlike its sibling rival, the Shannon bit, the qubit can
be both 0 and 1 at the same time. \ Moreover, unlike the Shannon bit, the
qubit can not be duplicated\footnote{This is a result of the no-cloning
theorem of Wootters and Zurek\cite{Wootters1}. \ A proof of the no-cloning
theorem is given in Section 10.8 of this paper.}. \ As we shall see, qubits
are like very slippery, irascible individuals, exceedingly difficult to deal with.

\index{Qubit}

\bigskip

One example of a qubit is a spin $\frac{1}{2}$ particle which can be in a
spin-up state $\left|  1\right\rangle $ which we label as ``$1$'', in a
spin-down state $\left|  0\right\rangle $ which we label as ``$0$'', or in a
\textbf{superposition} \textbf{
\index{Superposition}} of these states, which we interpret as being both $0$
and $1$ at the same time. \ (The term ``superposition'' will be explained
shortly.) \ 

\bigskip

Another example of a qubit is the polarization state of a photon. \ A photon
can be in a vertically polarized state $\left|  \updownarrow\right\rangle $.
We assign a label of ``$1$'' to this state. \ It can be in a horizontally
polarized state $\left|  \leftrightarrow\right\rangle $. \ We assign a label
of ``$0$'' to this state. Or, it can be in a superposition of these states. In
this case, we interpret its state as representing both $0$ and $1$ at the same
time. \ \bigskip

Anyone who has worn polarized sunglasses is familiar with the polarization
states of light. \ Polarized sunglasses eliminate glare by letting through
only vertically polarized light, while filtering out the horizontally
polarized light. \ For that reason, they are often used to eliminate road
glare, i.e., horizontally polarized light reflected from the road.

\bigskip

\subsection{Where do qubits live? -- But what is a qubit?}

\qquad\bigskip

But where do qubits live? \ They live in a Hilbert space $\mathcal{H}$. \ By a
Hilbert space, we mean:

\bigskip

A \textbf{Hilbert space
\index{Hilbert Space}} $\mathcal{H}$ is a vector space over the complex
numbers $\mathbb{C}$ with a complex valued inner product
\[
\left(  -,-\right)  :\mathcal{H}\times\mathcal{H\rightarrow}\mathbb{C}%
\]
which is complete with respect to the norm
\[
\left\|  u\right\|  =\sqrt{\left(  u,u\right)  }%
\]
induced by the inner product. \bigskip\ 

\begin{remark}
By a complex valued inner product, we mean a map
\[
\left(  -,-\right)  :\mathcal{H}\times\mathcal{H\rightarrow}\mathbb{C}%
\]
from $\mathcal{H\times H}$ into the complex numbers $\mathbb{C}$ such that:

\begin{itemize}
\item [1)]$\left(  u,u\right)  =0$ if and only if $u=0$

\item[2)] $(u,v)=(v,u)^{\ast}$

\item[3)] $(u,v+w)=(u,v)+(u,w)$

\item[4)] $(u,\lambda v)=\lambda(u,v)$
\end{itemize}

where `$^{\ast}$' denotes the complex conjugate.
\end{remark}

\bigskip

\begin{remark}
Please note that $(\lambda u,v)=\lambda^{\ast}(u,v)$.
\end{remark}

\bigskip

\subsection{\textbf{\bigskip}A qubit is ...}

\qquad\footnote{Barenco et al in \cite{Barenco1} define a qubit as a quantum
system with a two dimensional Hilbert space, capable of existing in a
superposition of Boolean states, and also capable of being entangled with the
states of other qubits. \ Their more functional definition will take on more
meaning as the reader progresses through this paper.}
\index{Qubit|)}\bigskip%

\[
\fbox{%
\begin{tabular}
[c]{c}%
A \textbf{qubit} is a quantum system $\mathcal{Q}$ whose\\
state lies in a two dimensional Hilbert space $\mathcal{H}$.
\end{tabular}
}%
\]

\bigskip

\part{{\protect\Large An Introduction to Quantum Mechanics}\bigskip}

\quad\bigskip

\section{\textbf{The beginnings of quantum mechanics}}

\bigskip

\subsection{A Rosetta stone for Dirac notation:\ Part I. Bras, kets, and
bra-(c)-kets
\index{Bra}
\index{Bracket}
\index{Ket}}

\qquad\medskip
\index{Dirac Notation|(}

\medskip\ The elements of a Hilbert space $\mathcal{H}$ will be called
\textbf{ket vectors}, \textbf{state kets}, or simply \textbf{kets}. They will
be denoted as:
\[
\left|  \,label\,\right\rangle
\]
where `$label$' denotes some label.\vspace*{0.3in}

Let $\mathcal{H}^{\ast}$ denote the Hilbert space of all Hilbert space
morphisms of $\mathcal{H}$ into the Hilbert space of all complex numbers
$\mathbb{C}$, i.e.,
\[
\mathcal{H}^{\ast}=Hom_{\mathbb{C}}\left(  \mathcal{H},\mathbb{C}\right)
\text{.}%
\]
The elements of $\mathcal{H}^{\ast}$ will be called \textbf{bra vectors},
\textbf{state bras}, or simply \textbf{bras}. They will be denoted as:
\[
\left\langle \,label\,\right|
\]
where once again `$label$' denotes some label.\vspace*{0.5in}\ 

Also please note that the complex number
\[
\left\langle \,label_{1}\,\right|  \left(  \left|  \,label_{2}\,\right\rangle
\right)
\]
will simply be denoted by
\[
\left\langle \,label_{1}\,\mid\,label_{2}\,\right\rangle
\]
and will be called the \textbf{bra-(c)-ket} product of the bra $\left\langle
\,label_{1}\,\right|  $ and the ket $\left|  \,label_{2}\,\right\rangle
$.\vspace*{0.5in}\ 

There is a monomorphism (which is an isomorphism if the underlying Hilbert
space is finite dimensional)
\[
\mathcal{H}\overset{\dagger}{\rightarrow}\mathcal{H}^{\ast}%
\]
defined by
\[
\left|  \,label\,\right\rangle \longmapsto\left(  \;\left|
\,label\,\right\rangle ,-\right)
\]
The bra $\left(  \;\left|  \,label\,\right\rangle ,-\right)  $ is denoted by
$\left\langle \,label\,\right|  $.\vspace*{0.5in}\ 

Hence,
\[
\left\langle \,label_{1}\,\mid\,label_{2}\,\right\rangle =\left(  \left|
\,label_{1}\,\right\rangle ,\left|  \,label_{2}\,\right\rangle \right)
\]
\medskip\ 

\begin{remark}
Please note that $\left(  \lambda\left|  \,label\,\right\rangle \right)
^{\dagger}=\lambda^{\ast}\left\langle label\right|  $.\medskip\ 
\end{remark}

The \textbf{tensor product}\footnote{Readers well versed in homological
algebra will recognize this informal definition as a slightly disguised
version of the more rigorous universal definition of the tensor product. For
more details, please refer to \cite{Cartan1}, or any other standard reference
on homological algebra.}
\index{Tensor Product} $\mathcal{H}\otimes\mathcal{K}$ of two Hilbert spaces
$\mathcal{H}$ and $\mathcal{K}$ is simply the ``simplest'' Hilbert space such that\bigskip

\begin{enumerate}
\item [1)]$\left(  h_{1}+h_{2}\right)  \otimes k=h_{1}\otimes k+h_{2}\otimes k
$, for all $h_{1}$, $h_{2}\in\mathcal{H}$ and for all $k\in\mathcal{K}$, and

\item[2)] $h\otimes\left(  k_{1}+k_{2}\right)  =h\otimes k_{1}+h\otimes k_{2}
$ for all $h\in\mathcal{H}$ and for all $k_{1}$, $k_{2}\in\mathcal{K} $.

\item[3)] $\lambda\left(  h\otimes k\right)  \equiv\left(  \lambda h\right)
\otimes k=h\otimes\left(  \lambda k\right)  $ for all $\lambda\in\mathbb{C}$,
$h\in\mathcal{H}$, $k\in\mathcal{K}$.
\end{enumerate}

\bigskip

\begin{remark}
Hence, $\left\|  \ \left|  label\right\rangle \ \right\|  =\sqrt{\left\langle
\ label\mid label\ \right\rangle }$ and $\left\langle \ label_{1}\mid
label_{2}\ \right\rangle =\left(  \left|  \,label_{1}\,\right\rangle ,\left|
\,label_{2}\,\right\rangle \right)  $ .
\end{remark}

\vspace*{0.3in}

It follows that, if $\left\{  \ e_{1},e_{2},\ldots,e_{m}\ \right\}  $ and
$\left\{  \ f_{1},f_{2},\ldots,f_{n}\ \right\}  $ are respectively bases of
the Hilbert spaces $\mathcal{H}$ and $\mathcal{K}$, then $\left\{
\ e_{i}\otimes f_{j}\mid1\leq i\leq m\text{, }1\leq j\leq n\ \right\}  $ is a
basis of $\mathcal{H}\otimes\mathcal{K}$. \ Hence, the dimension of the
Hilbert space $\mathcal{H}\otimes\mathcal{K}$ is the product of the dimensions
of the Hilbert spaces $\mathcal{H}$ and $\mathcal{K}$, i.e.,
\[
Dim\left(  \mathcal{H}\otimes\mathcal{K}\right)  =Dim\left(  \mathcal{H}%
\right)  \cdot Dim\left(  \mathcal{K}\right)  \text{ .}%
\]
\vspace*{0.3in}

Finally, if $\left|  \,label_{1}\,\right\rangle $ and $\left|  \,label_{2}%
\,\right\rangle $ are kets respectively in Hilbert spaces $\mathcal{H}_{1}$
and $\mathcal{H}_{2}$, then their tensor product will be written in any one of
the following three ways:
\[%
\begin{array}
[c]{c}%
\left|  \,label_{1}\,\right\rangle \otimes\left|  \,label_{2}\,\right\rangle
\\
\\
\left|  \,label_{1}\,\right\rangle \left|  \,label_{2}\,\right\rangle \\
\\
\left|  \,label_{1}\,,\,label_{2}\,\right\rangle
\end{array}
\]
$\,$\medskip

\subsection{Quantum mechanics: Part I. The state of a quantum system}

\qquad\bigskip

The states of a quantum system $\mathcal{Q}$ are represented by state kets in
a Hilbert space $\mathcal{H}$. Two kets $\left|  \alpha\right\rangle $ and
$\left|  \beta\right\rangle $ represent the same state of a quantum system
$\mathcal{Q}$ if they differ by a non-zero multiplicative constant. In other
words, $\left|  \alpha\right\rangle $ and $\left|  \beta\right\rangle $
represent the same quantum state $\mathcal{Q}$ if there exists a non-zero
$\lambda\in\mathbb{C}$ such that
\[
\left|  \alpha\right\rangle =\lambda\left|  \beta\right\rangle
\]
Hence, quantum states are simply elements of the manifold
\[
\mathcal{H}/\symbol{126}=\mathbb{C}P^{n-1}%
\]
where $n$ denotes the dimension of $\mathcal{H}$, and $\mathbb{C}P^{n-1}$
denotes \textbf{complex projective }$\left(  n-1\right)  $\textbf{-space}
\index{Complex Projective Space@Complex Projective Space $\mathbb{C}P^{n-1}$}.\bigskip

\begin{description}
\item [Convention]Since a quantum mechanical state is represented by a state
ket up to a multiplicative constant, we will, unless stated otherwise, choose
those kets $\left|  \alpha\right\rangle $ which are of unit length, i.e., such
that
\[
\left\langle \alpha\mid\alpha\right\rangle =1\Longleftrightarrow\left\|
\,\left|  \alpha\right\rangle \right\|  =1
\]
\end{description}

\bigskip

\subsubsection{Polarized light: Part II. The quantum mechanical perspective\ }

\qquad\bigskip

As an illustration of the above concepts, we consider the polarization states
of a photon.

\bigskip

The polarization states of a photon are represented as state kets in a two
dimensional Hilbert space $\mathcal{H}$. One orthonormal basis of
$\mathcal{H}$ consists of the kets
\[
\left|  \circlearrowleft\right\rangle \text{ and }\left|  \circlearrowright
\right\rangle
\]
which represent respectively the quantum mechanical states of left- and
right-circularly polarized photons. Another orthonormal basis consists of the
kets
\[
\left|  \updownarrow\right\rangle \text{ and }\left|  \leftrightarrow
\right\rangle
\]
representing respectively vertically and horizontally linearly polarized
photons. And yet another orthonormal basis consists of the kets
\[
\left|  \nearrow\right\rangle \text{ and }\left|  \searrow\right\rangle
\]
for linearly polarized photons at the angles $\theta=\pi/4$ and $\theta
=-\pi/4$ off the vertical, respectively.\medskip\ 

These orthonormal bases are related as follows:
\begin{align*}
&  \left\{
\begin{array}
[c]{ccc}%
\left|  \nearrow\right\rangle  & = & \frac{1}{\sqrt{2}}\left(  \left|
\updownarrow\right\rangle +\left|  \leftrightarrow\right\rangle \right) \\
&  & \\
\left|  \searrow\right\rangle  & = & \frac{1}{\sqrt{2}}\left(  \left|
\updownarrow\right\rangle -\left|  \leftrightarrow\right\rangle \right)
\end{array}
\right.  \qquad\qquad\qquad\left\{
\begin{array}
[c]{ccc}%
\left|  \nearrow\right\rangle  & = & \frac{1+i}{2}\left|  \circlearrowright
\right\rangle +\frac{1-i}{2}\left|  \circlearrowleft\right\rangle \\
&  & \\
\left|  \searrow\right\rangle  & = & \frac{1-i}{2}\left|  \circlearrowright
\right\rangle +\frac{1+i}{2}\left|  \circlearrowleft\right\rangle
\end{array}
\right. \\
&  \left\{
\begin{array}
[c]{ccc}%
\left|  \updownarrow\right\rangle  & = & \frac{1}{\sqrt{2}}\left(  \left|
\nearrow\right\rangle +\left|  \searrow\right\rangle \right) \\
&  & \\
\left|  \leftrightarrow\right\rangle  & = & \frac{1}{\sqrt{2}}\left(  \left|
\nearrow\right\rangle -\left|  \searrow\right\rangle \right)
\end{array}
\right.  \qquad\qquad\qquad\left\{
\begin{array}
[c]{ccc}%
\left|  \updownarrow\right\rangle  & = & \frac{1}{\sqrt{2}}\left(  \left|
\circlearrowright\right\rangle +\left|  \circlearrowleft\right\rangle \right)
\\
&  & \\
\left|  \leftrightarrow\right\rangle  & = & \frac{i}{\sqrt{2}}\left(  \left|
\circlearrowright\right\rangle -\left|  \circlearrowleft\right\rangle \right)
\end{array}
\right. \\
&  \left\{
\begin{array}
[c]{ccc}%
\left|  \circlearrowright\right\rangle  & = & \frac{1}{\sqrt{2}}\left(
\left|  \updownarrow\right\rangle -i\left|  \leftrightarrow\right\rangle
\right) \\
&  & \\
\left|  \circlearrowleft\right\rangle  & = & \frac{1}{\sqrt{2}}\left(  \left|
\updownarrow\right\rangle +i\left|  \leftrightarrow\right\rangle \right)
\end{array}
\right.  \qquad\qquad\qquad\left\{
\begin{array}
[c]{ccc}%
\left|  \circlearrowright\right\rangle  & = & \frac{1-i}{2}\left|
\nearrow\right\rangle +\frac{1+i}{2}\left|  \searrow\right\rangle \\
&  & \\
\left|  \circlearrowleft\right\rangle  & = & \frac{1+i}{2}\left|
\nearrow\right\rangle +\frac{1-i}{2}\left|  \searrow\right\rangle
\end{array}
\right.
\end{align*}

\bigskip

The bracket products of the various polarization kets are given in the table
below:\
\[%
\begin{tabular}
[c]{|c||c|c||c|c||c|c|}\hline
& $\left|  \updownarrow\right\rangle $ & $\left|  \leftrightarrow\right\rangle
$ & $\left|  \nearrow\right\rangle $ & $\left|  \searrow\right\rangle $ &
$\left|  \circlearrowright\right\rangle $ & $\left|  \circlearrowleft
\right\rangle $\\\hline\hline
$\left\langle \updownarrow\right|  $ & $1$ & $0$ & $\frac{1}{\sqrt{2}}$ &
$\frac{1}{\sqrt{2}}$ & $\frac{1}{\sqrt{2}}$ & $\frac{1}{\sqrt{2}}$\\\hline
$\left\langle \leftrightarrow\right|  $ & $0$ & $1$ & $\frac{1}{\sqrt{2}}$ &
$-\frac{1}{\sqrt{2}}$ & $-\frac{i}{\sqrt{2}}$ & $\frac{i}{\sqrt{2}}%
$\\\hline\hline
$\left\langle \nearrow\right|  $ & $\frac{1}{\sqrt{2}}$ & $\frac{1}{\sqrt{2}}
$ & $1$ & $0$ & $\frac{1-i}{2}$ & $\frac{1+i}{2}$\\\hline
$\left\langle \searrow\right|  $ & $\frac{1}{\sqrt{2}}$ & $-\frac{1}{\sqrt{2}%
}$ & $0$ & $1$ & $\frac{1+i}{2}$ & $\frac{1-i}{2}$\\\hline\hline
$\left\langle \circlearrowright\right|  $ & $\frac{1}{\sqrt{2}}$ & $\frac
{i}{\sqrt{2}}$ & $\frac{1+i}{2}$ & $\frac{1-i}{2}$ & $1$ & $0$\\\hline
$\left\langle \circlearrowleft\right|  $ & $\frac{1}{\sqrt{2}}$ & $-\frac
{i}{\sqrt{2}}$ & $\frac{1-i}{2}$ & $\frac{1+i}{2}$ & $0$ & $1$\\\hline
\end{tabular}
\]
\bigskip

In terms of the basis $\left\{  \left|  \updownarrow\right\rangle ,\left|
\leftrightarrow\right\rangle \right\}  $ and the dual basis $\left\{
\left\langle \updownarrow\right|  ,\left\langle \leftrightarrow\right|
\right\}  $, these kets and bras can be written as matrices as indicated
below:
\begin{align*}
&  \left\{
\begin{array}
[c]{ccccccc}%
\left\langle \updownarrow\right|  & = & \left(
\begin{array}
[c]{cc}%
1 & 0
\end{array}
\right)  , & \qquad\qquad & \left|  \updownarrow\right\rangle  & = & \left(
\begin{array}
[c]{c}%
1\\
0
\end{array}
\right) \\
&  &  &  &  &  & \\
\left\langle \leftrightarrow\right|  & = & \left(
\begin{array}
[c]{cc}%
0 & 1
\end{array}
\right)  , &  & \left|  \leftrightarrow\right\rangle  & = & \left(
\begin{array}
[c]{c}%
0\\
1
\end{array}
\right)
\end{array}
\right. \\
&  \left\{
\begin{array}
[c]{ccccccc}%
\left\langle \nearrow\right|  & = & \frac{1}{\sqrt{2}}\left(
\begin{array}
[c]{cc}%
1 & 1
\end{array}
\right)  , &  & \left|  \nearrow\right\rangle  & = & \frac{1}{\sqrt{2}%
}\left(
\begin{array}
[c]{c}%
1\\
1
\end{array}
\right) \\
&  &  &  &  &  & \\
\left\langle \searrow\right|  & = & \frac{1}{\sqrt{2}}\left(
\begin{array}
[c]{cc}%
1 & -1
\end{array}
\right)  , &  & \left|  \searrow\right\rangle  & = & \frac{1}{\sqrt{2}%
}\left(
\begin{array}
[c]{r}%
1\\
-1
\end{array}
\right)
\end{array}
\right. \\
&  \left\{
\begin{array}
[c]{ccccccc}%
\left\langle \circlearrowright\right|  & = & \frac{1}{\sqrt{2}}\left(
\begin{array}
[c]{cc}%
1 & i
\end{array}
\right)  , &  & \left|  \circlearrowright\right\rangle  & = & \frac{1}%
{\sqrt{2}}\left(
\begin{array}
[c]{r}%
1\\
-i
\end{array}
\right) \\
&  &  &  &  &  & \\
\left\langle \circlearrowleft\right|  & = & \frac{1}{\sqrt{2}}\left(
\begin{array}
[c]{cc}%
1 & -i
\end{array}
\right)  , &  & \left|  \circlearrowleft\right\rangle  & = & \frac{1}{\sqrt
{2}}\left(
\begin{array}
[c]{c}%
1\\
i
\end{array}
\right)
\end{array}
\right.
\end{align*}
\bigskip\bigskip In this basis, for example, the tensor product $\left|
\nearrow\circlearrowright\right\rangle $ is
\[
\left|  \nearrow\circlearrowright\right\rangle =\left(
\begin{array}
[c]{c}%
\frac{1}{\sqrt{2}}\\
\frac{1}{\sqrt{2}}%
\end{array}
\right)  \otimes\left(
\begin{array}
[c]{r}%
\frac{1}{\sqrt{2}}\\
-\frac{i}{\sqrt{2}}%
\end{array}
\right)  =\frac{1}{2}\left(
\begin{array}
[c]{r}%
1\\
-i\\
1\\
-i
\end{array}
\right)
\]
and the projection operator $\left|  \circlearrowleft\right\rangle
\left\langle \circlearrowleft\right|  $ is:
\[
\left|  \circlearrowleft\right\rangle \left\langle \circlearrowleft\right|
=\frac{1}{\sqrt{2}}\left(
\begin{array}
[c]{c}%
1\\
i
\end{array}
\right)  \otimes\frac{1}{\sqrt{2}}\left(
\begin{array}
[c]{cc}%
1 & -i
\end{array}
\right)  =\frac{1}{2}\left(
\begin{array}
[c]{rr}%
1 & -i\\
i & 1
\end{array}
\right)
\]
\medskip\ 

\subsection{A Rosetta stone for Dirac notation:\ Part II. Operators}

\qquad\medskip\ 

An \textbf{(linear) operator} or \textbf{transformation} $\mathcal{O}$ on a
ket space $\mathcal{H}$ is a Hilbert space morphism of $\mathcal{H}$ into
$\mathcal{H}$, i.e., is an element of
\[
Hom_{\mathbb{C}}\left(  \mathcal{H},\mathcal{H}\right)
\]
\medskip\ 

The \textbf{adjoint}
\index{Adjoint} $\mathcal{O}^{\dagger}$ of an operator $\mathcal{O}$ is that
operator such that
\[
\left(  \mathcal{O}^{\dagger}\left|  \,label_{1}\,\right\rangle ,\left|
\,label_{2}\,\right\rangle \right)  =\left(  \left|  \,label_{1}%
\,\right\rangle ,\mathcal{O}\left|  \,label_{2}\,\right\rangle \right)
\]
for all kets $\left|  \,label_{1}\,\right\rangle $ and $\left|  \,label_{2}%
\,\right\rangle $.\bigskip\ 

In like manner, an (linear) operator or transformation on a bra space
$\mathcal{H}^{\ast}$ is an element of
\[
Hom_{\mathbb{C}}\left(  \mathcal{H}^{\ast},\mathcal{H}^{\ast}\right)
\]
Moreover, each operator $\mathcal{O}$ on $\mathcal{H}$ can be identified with
an operator, also denoted by $\mathcal{O}$, on $\mathcal{H}^{\ast}$ defined
by
\[
\left\langle \,label_{1}\,\right|  \longmapsto\left\langle \,label_{1}%
\,\right|  \mathcal{O}%
\]
where $\left\langle \,label_{1}\,\right|  \mathcal{O}$ is the bra defined by
\[
\left(  \left\langle \,label_{1}\,\right|  \mathcal{O}\right)  \left(  \left|
\,label_{2}\right\rangle \right)  =\left\langle \,label_{1}\,\right|  \left(
\mathcal{O}\left|  \,label_{2}\right\rangle \right)
\]
(This is sometimes called Dirac's associativity law.) Hence, the expression
\[
\left\langle \,label_{1}\,\right|  \mathcal{O}\left|  \,label_{2}%
\right\rangle
\]
is unambiguous.\bigskip\ 

\begin{remark}
Please note that
\[
\left(  \mathcal{O}\left|  \,label\right\rangle \right)  ^{\dagger
}=\left\langle label\right|  \mathcal{O}^{\dagger}%
\]
\end{remark}

\index{Dirac Notation|)}\bigskip

\subsection{Quantum mechanics: Part II. Observables}

\qquad\bigskip

In quantum mechanics, an \textbf{observable
\index{Observable}} is simply a \textbf{Hermitian
\index{Hermitian Operator}
\index{Operator!Hermitian}} (also called \textbf{self-adjoint})
\index{Operator!Self-Adjoint}
\index{Self-Adjoint Operator} operator on a Hilbert space $\mathcal{H}$, i.e.,
an operator $\mathcal{O}$ such that
\[
\mathcal{O}^{\dagger}=\mathcal{O}\text{ .}%
\]
\bigskip An \textbf{eigenvalue}
\index{Eigenvalue} $a$ of an operator $A$ is a complex number for which there
is a ket $\left|  label\right\rangle $ such that
\[
A\left|  label\right\rangle =a\left|  label\right\rangle \text{ .}%
\]
The ket $\left|  label\right\rangle $ is called an \textbf{eigenket}
\index{Eigenket} of $A$ corresponding to the eigenvalue $a$. \bigskip\ 

An important theorem about observables is given below:

\begin{theorem}
The eigenvalues $a_{i}$ of an observable $A$ are all real numbers. Moreover,
the eigenkets for distinct eigenvalues of an observable are
orthogonal.\bigskip\ 
\end{theorem}

\begin{definition}
An eigenvalue is \textbf{degenerate}
\index{Eigenvalue!Degenerate} if there are at least two linearly independent
eigenkets for that eigenvalue. Otherwise, it is \textbf{non-degenerate}
\index{Eigenvalue!Non-degenerate}.\medskip\ 
\end{definition}

\begin{description}
\item [Notational Convention]If all the eigenvalues $a_{i}$ of an observable
$A$ are nondegenerate, then we can and do label the eigenkets of $A$ with the
corresponding eigenvalues $a_{i}$. Thus, we can write:
\[
A\left|  a_{i}\right\rangle =a_{i}\left|  a_{i}\right\rangle
\]
for each eigenvalue $a_{i}$. \medskip

\item[Convention] In this paper, unless stated otherwise, we assume that the
eigenvalues of observables are non-degenerate.\medskip\ 
\end{description}

One notable exception to the above convention is the \textbf{measurement
operator
\index{Observable, Measurement}
\index{Operator!Measurement}}
\[
\left|  a_{i}\right\rangle \left\langle a_{i}\right|
\]
for the eigenvalue $a_{i}$, which is the outer product of ket $\left|
a_{i}\right\rangle $ with its adjoint the bra $\left\langle a_{i}\right|  $,
where we have assumed that $\left|  a_{i}\right\rangle $ (and hence,
$\left\langle a_{i}\right|  $) is of unit length. It has two eigenvalues $0$
and $1$. $1$ is a nondegenerate eigenvalue with eigenket $\left|
a_{i}\right\rangle $. $0$ is a degenerate eigenvalue with corresponding
eigenkets $\left\{  \,\left|  a_{j}\right\rangle \,\right\}  _{j\neq i}$ .

\bigskip

An observable $A$ is said to be \textbf{complete
\index{Observable!Complete}} if its eigenkets $\left|  a_{i}\right\rangle $
form a basis of the Hilbert space $\mathcal{H}$. Since by convention all the
eigenkets are chosen to be of unit length, it follows that the eigenkets of a
complete nondegenerate observable $A$ form an orthonormal basis of the
underlying Hilbert space. \ 

\bigskip

Moreover, given a complete nondegenerate observable $A$, every ket $\left|
\psi\right\rangle $ in $\mathcal{H}$ can be written as:
\[
\left|  \psi\right\rangle =\sum_{i}\left|  a_{i}\right\rangle \left\langle
a_{i}\mid\psi\right\rangle
\]
Thus, for a complete nondegenerate observable $A$, we have the following
operator equation which expresses the completeness of $A$,
\[
\sum_{i}\left|  a_{i}\right\rangle \left\langle a_{i}\right|  =1
\]
In this notation, we also have
\[
A=\sum_{i}a_{i}\left|  a_{i}\right\rangle \left\langle a_{i}\right|  \text{ ,}%
\]
where once again we have assumed that $\left|  a_{i}\right\rangle $ and
$\left\langle a_{i}\right|  $ are of unit length for all $i$.

\bigskip

\begin{example}
The Pauli spin matrices
\index{Pauli Spin Matrices}
\[
\sigma_{1}=\left(
\begin{array}
[c]{cc}%
0 & 1\\
1 & 0
\end{array}
\right)  \text{,\qquad}\sigma_{2}=\left(
\begin{array}
[c]{rr}%
0 & -i\\
i & 0
\end{array}
\right)  \text{,\qquad}\sigma_{3}=\left(
\begin{array}
[c]{rr}%
1 & 0\\
0 & -1
\end{array}
\right)
\]
are examples of observables that frequently appear in quantum mechanics and
quantum computation. \ Their eigenvalues and eigenkets are given in the
following table:
\[%
\begin{tabular}
[c]{||c||c||}\hline\hline
Pauli Matrices & Eigenvalue/Eigenket\\\hline\hline
$\sigma_{1}=\left(
\begin{array}
[c]{cc}%
0 & 1\\
1 & 0
\end{array}
\right)  $ &
\begin{tabular}
[c]{|l|l|}\hline
+1 & $\frac{\left|  0\right\rangle +\left|  1\right\rangle }{\sqrt{2}}%
=\frac{1}{\sqrt{2}}\left(
\begin{array}
[c]{c}%
1\\
1
\end{array}
\right)  $\\\hline
-1 & $\frac{\left|  0\right\rangle -\left|  1\right\rangle }{\sqrt{2}}%
=\frac{1}{\sqrt{2}}\left(
\begin{array}
[c]{r}%
1\\
-1
\end{array}
\right)  $\\\hline
\end{tabular}
\\\hline\hline
$\sigma_{2}=\left(
\begin{array}
[c]{rr}%
0 & -i\\
i & 0
\end{array}
\right)  $ &
\begin{tabular}
[c]{|l|l|}\hline
+1 & $\frac{\left|  0\right\rangle +i\left|  1\right\rangle }{\sqrt{2}}%
=\frac{1}{\sqrt{2}}\left(
\begin{array}
[c]{r}%
1\\
i
\end{array}
\right)  $\\\hline
-1 & $\frac{\left|  0\right\rangle -i\left|  1\right\rangle }{\sqrt{2}}%
=\frac{1}{\sqrt{2}}\left(
\begin{array}
[c]{r}%
1\\
-i
\end{array}
\right)  $\\\hline
\end{tabular}
\\\hline\hline
$\sigma_{3}=\left(
\begin{array}
[c]{rr}%
1 & 0\\
0 & -1
\end{array}
\right)  $ &
\begin{tabular}
[c]{|l|l|}\hline
+1 & $\left|  0\right\rangle =\left(
\begin{array}
[c]{r}%
1\\
0
\end{array}
\right)  $\\\hline
-1 & $\left|  1\right\rangle =\left(
\begin{array}
[c]{r}%
0\\
1
\end{array}
\right)  $\\\hline
\end{tabular}
\\\hline\hline
\end{tabular}
\]
\end{example}

\subsection{Quantum mechanics: Part III. Quantum measurement -- General principles}

\qquad\bigskip

In this section, $A$ will denote a complete nondegenerate observable with
eigenvalues $a_{i}$ and eigenkets $\left|  a_{i}\right\rangle $. \ We will, on
occasion, refer to $\left\{  \left|  a_{i}\right\rangle \right\}  $ as the
\textbf{frame} (or \textbf{the basis}) of the observable $A$. \ 

\index{Measurement}

According to quantum measurement theory, the \textbf{measurement of an
observable} $A$ of a quantum system $\mathcal{Q}$ in the state $\left|
\psi\right\rangle $ produces the eigenvalue $a_{i}$ as the measured result
with probability
\[
Prob\left(  \text{Value\quad}a_{i}\text{\quad is\quad observed}\right)
=\left\|  \left\langle a_{i}\mid\psi\right\rangle \right\|  ^{2}\text{ ,}%
\]
and forces the state of the quantum system $\mathcal{Q}$ into the state of the
corresponding eigenket $\left|  a_{i}\right\rangle $.\bigskip\ 

Since quantum measurement is such a hotly debated topic among physicists, we
(in self-defense) quote P.A.M. Dirac\cite{Dirac1}:

\begin{quote}
``A measurement always causes the (quantum mechanical) system to jump into an
eigenstate of the dynamical variable that is being measured.''\bigskip\ 
\end{quote}

Thus, the result of the above mentioned measurement of observable $A$ of a
quantum system $\mathcal{Q}$ which is in the state $\left|  \psi\right\rangle
$ before the measurement can be diagrammatically represented as follows:
\[
\fbox{$\left|  \psi\right\rangle =\sum_i\left|  a_i\right\rangle \left\langle
a_i\mid\psi\right\rangle \quad%
\begin{array}
[c]{c}%
\text{First}\\
\text{Meas. of }A\\
\Longrightarrow\\
Prob=\left\|  \left\langle a_{j}\mid\psi\right\rangle \right\|  ^{2}%
\end{array}
\quad a_j\left|  a_j\right\rangle \quad\approx\quad\left|  a_j\right\rangle
\quad%
\begin{array}
[c]{c}%
\text{Second}\\
\text{Meas. of }A\\
\Longrightarrow\\
Prob=1
\end{array}
\quad\left|  a_j\right\rangle $}%
\]
\ Please note that the measured value is the eigenvalue $a_{j}$ with
probability $\left\|  \ \left\langle \ a_{j}\mid\psi\ \right\rangle
\ \right\|  ^{2}$ . \ If the same measurement is repeated on the quantum
system $\mathcal{Q}$ after the first measurement, then the result of the
second measurement is no longer stochastic. \ It produces the previous
measured value $a_{j}$ and the state of $\mathcal{Q}$ remains the same, i.e.,
$\left|  a_{j}\right\rangle $ . \vspace*{0.3in}\ 

The observable
\[
\left|  a_{i}\right\rangle \left\langle a_{i}\right|
\]
is frequently called a \textbf{selective measurement operator
\index{Observable!Selective Measurement}
\index{Selective Measurement Operator}} (or a \textbf{filtration})
\index{Filtration} for $a_{i}$. As mentioned earlier, it has two eigenvalues
$0$ and $1$. $1$ is a nondegenerate eigenvalue with eigenket $\left|
a_{j}\right\rangle $, and $0$ is a degenerate eigenvalue with eigenkets
$\left\{  \left|  a_{j}\right\rangle \right\}  _{j\neq i}$.\bigskip\ 

Thus,
\[
\fbox{$\left|  \psi\right\rangle \quad%
\begin{array}
[c]{c}%
\text{ Meas. of }\left|  a_{i}\right\rangle \left\langle a_{i}\right| \\
\Longrightarrow\\
Prob=\left\|  \left\langle a_{i}\mid\psi\right\rangle \right\|  ^{2}%
\end{array}
\quad1\cdot\left|  a_i\right\rangle =\left|  a_i\right\rangle $ ,}%
\]
\medskip\ but for $j\neq i$,
\[
\fbox{$\left|  \psi\right\rangle \quad%
\begin{array}
[c]{c}%
\text{Meas. of }\left|  a_{i}\right\rangle \left\langle a_{i}\right| \\
\Longrightarrow\\
Prob=\left\|  \left\langle a_{j}\mid\psi\right\rangle \right\|  ^{2}%
\end{array}
\quad0\cdot\left|  a_j\right\rangle =0$}%
\]

\bigskip

The above description of quantum measurement is not the most general possible.
\ For the more advanced quantum measurement theory of \textbf{probabilistic
operator valued measures
\index{Probabilistic Operator Valued Measure}} (\textbf{POVM}s)
\index{POVM} (a.k.a., \textbf{positive operator valued measures}),
\index{Positive Operator Valued Measure} please refer to such books as for
example \cite{Helstrom1} and \cite{Peres1}.

\subsection{Polarized light: \ Part III. \ \ Three examples of quantum measurement}

\qquad\bigskip

We can now apply the above general principles of quantum measurement to
polarized light. Three examples are given below:\footnote{The last two
examples can easily be verified experimentally with at most three pair of
polarized sunglasses.}

\begin{example}%
\[
\fbox{$%
\begin{array}
[c]{c}%
\text{Rt. Circularly}\\
\text{polarized photon}\\
\qquad\\
\left|  \circlearrowleft\right\rangle =\frac{1}{\sqrt{2}}\left(  \left|
\updownarrow\right\rangle +i\left|  \leftrightarrow\right\rangle \right) \\
\qquad
\end{array}
\qquad\Longrightarrow\qquad%
\begin{array}
[c]{c}%
\text{Vertical}\\
\text{Polaroid}\\
\text{filter}\\%
{\includegraphics[
height=0.5604in,
width=0.3355in
]%
{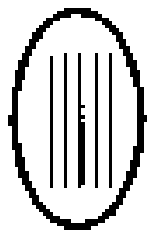}%
}%
\\
\text{Measurement op.}\\
\left|  \updownarrow\right\rangle \left\langle \updownarrow\right|
\end{array}
\qquad%
\begin{array}
[c]{ccc}%
&  & \text{Vertically}\\
&  & \text{polarized}\\
Prob=\frac{1}{2} &  & \text{photon}\\
\Longrightarrow &  & \left|  \updownarrow\right\rangle \\
&  & \\
&  & \\
\Longrightarrow &  & 0\\
Prob=\frac{1}{2} &  & \text{No photon}\\
&  & \\
&  &
\end{array}
$}%
\]
\bigskip\ 
\end{example}

\begin{example}
A vertically polarized filter followed by a horizontally polarized filter.
\end{example}%

\[
\fbox{$%
\begin{array}
[c]{c}%
\\
\text{photon}\\
\qquad\\
\alpha\left|  \updownarrow\right\rangle +\beta\left|  \leftrightarrow
\right\rangle \\
\qquad\\
\text{Normalized so that}\\
\left\|  \alpha\right\|  ^{2}+\left\|  \beta\right\|  ^{2}=1
\end{array}
\Longrightarrow%
\begin{array}
[c]{c}%
\text{Vert.}\\
\text{polar.}\\
\text{filter}\\%
{\includegraphics[
height=0.5604in,
width=0.3355in
]%
{vertical.ps}%
}%
\\
\qquad\\
\left|  \updownarrow\right\rangle \left\langle \updownarrow\right|
\end{array}
$.$%
\begin{array}
[c]{c}%
Prob=\left\|  \alpha\right\|  ^{2}\\
\qquad\\
\Longrightarrow
\end{array}%
\begin{array}
[c]{c}%
\text{Vert.}\\
\text{polar.}\\
\text{photon}\\
\qquad\\
\left|  \updownarrow\right\rangle
\end{array}
\Longrightarrow%
\begin{array}
[c]{c}%
\text{Horiz.}\\
\text{polar.}\\
\text{filter}\\%
{\includegraphics[
height=0.5526in,
width=0.3355in
]%
{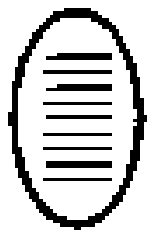}%
}%
\\
\qquad\\
\left|  \leftrightarrow\right\rangle \left\langle \leftrightarrow\right|
\end{array}
\quad%
\begin{array}
[c]{cc}%
& \text{No}\\
& \text{photon}\\
Prob=1 & \\
\Longrightarrow & 0\\
& \\
& \\
. &
\end{array}
$}%
\]
\bigskip\ 

\begin{example}
But if we insert a diagonally polarized filter (by $45^{o}$ off the vertical)
between the two polarized filters in the above example, we have:
\[
\fbox{$%
\begin{array}
[c]{c}%
\\
\\%
{\includegraphics[
height=0.5604in,
width=0.3355in
]%
{vertical.ps}%
}%
\\
\\
\left|  \updownarrow\right\rangle \left\langle \updownarrow\right|
\end{array}%
\begin{array}
[c]{c}%
\left\|  \alpha\right\|  ^{2}\\
\\
\Rightarrow
\end{array}
\left|  \updownarrow\right\rangle =\frac{1}{\sqrt{2}}\left(  \left|
\nearrow\right\rangle +\left|  \nwarrow\right\rangle \right)
\begin{array}
[c]{c}%
\\
\\%
{\includegraphics[
height=0.5613in,
width=0.3416in
]%
{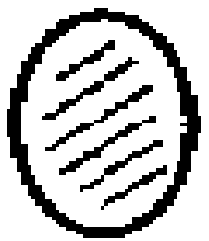}%
}%
\\
\\
\left|  \nearrow\right\rangle \left\langle \nearrow\right|
\end{array}%
\begin{array}
[c]{c}%
\frac{1}{2}\\
\\
\Rightarrow
\end{array}
\left|  \nearrow\right\rangle =\frac{1}{\sqrt{2}}\left(  \left|
\updownarrow\right\rangle +\left|  \leftrightarrow\right\rangle \right)
\begin{array}
[c]{c}%
\\
\\%
{\includegraphics[
height=0.5526in,
width=0.3355in
]%
{horizon.ps}%
}%
\\
\\
\left|  \leftrightarrow\right\rangle \left\langle \leftrightarrow\right|
\end{array}%
\begin{array}
[c]{c}%
\frac{1}{2}\\
\\
\Rightarrow
\end{array}
\left|  \leftrightarrow\right\rangle $}%
\]
\medskip\ 
\end{example}

where the input to the first filter is $\alpha\left|  \updownarrow
\right\rangle +\beta\left|  \leftrightarrow\right\rangle $. \ 

\bigskip

\index{Measurement}

\subsection{A Rosetta stone for Dirac notation: Part III. Expected values}

\qquad
\index{Dirac Notation|(}\bigskip

The \textbf{average value} (\textbf{expected value})
\index{Expected Value} of a measurement of an observable $A$ on a state
$\left|  \alpha\right\rangle $ is:
\[
\left\langle A\right\rangle =\left\langle \alpha\right|  A\left|
\alpha\right\rangle
\]
For, since
\[
\sum_{i}\left|  a_{i}\right\rangle \left\langle a_{i}\right|  =1\text{ ,}%
\]
we have%

\[
\left\langle A\right\rangle =\left\langle \alpha\right|  A\left|
\alpha\right\rangle =\left\langle \alpha\right|  \left(  \sum_{i}\left|
a_{i}\right\rangle \left\langle a_{i}\right|  \right)  A\left(  \sum
_{j}\left|  a_{j}\right\rangle \left\langle a_{j}\right|  \right)  \left|
\alpha\right\rangle =\sum_{i,j}\left\langle \alpha\mid a_{i}\right\rangle
\left\langle a_{i}\right|  A\left|  a_{j}\right\rangle \left\langle a_{j}%
\mid\alpha\right\rangle
\]
But on the other hand,
\[
\left\langle a_{i}\right|  A\left|  a_{j}\right\rangle =a_{j}\left\langle
a_{i}\mid a_{j}\right\rangle =a_{i}\delta_{ij}%
\]
Thus,
\[
\left\langle A\right\rangle =\sum_{i}\left\langle \alpha\mid a_{i}%
\right\rangle a_{i}\left\langle a_{i}\mid\alpha\right\rangle =\sum_{i}%
a_{i}\left\|  \left\langle a_{i}\mid\alpha\right\rangle \right\|  ^{2}%
\]
Hence, we have the standard expected value formula,
\[
\left\langle A\right\rangle =\sum_{i}a_{i}Prob\left(  \text{Observing }%
a_{i}\text{ on input }\left|  \alpha\right\rangle \right)  \text{ .}%
\]

\bigskip

\index{Dirac Notation}

\medskip

\subsection{Quantum Mechanics: Part IV. The Heisenberg uncertainty principle}

\qquad\bigskip

There
\index{Heisenberg!Uncertainty Principle} is, surprisingly enough, a limitation
of what we can observe in the quantum world.\bigskip

From classical probability theory, we know that one yardstick of uncertainty
is the \textbf{standard deviation},
\index{Standard Deviation} which measures the average fluctuation about the
mean. \ Thus, the \textbf{uncertainty} involved in the measurement of a
quantum observable $A$ is defined as the standard deviation of the observed
eigenvalues. \ This standard deviation is given by the expression
\[
Uncertainty(A)=\sqrt{\left\langle \left(  \triangle A\right)  ^{2}%
\right\rangle }%
\]
where
\[
\triangle A=A-\left\langle A\right\rangle
\]
\bigskip

Two observables $A$ and $B$ are said to be \textbf{compatible
\index{Observables!Compatible Operators}
\index{Operator!Compatible}} if they commute, i.e., if
\[
AB=BA\text{.}%
\]
Otherwise, they are said to be \textbf{incompatible}.
\index{Observable!Incompatible Operators}
\index{Operator!Incompatible}

\bigskip

Let $\left[  A,B\right]  $, called the \textbf{commutator}
\index{Commutator} of $A$ and $B$, denote the expression
\[
\left[  A,B\right]  =AB-BA
\]
In this notation, two operators $A$ and $B$ are compatible if and only if
$\left[  A,B\right]  =0$.

\bigskip\bigskip

The following principle is one expression of how quantum mechanics places
limits on what can be observed:

\bigskip\bigskip

\noindent\textbf{Heisenberg's Uncertainty Principle}\footnote{We have assumed
units have been chosen such that $\hslash=1$.}
\[
\left\langle \left(  \triangle A\right)  ^{2}\right\rangle \left\langle
\left(  \triangle B\right)  ^{2}\right\rangle \geq\frac{1}{4}\left|
\left\langle \left[  A,B\right]  \right\rangle \right|  ^{2}%
\]

\bigskip

Thus, if $A$ and $B$ are incompatible, i.e., do not commute, then, by
measuring $A$ more precisely, we are forced to measure $B$ less precisely, and
vice versa. \ We can not simultaneously measure both $A$ and $B$ to unlimited
precision. Measurement of $A$ somehow has an impact on the measurement of $B$,
and vice versa.\bigskip

\subsection{Quantum mechanics: Part V. Dynamics of closed quantum systems:
Unitary transformations, the Hamiltonian, and Schr\"{o}dinger's equation}

\qquad\bigskip

An operator $U$ on a Hilbert space $\mathcal{H}$ is \textbf{unitary
\index{Operator!Unitary}
\index{Unitary!Operator}
\index{Unitary!Transformation}} if
\[
U^{\dagger}=U^{-1}\text{ .}%
\]
Unitary operators are of central importance in quantum mechanics for many
reasons. We list below only two:

\begin{itemize}
\item  Closed quantum mechanical systems transform only via unitary transformations

\item  Unitary transformations preserve quantum probabilities
\end{itemize}

\bigskip

Let $\left|  \psi(t)\right\rangle $ denote the state as a function of time $t
$ of a closed quantum mechanical system $\mathcal{Q}$ . Then the dynamical
behavior of the state of $\mathcal{Q}$ is determined by the \textbf{Schr\"{o}%
dinger equation}
\[
i\hslash\frac{\partial}{\partial t}\left|  \psi(t)\right\rangle =H\left|
\psi(t)\right\rangle \text{ ,}%
\]
where $\hslash$ denotes \textbf{Planck's constant
\index{Planck's Constant}} divided by $2\pi$, and where $H$ denotes an
observable of $\mathcal{Q}$ called the \textbf{Hamiltonian}.
\index{Hamiltonian} The Hamiltonian is the quantum mechanical analog of the
Hamiltonian of classical mechanics. In classical physics, the Hamiltonian is
the total energy of the system. \ 

\bigskip

\subsection{\bigskip The mathematical perspective}

\qquad\bigskip

From the mathematical perspective, Schr\"{o}dinger's equation is written as:
\[
\frac{\partial}{\partial t}U(t)=-\frac{i}{\hslash}H(t)U(t)\text{ ,}%
\]
where
\[
\left|  \psi(t)\right\rangle =U\left|  \psi(0)\right\rangle \text{ ,}%
\]
and where $-\frac{i}{\hslash}H(t)$ is a skew-Hermitian operator lying in the
Lie algebra of the unitary group. \ The solution is given by a multiplicative
integral, called the \textbf{path-ordered integral},
\[
U(t)=\ _{\overset{\vspace{5pt}}{t}}%
\raisebox{-4pt}{\includegraphics[
trim=0.000000in -0.042019in -0.046161in 0.000000in,
height=22.875pt,
width=10.25pt
]%
{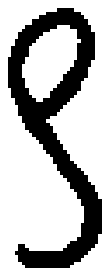}%
}%
_{\ 0}e^{-\frac{i}{\hslash}H(t)dt},
\]
which is taken over the path $-\frac{i}{\hslash}H(t)$ in the Lie algebra of
the unitary group. \ The path-ordered integral is given by:
\begin{align*}
\ _{\overset{\vspace{5pt}}{t}}%
\raisebox{-4pt}{\includegraphics[
trim=0.000000in -0.042019in -0.046161in 0.000000in,
height=22.875pt,
width=10.25pt
]%
{prodint.ps}%
}%
_{\ 0}e^{-\frac{i}{\hslash}H(t)dt}  &  =\lim_{n\rightarrow\infty}%
{\displaystyle\prod\limits_{k=n}^{0}}
e^{-\frac{i}{\hslash}H(k\frac{t}{n})\frac{t}{n}}\\
&  =\lim_{n\rightarrow\infty}\left[  e^{-\frac{i}{\hslash}H(n\cdot\frac{t}%
{n})}\cdot e^{-\frac{i}{\hslash}H((n-1)\cdot\frac{t}{n})}\cdot\ \cdots\ \cdot
e^{-\frac{i}{\hslash}H(1\cdot\frac{t}{n})}\cdot e^{-\frac{i}{\hslash}%
H(0\cdot\frac{t}{n})}\right]
\end{align*}
\bigskip

\bigskip

\index{Path-Ordered Integral}

\bigskip

\begin{remark}
The standard notation for the above path-ordered integral is
\[
\mathbf{P}\exp\left(  -\frac{i}{\hslash}%
{\displaystyle\int\limits_{0}^{t}}
H(t)dt\right)
\]
\end{remark}

\bigskip

If the Hamiltonian $H(t)=H$ is independent of time, then all matrices commute
and the above path-ordered integral simplifies to
\[
\ _{\overset{\vspace{5pt}}{t}}%
\raisebox{-4pt}{\includegraphics[
trim=0.000000in -0.042019in -0.046161in 0.000000in,
height=22.875pt,
width=10.25pt
]%
{prodint.ps}%
}%
_{\ 0}e^{-\frac{i}{\hslash}Hdt}=e^{\int_{0}^{t}-\frac{i}{\hslash}%
Hdt}=e^{-\frac{i}{\hslash}Ht}%
\]
Thus, in this case, $U(t)$ is nothing more than a one parameter subgroup of
the unitary group.

\bigskip

\index{Path-Ordered Integral}

\bigskip

\section{\textbf{The Density Operator}}

\index{Density Operator|(}

\subsection{Introducing the density operator}

\qquad\bigskip

John von Neumann suggested yet another way of representing the state of a
quantum system.

\bigskip

Let $\left|  \psi\right\rangle $ be a unit length ket (i.e., $\left\langle
\ \psi\mid\psi\ \right\rangle =1$) in the Hilbert space $\mathcal{H}$
representing the state of a quantum system\footnote{Please recall that each of
the kets in the set $\left\{  \ \lambda\left|  \psi\right\rangle \mid
\lambda\in\mathbb{C}\text{, }\lambda\neq0\ \right\}  $ represent the same
state of a quantum system. \ Hence, we can always (and usually do) represent
the state of a quantum system as a unit normal ket, i.e., as a ket such that
$\left\langle \ \psi\mid\psi\ \right\rangle =1$ .}. \ The \textbf{density
operator} $\rho$ associated with the state ket $\left|  \psi\right\rangle $ is
defined as the outer product of the ket $\left|  \psi\right\rangle $ (which
can be thought of as a column vector) with the bra $\left\langle \psi\right|
$ (which can be thought of as a row vector), i.e.,
\[
\rho=\left|  \psi\right\rangle \left\langle \psi\right|
\]

The density operator formalism has a number of advantages over the ket state
formalism. \ One advantage is that the density operator can also be used to
represent hybrid quantum/classical states, i.e., states which are a classical
statistical mixture of quantum states. \ Such hybrid states may also be
thought of as quantum states for which we have incomplete information.\bigskip

For example, consider a quantum system which is in the states (each of unit
length)
\[
\left|  \psi_{1}\right\rangle ,\left|  \psi_{2}\right\rangle ,\ldots,\left|
\psi_{n}\right\rangle
\]
with probabilities
\[
p_{1},p_{2},\ldots,p_{n}%
\]
respectively, where
\[
p_{1}+p_{2}+\ldots+p_{n}=1
\]
(Please note that the states $\left|  \psi_{1}\right\rangle ,\left|  \psi
_{2}\right\rangle ,\ldots,\left|  \psi_{n}\right\rangle $ need not be
orthogonal.) Then the density operator representation of this state is defined
as
\[
\rho=p_{1}\left|  \psi_{1}\right\rangle \left\langle \psi_{1}\right|
+p_{2}\left|  \psi_{2}\right\rangle \left\langle \psi_{2}\right|
+\ldots+p_{n}\left|  \psi_{n}\right\rangle \left\langle \psi_{n}\right|
\]

\bigskip\bigskip

If a density operator $\rho$ can be written in the form
\[
\rho=\left|  \psi\right\rangle \left\langle \psi\right|  \text{ ,}%
\]
it is said to represent a \textbf{pure ensemble
\index{Ensemble!Pure}}. \ Otherwise, it is said to represent a \textbf{mixed
ensemble
\index{Ensemble!Mixed}}.

\bigskip

\subsection{Properties of density operators}

\qquad\bigskip

It can be shown that all density operators are positive semi-definite
Hermitian operators of trace 1, and vice versa. \ As a result, we have the
following crisp mathematical definition:\bigskip

\begin{definition}
An linear operator on a Hilbert space $\mathcal{H}$ is a \textbf{density
operator} if it is a positive semi-definite Hermitian operator of trace 1.
\end{definition}

\bigskip

It can be shown that a density operator represents a pure ensemble if and only
if $\rho^{2}=\rho$, or equivalently, if and only if $Trace(\rho^{2})=1$. \ For
all ensembles, both pure and mixed, $Trace(\rho^{2})\leq1$.\bigskip

From standard theorems in linear algebra, we know that, for every density
operator $\rho$, there exists a unitary matrix $U$ which
\textbf{diagonalizes}
\index{Diagonalization} $\rho$, i.e., such that $U\rho U^{\dagger}$ is a
diagonal matrix. \ The diagonal entries in this matrix are, of course, the
eigenvalues of $\rho$. \ These are non-negative real numbers which all sum to 1.\bigskip

Finally, if we let $\mathcal{D}$ denote the set of all density operators for a
Hilbert space $\mathcal{H}$, then $i\mathcal{D}$ is a convex subset of the Lie
algebra of the unitary group associated with $\mathcal{H}$.

\bigskip

\subsection{Quantum measurement in terms of density operators}

\qquad

\bigskip

Let $\left\{  a_{i}\right\}  $ denote the set of distinct eigenvalues $a_{i}$
of an observable $A$. \ Let $P_{a_{i}}$ denote the projection operator that
projects the underlying Hilbert space onto the eigenspace determined by the
eigenvalue $a_{i}$. \ For example, if $a_{i}$ is a non-degenerate eigenvalue,
then
\[
P_{a_{i}}=\left|  a_{i}\right\rangle \left\langle a_{i}\right|
\]
Finally, let $\mathcal{Q}$ be a quantum system with state given by the density
operator $\rho$.

\bigskip

\index{Measurement}

\bigskip

If the quantum system $\mathcal{Q}$ is measured with respect to the observable
$A$, then with probability
\[
p_{i}=Trace\left(  P_{a_{i}}\rho\right)
\]
the resulting measured eigenvalue is $a_{i}$, and the resulting state of
$\mathcal{Q}$ is given by the density operator
\[
\rho_{i}=\frac{P_{a_{i}}\rho P_{a_{i}}}{Trace\left(  P_{a_{i}}\rho\right)
}\text{ .}%
\]

\bigskip

Moreover, for an observable $A$, the averaged observed eigenvalue expressed in
terms of the density operator is:
\[
\left\langle A\right\rangle =trace(\rho A)
\]
Thus, we have extended the definition of $\left\langle A\right\rangle $ so
that it applies to mixed as well as pure ensembles, i.e., generalized the
following formula to mixed ensembles:
\[
\left\langle A\right\rangle =\left\langle \psi\mid A\mid\psi\right\rangle
=trace\left(  \left|  \psi\right\rangle \left\langle \psi\right|  A\right)
=trace(\rho A)\text{ .}%
\]

\bigskip

\index{Measurement}

\bigskip

\subsection{Some examples of density operators}

\qquad\bigskip

For example, consider the following mixed ensemble
\index{Ensemble!Mixed} of the polarization state of a photon:

\begin{example}%
\[%
\begin{tabular}
[c]{|l||l|l|}\hline
Ket & $\overset{}{\underset{}{\left|  \updownarrow\right\rangle }}$ & $\left|
\nearrow\right\rangle $\\\hline
Prob. & $\overset{}{\underset{}{\frac{3}{4}}}$ & $\frac{1}{4}$\\\hline
\end{tabular}
\]
In terms of the basis $\left|  \leftrightarrow\right\rangle $, $\left|
\updownarrow\right\rangle $ of the two dimensional Hilbert space $\mathcal{H}%
$, the density operator $\rho$ of the above mixed ensemble can be written as:
\[%
\begin{array}
[c]{ccl}%
\rho & = & \frac{3}{4}\left|  \updownarrow\right\rangle \left\langle
\updownarrow\right|  +\frac{1}{4}\left|  \nearrow\right\rangle \left\langle
\nearrow\right| \\
&  & \\
& = & \frac{3}{4}\left(
\begin{array}
[c]{c}%
1\\
0
\end{array}
\right)  \left(
\begin{array}
[c]{cc}%
1 & 0
\end{array}
\right)  +\frac{1}{4}\left(
\begin{array}
[c]{c}%
1/\sqrt{2}\\
1/\sqrt{2}%
\end{array}
\right)  \left(
\begin{array}
[c]{cc}%
1/\sqrt{2} & 1/\sqrt{2}%
\end{array}
\right) \\
&  & \\
& = & \frac{3}{4}\left(
\begin{array}
[c]{cc}%
1 & 0\\
0 & 0
\end{array}
\right)  +\frac{1}{8}\left(
\begin{array}
[c]{cc}%
1 & 1\\
1 & 1
\end{array}
\right)  =\left(
\begin{array}
[c]{cc}%
\overset{}{\underset{}{\frac{7}{8}}} & \frac{1}{8}\\
\overset{}{\underset{}{\frac{1}{8}}} & \frac{1}{8}%
\end{array}
\right)
\end{array}
\]
\end{example}

\bigskip

\begin{example}
The following two \textbf{preparations} produce mixed ensembles
\index{Ensemble!Mixed} with the same density operator:
\[%
\begin{tabular}
[c]{|l||l|l|}\hline
Ket & $\overset{}{\underset{}{\left|  \updownarrow\right\rangle }}$ & $\left|
\leftrightarrow\right\rangle $\\\hline
Prob. & $\overset{}{\underset{}{\frac{1}{2}}}$ & $\frac{1}{2}$\\\hline
\end{tabular}
\qquad\text{and}\qquad%
\begin{tabular}
[c]{|l||l|l|}\hline
Ket & $\overset{}{\underset{}{\left|  \nearrow\right\rangle }}$ & $\left|
\nwarrow\right\rangle $\\\hline
Prob. & $\overset{}{\underset{}{\frac{1}{2}}}$ & $\frac{1}{2}$\\\hline
\end{tabular}
\]
\bigskip For, for the left preparation, we have
\[%
\begin{array}
[c]{ccl}%
\rho & = & \frac{1}{2}\left|  \updownarrow\right\rangle \left\langle
\updownarrow\right|  +\frac{1}{2}\left|  \leftrightarrow\right\rangle
\left\langle \leftrightarrow\right| \\
&  & \\
& = & \frac{1}{2}\left(
\begin{array}
[c]{c}%
1\\
0
\end{array}
\right)  \left(
\begin{array}
[c]{cc}%
1 & 0
\end{array}
\right)  +\frac{1}{2}\left(
\begin{array}
[c]{c}%
0\\
1
\end{array}
\right)  \left(
\begin{array}
[c]{cc}%
0 & 1
\end{array}
\right) \\
&  & \\
& = & \frac{1}{2}\left(
\begin{array}
[c]{cc}%
1 & 0\\
0 & 1
\end{array}
\right)
\end{array}
\]
\bigskip And for the right preparation, we have
\[%
\begin{array}
[c]{ccl}%
\rho & = & \frac{1}{2}\left|  \nearrow\right\rangle \left\langle
\nearrow\right|  +\frac{1}{2}\left|  \nwarrow\right\rangle \left\langle
\nwarrow\right| \\
&  & \\
& = & \frac{1}{2}\frac{1}{\sqrt{2}}\left(
\begin{array}
[c]{c}%
1\\
1
\end{array}
\right)  \frac{1}{\sqrt{2}}\left(
\begin{array}
[c]{cc}%
1 & 1
\end{array}
\right)  +\frac{1}{2}\frac{1}{\sqrt{2}}\left(
\begin{array}
[c]{r}%
1\\
-1
\end{array}
\right)  \frac{1}{\sqrt{2}}\left(
\begin{array}
[c]{cc}%
1 & -1
\end{array}
\right) \\
&  & \\
& = & \frac{1}{4}\left(
\begin{array}
[c]{cc}%
1 & 1\\
1 & 1
\end{array}
\right)  +\frac{1}{4}\left(
\begin{array}
[c]{rr}%
1 & -1\\
-1 & 1
\end{array}
\right)  =\frac{1}{2}\left(
\begin{array}
[c]{cc}%
1 & 0\\
0 & 1
\end{array}
\right)
\end{array}
\]
\bigskip

There is no way of physically distinquishing the above two mixed ensembles
\index{Ensemble!Mixed} which were prepared in two entirely different ways.
\ For the density operator represents all that can be known about the state of
the quantum system.
\end{example}

\bigskip

\subsection{The partial trace of a linear operator}

\qquad\bigskip\bigskip

In order to deal with a quantum system composed of many quantum subsystems, we
need to define the partial trace.

\bigskip

\index{Partial Trace}

Let
\[
\mathcal{O}:\mathcal{H}\longrightarrow\mathcal{H}\in Hom_{\mathbb{C}}\left(
\mathcal{H},\mathcal{H}\right)
\]
be a linear operator on the Hilbert space $\mathcal{H}$.

\bigskip\bigskip

Since Hilbert spaces are free algebraic objects, it follows from standard
results in abstract algebra\footnote{See for example \cite{Lang1}.} that
\[
Hom_{\mathbb{C}}\left(  \mathcal{H},\mathcal{H}\right)  \cong\mathcal{H}%
\otimes\mathcal{H}^{\ast}\text{ ,}%
\]
where we recall that
\[
\mathcal{H}^{\ast}=Hom_{\mathbb{C}}\left(  \mathcal{H},\mathbb{C}\right)
\text{ .}%
\]
\bigskip\bigskip

Hence, such an operator $\mathcal{O}$ can be written in the form
\[
\mathcal{O}=\sum_{\alpha}a_{\alpha}\left|  h_{\alpha}\right\rangle
\otimes\left\langle k_{\alpha}\right|  \text{ ,}%
\]
where the kets $\left|  h_{\alpha}\right\rangle $ lie in $\mathcal{H}$ and the
bras $\left\langle k_{\alpha}\right|  $ lie in $\mathcal{H}^{\dagger}%
$.\bigskip\bigskip\bigskip

Thus, the standard \textbf{trace
\index{Trace}} of a linear operator
\[
Trace:Hom_{\mathbb{C}}\left(  \mathcal{H},\mathcal{H}\right)  \longrightarrow
\mathbb{C}%
\]
is nothing more than a contraction, i.e.,
\[
Trace(\mathcal{O})=\sum_{\alpha}a_{\alpha}\left\langle \ k_{\alpha}\mid
h_{\alpha}\ \right\rangle \text{ ,}%
\]
\bigskip\bigskip i.e., a replacement of each outer product $\left|  h_{\alpha
}\right\rangle \otimes\left\langle k_{\alpha}\right|  $ by the corresponding
bracket $\left\langle \ k_{\alpha}\mid h_{\alpha}\ \right\rangle $.

We can generalize the $Trace$ as follows:\bigskip

Let $\mathcal{H}$ now be the tensor product of Hilbert spaces $\mathcal{H}%
_{1}$, $\mathcal{H}_{2}$, $\ldots$\ ,$\mathcal{H}_{n}$, i.e.,
\[
\mathcal{H}=%
{\displaystyle\bigotimes_{j=1}^{n}}
\mathcal{H}_{j}\text{ .}%
\]
Then it follows once again from standard results in abstract algebra that
\[
Hom_{\mathbb{C}}\left(  \mathcal{H},\mathcal{H}\right)  \cong%
{\displaystyle\bigotimes_{j=1}^{n}}
\left(  \mathcal{H}_{j}\otimes\mathcal{H}_{j}^{\ast}\right)  \text{ .}%
\]
Hence, the operator $\mathcal{O}$ can be written in the form
\[
\mathcal{O}=\sum_{\alpha}a_{\alpha}%
{\displaystyle\bigotimes_{j=1}^{n}}
\left|  h_{\alpha,j}\right\rangle \otimes\left\langle k_{\alpha,j}\right|
\text{ ,}%
\]
where, for each $j$, the kets $\left|  h_{\alpha,j}\right\rangle $ lie in
$\mathcal{H}_{j}$ and the bras $\left\langle k_{\alpha,j}\right|  $ lie in
$\mathcal{H}_{j}^{\ast}$ for all $\alpha$. \bigskip\bigskip

Next we note that for every subset $\mathcal{I}$ of the set of indices
$\mathcal{J}=\left\{  1,2,\ldots,n\right\}  $, we can define the
\textbf{partial trace} over $\mathcal{I}$, written
\[
Trace_{\mathcal{I}}:Hom_{\mathbb{C}}\left(  \bigotimes\limits_{j\in
\mathcal{J}}\mathcal{H}_{j},\bigotimes\limits_{j\in\mathcal{J}}\mathcal{H}%
_{j}\right)  \longrightarrow Hom_{\mathbb{C}}\left(  \bigotimes\limits_{j\in
\mathcal{J-I}}\mathcal{H}_{j},\bigotimes\limits_{j\in\mathcal{J-I}}%
\mathcal{H}_{j}\right)
\]
as the contraction on the indices $\mathcal{I}$, i.e.,
\[
Trace_{\mathcal{I}}\left(  \mathcal{O}\right)  =\sum_{\alpha}a_{\alpha
}\left(
{\displaystyle\prod\limits_{j\in\mathcal{I}}}
\left\langle \ k_{\alpha,j}\mid h_{\alpha,j}\ \right\rangle \right)
{\displaystyle\bigotimes\limits_{j\in\mathcal{J}-\mathcal{I}}}
\left|  \ h_{\alpha,j}\ \right\rangle \left\langle \ k_{\alpha,j}\ \right|
\text{ .}%
\]

\bigskip\bigskip

For example, let $\mathcal{H}_{1}$ and $\mathcal{H}_{0}$ be two dimensional
Hilbert spaces with selected orthonormal bases $\left\{  \left|
0_{1}\right\rangle ,\left|  1_{1}\right\rangle \right\}  $ and $\left\{
\left|  0_{0}\right\rangle ,\left|  1_{0}\right\rangle \right\}  $,
respectively. \ Thus, $\left\{  \left|  0_{1}0_{0}\right\rangle ,\left|
0_{1}1_{0}\right\rangle ,\left|  1_{1}0_{0}\right\rangle ,\left|  1_{1}%
1_{0}\right\rangle \right\}  $ is an orthonormal basis of $\mathcal{H}%
=\mathcal{H}_{1}\otimes\mathcal{H}_{0}$ .\bigskip\bigskip

Let $\rho\in Hom_{\mathbb{C}}\left(  \mathcal{H},\mathcal{H}\right)  $ be the
operator
\begin{align*}
\rho &  =\left(  \frac{\left|  0_{1}0_{0}\right\rangle -\left|  1_{1}%
1_{0}\right\rangle }{\sqrt{2}}\right)  \otimes\left(  \frac{\left\langle
0_{1}0_{0}\right|  -\left\langle 1_{1}1_{0}\right|  }{\sqrt{2}}\right) \\
&  =\frac{1}{2}\left(  \left|  0_{1}0_{0}\right\rangle \left\langle 0_{1}%
0_{0}\right|  -\left|  0_{1}0_{0}\right\rangle \left\langle 1_{1}1_{0}\right|
-\left|  1_{1}1_{0}\right\rangle \left\langle 0_{1}0_{0}\right|  +\left|
1_{1}1_{0}\right\rangle \left\langle 1_{1}1_{0}\right|  \right)
\end{align*}
which in terms of the basis $\left\{  \left|  0_{1}0_{0}\right\rangle ,\left|
0_{1}1_{0}\right\rangle ,\left|  1_{1}0_{0}\right\rangle ,\left|  1_{1}%
1_{0}\right\rangle \right\}  $ can be written as the matrix
\[
\rho=\frac{1}{2}\left(
\begin{array}
[c]{rrrr}%
1 & 0 & 0 & -1\\
0 & 0 & 0 & 0\\
0 & 0 & 0 & 0\\
-1 & 0 & 0 & 1
\end{array}
\right)  \text{ ,}%
\]
where the rows and columns are listed in the order $\left|  0_{1}%
0_{0}\right\rangle $, $\left|  0_{1}1_{0}\right\rangle $, $\left|  1_{1}%
0_{0}\right\rangle $, $\left|  1_{1}1_{0}\right\rangle $\ 

\bigskip\bigskip

The partial trace $Trace_{0}$ with respect to $\mathcal{I}=\left\{  0\right\}
$ of $\rho$ is
\begin{align*}
\rho_{1}  &  =Trace_{0}\left(  \rho\right) \\
&  =\frac{1}{2}Trace_{0}\left(  \left|  0_{1}0_{0}\right\rangle \left\langle
0_{1}0_{0}\right|  -\left|  0_{1}0_{0}\right\rangle \left\langle 1_{1}%
1_{0}\right|  -\left|  1_{1}1_{0}\right\rangle \left\langle 0_{1}0_{0}\right|
+\left|  1_{1}1_{0}\right\rangle \left\langle 1_{1}1_{0}\right|  \right) \\
&  =\frac{1}{2}\left(  \left\langle 0_{0}\!\mid\!0_{0}\right\rangle \left|
0_{1}\right\rangle \left\langle 0_{1}\right|  -\left\langle 1_{0}\!\mid
\!0_{0}\right\rangle \left|  0_{1}\right\rangle \left\langle 1_{1}\right|
-\left\langle 0_{0}\!\mid\!1_{0}\right\rangle \left|  1_{1}\right\rangle
\left\langle 0_{1}\right|  +\left\langle 1_{0}\!\mid\!1_{0}\right\rangle
\left|  1_{1}\right\rangle \left\langle 1_{1}\right|  \right) \\
&  =\frac{1}{2}\left(  \left|  0_{1}\right\rangle \left\langle 0_{1}\right|
-\left|  0_{1}\right\rangle \left\langle 1_{1}\right|  \right)
\end{align*}
which in terms of the basis $\left\{  \left|  0_{1}\right\rangle ,\left|
1_{1}\right\rangle \right\}  $ becomes
\[
\rho_{1}=Trace_{0}(\rho)=\frac{1}{2}\left(
\begin{array}
[c]{cc}%
1 & 0\\
0 & 1
\end{array}
\right)  \text{ ,}%
\]
where the rows and columns are listed in the order $\left|  0_{1}\right\rangle
$, $\left|  1_{1}\right\rangle $ .

\index{Partial Trace}

\subsection{Multipartite quantum systems}

\qquad\bigskip

One advantage density operators have over kets is that they provide us with a
means for dealing with multipartite quantum systems.
\index{Multipartite Quantum System}\bigskip

\begin{definition}
Let $\mathcal{Q}_{1}$, $\mathcal{Q}_{2}$, $\ldots$ , $\mathcal{Q}_{n}$ be
quantum systems with underlying Hilbert spaces $\mathcal{H}_{1}$,
$\mathcal{H}_{2}$, $\ldots$ , $\mathcal{H}_{n}$, respectively. \ The global
quantum system
\index{Global Quantum System} $\mathcal{Q}$ consisting of the quantum systems
$\mathcal{Q}_{1}$, $\mathcal{Q}_{2}$, $\ldots$ , $\mathcal{Q}_{n}$ is called a
\textbf{multipartite quantum system}. \ Each of the quantum systems
$\mathcal{Q}_{j}$ ($j=1,2,\ \ldots\ ,n$) is called a \textbf{constituent
``part''}
\index{Constituent Part@Constituent ``Part''} of $\mathcal{Q}$ . The
underlying Hilbert space $\mathcal{H}$ of $\mathcal{Q}$ is the tensor product
of the Hilbert spaces of the constituent ``parts,'' i.e.,
\[
\mathcal{H}=\bigotimes_{j=1}^{n}\mathcal{H}_{j}\text{ .}%
\]
\end{definition}

\bigskip If the density operator $\rho$ is the state of a multipartite quantum
system $\mathcal{Q}$, then the state of each constituent ``part''
$\mathcal{Q}_{j}$ is the density operator $\rho_{j}$ given by the partial
trace
\[
\rho_{j}=Trace_{\mathcal{J}-\left\{  j\right\}  }\left(  \rho\right)  \text{
,}%
\]
where $\mathcal{J}=\left\{  1,2,\ \ldots\ ,n\right\}  $ is the set of indices.\bigskip

Obviously, much more can be said about the states of multipartite systems and
their constituent parts. \ However, we will forego that discussion until after
we have had an opportunity introduce the concepts of quantum entanglement and
von Neumann entropy. \ 

\subsection{Quantum dynamics in density operator formalism}

\qquad\bigskip

Under a unitary transformation $U$, a density operator $\rho$ transforms
according to the rubric:
\[
\rho\longmapsto U\rho U^{\dagger}%
\]
Moreover, in terms of the density operator, Schr\"{o}dinger's
equation\footnote{Schr\"{o}dinger's equation determines the dynamics of closed
quantum systems. \ However, non-closed quantum systems are also of importance
in quantum computation and quantum information theory. \ See for example the
Schumacher's work on superoperators, e.g., \cite{Schumacher1}.} becomes:
\[
i\hslash\frac{\partial\rho}{\partial t}=\left[  H,\rho\right]  \text{ ,}%
\]
where $\left[  H,\rho\right]  $ denotes the \textbf{commutator}
\index{Commutator} of $H$ and $\rho$, i.e.,
\[
\left[  H,\rho\right]  =H\rho-\rho H
\]

\subsection{\bigskip The mathematical perspective}

\qquad\bigskip

From the mathematical perspective, one works with $i\rho$ instead of $\rho$
because $i\rho$ lies in the Lie algebra of the unitary group. Thus, the
density operator transforms under a unitary transformation $U$ according to
the rubric:
\[
i\rho\longmapsto Ad_{U}(i\rho)\text{ ,}%
\]
where $Ad_{U}$ denotes the \textbf{big adjoint
\index{Adjoint!Big} representation}.

\bigskip

From the mathematical perspective, Schr\"{o}dinger's equation is in this case
more informatively written as:
\[
\frac{\partial(i\rho)}{\partial t}=-\frac{1}{\hslash}ad_{iH}(i\rho)\text{ ,}%
\]
where $ad_{-\frac{i}{\hslash}H}$ denotes the \textbf{little adjoint
representation
\index{Adjoint!Little}}. Thus, the solution to the above form of
Schr\"{o}dinger's equation is given by the path ordered integral:
\[
\rho=\left(  \ _{\overset{\vspace{5pt}}{t}}%
\raisebox{-4pt}{\includegraphics[
trim=0.000000in -0.042019in -0.046161in 0.000000in,
height=22.875pt,
width=10.25pt
]%
{prodint.ps}%
}%
_{\ 0}e^{-\frac{1}{\hslash}\left(  ad_{iH(t)}\right)  dt}\right)  \rho_{0}%
\]
where $\rho_{0}$ denotes the density operator at time $t=0$.

\index{Density Operator|)}\bigskip

\section{\bigskip\textbf{The Heisenberg model of quantum mechanics}}

\qquad\bigskip

Consider a computing device with inputs and outputs for which we have no
knowledge of the internal workings of the device. \ We are allowed to probe
the device with inputs and observe the corresponding outputs. \ But we are
given no information as to how the device performs its calculation. \ We call
such a device a \textbf{blackbox} computing device.

\index{Heisenberg, Picture}

For such blackboxes, we say that two theoretical models for blackboxes are
\textbf{equivalent} provided both predict the same input/output behavior. \ We
may prefer one model over the other for various reasons, such as simplicity,
aesthetics, or whatever meets our fancy. \ But the basic fact is that each of
the two equivalent models is just as ``correct'' as the other.

\bigskip

In like manner, two theoretical models of the quantum world are said to be
\textbf{equivalent} if they both predict the same results in regard to quantum measurements.

\bigskip

Up to this point, we have been describing the Schr\"{o}dinger model of quantum
mechanics. \ However, shortly after Schr\"{o}dinger proposed his model for the
quantum world, called the \textbf{Schr\"{o}dinger picture}, Heisenberg
proposed yet another, called the \textbf{Heisenberg picture}. \ Both models
were later proven to be equivalent.

\index{Schrodinger Picture}

\bigskip

In the Heisenberg picture, state kets remain stationary with time, but
observables move with time. \ While state kets, and hence density operators,
remain fixed with respect to time, the observables $A$ change dynamically as:
\[
A\longmapsto U^{\dagger}AU
\]
under a unitary transformation $U=U(t)$, where the unitary transformation is
determined by the equation \
\[
i\hslash\frac{\partial U}{\partial t}=HU
\]
\ It follows that the equation of motion of observables is according to the
following equation
\[
i\hslash\frac{\partial A}{\partial t}=\left[  A,H\right]
\]
One advantage the Heisenberg picture has over the Schr\"{o}dinger picture is
that the equations appearing in it are similar to those found in classical mechanics.

\bigskip\pagebreak 

In summary, we have the following table which contrasts the two pictures:

\bigskip%
\[%
\begin{tabular}
[c]{|l|c|c|}\hline\hline
\multicolumn{1}{||l|}{} & \multicolumn{1}{||c|}{\ Schr\"{o}dinger Picture} &
\multicolumn{1}{||c||}{\ Heisenberg Picture}\\\hline\hline
State ket &
\begin{tabular}
[c]{c}%
Moving\\
\multicolumn{1}{l}{$\overset{\mathstrut}{\underset{\mathstrut}{\left|
\psi_{0}\right\rangle \longmapsto\left|  \psi\right\rangle =U\left|  \psi
_{0}\right\rangle }}$}%
\end{tabular}
&
\begin{tabular}
[c]{l}%
Stationary\\
\multicolumn{1}{c}{$\overset{\mathstrut}{\underset{\mathstrut}{\left|
\psi_{0}\right\rangle }}$}%
\end{tabular}
\\\hline%
\begin{tabular}
[c]{l}%
Density\\
Operator
\end{tabular}
&
\begin{tabular}
[c]{c}%
Moving\\
\multicolumn{1}{l}{$\overset{\mathstrut}{\underset{\mathstrut}{\rho
_{0}\longmapsto\rho=U\rho_{0}U^{\dagger}=Ad_{U}\left(  \rho_{0}\right)  }}$}%
\end{tabular}
&
\begin{tabular}
[c]{l}%
Stationary\\
\multicolumn{1}{c}{$\overset{\mathstrut}{\underset{\mathstrut}{\rho_{0}}}$}%
\end{tabular}
\\\hline
Observable &
\begin{tabular}
[c]{l}%
Stationary\\
\multicolumn{1}{c}{$\overset{\mathstrut}{\underset{\mathstrut}{A_{0}}}$}%
\end{tabular}
&
\begin{tabular}
[c]{c}%
Moving\\
\multicolumn{1}{l}{$\overset{\mathstrut}{\underset{\mathstrut}{A_{0}%
\longmapsto A=U^{\dagger}A_{0}U=Ad_{U^{\dagger}}\left(  A_{0}\right)  }}$}%
\end{tabular}
\\\hline%
\begin{tabular}
[c]{l}%
Observable\\
Eigenvalues
\end{tabular}
&
\begin{tabular}
[c]{l}%
Stationary\\
\multicolumn{1}{c}{$\overset{\mathstrut}{\underset{\mathstrut}{a_{j}}}$}%
\end{tabular}
&
\begin{tabular}
[c]{l}%
Stationary\\
\multicolumn{1}{c}{$\overset{\mathstrut}{\underset{\mathstrut}{a_{j}}}$}%
\end{tabular}
\\\hline%
\begin{tabular}
[c]{l}%
Observable\\
Frame
\end{tabular}
&
\begin{tabular}
[c]{c}%
Stationary\\
\multicolumn{1}{l}{$\overset{\mathstrut}{\underset{\mathstrut}{A_{0}=\sum
_{j}a_{j}\left|  a_{j}\right\rangle _{0}\left\langle a_{j}\right|  _{0}}}$}%
\end{tabular}
&
\begin{tabular}
[c]{c}%
Moving\\
$\overset{\mathstrut}{\underset{\mathstrut}{A_{0}=\sum_{j}a_{j}\left|
a_{j}\right\rangle _{0}\left\langle a_{j}\right|  _{0}}}$\\
$\longmapsto$\\
$\overset{\mathstrut}{\underset{\mathstrut}{A_{t}=\sum_{j}a_{j}\left|
a_{j}\right\rangle _{t}\left\langle a_{j}\right|  _{t}}}$\\
where $\overset{\mathstrut}{\underset{\mathstrut}{\left|  a_{j}\right\rangle
_{t}=U^{\dagger}\left|  a_{j}\right\rangle _{0}}}$%
\end{tabular}
\\\hline%
\begin{tabular}
[c]{l}%
Dynamical\\
Equations
\end{tabular}
&
\begin{tabular}
[c]{c}%
$\overset{\mathstrut}{\underset{\mathstrut}{i\hslash\frac{\partial U}{\partial
t}=H^{(S)}U}}$\\
\multicolumn{1}{l}{$\overset{\mathstrut}{\underset{\mathstrut}{i\hslash
\frac{\partial}{\partial t}\left|  \psi\right\rangle =H^{(S)}\left|
\psi\right\rangle }}$}%
\end{tabular}
&
\begin{tabular}
[c]{c}%
$\overset{\mathstrut}{\underset{\mathstrut}{i\hslash\frac{\partial U}{\partial
t}=H^{(H)}U}}$\\
\multicolumn{1}{l}{$\overset{\mathstrut}{\underset{\mathstrut}{i\hslash
\frac{\partial A}{\partial t}=\left[  A,H^{(H)}\right]  }}$}%
\end{tabular}
\\\hline
Measurement &
\begin{tabular}
[c]{l}%
Measurement of observable $A_{0}$\\
produces eigenvalue $a_{j}$ with\\
probability\\
\multicolumn{1}{c}{$\overset{\mathstrut}{\underset{\mathstrut}{\left|  \left(
\left\langle a_{j}\right|  _{0}\right)  \left|  \psi\right\rangle \right|
^{2}=\left|  \left(  \left\langle a_{j}\right|  _{0}\right)  \left|
\psi\right\rangle \right|  ^{2}}}$}%
\end{tabular}
&
\begin{tabular}
[c]{l}%
Measurement of observable $A$\\
produces eigenvalue $a_{j}$ with\\
probability\\
\multicolumn{1}{c}{$\overset{\mathstrut}{\underset{\mathstrut}{\left|  \left(
\left\langle a_{j}\right|  _{t}\right)  \left|  \psi_{0}\right\rangle \right|
^{2}=\left|  \left(  \left\langle a_{j}\right|  _{0}\right)  \left|
\psi\right\rangle \right|  ^{2}}}$}%
\end{tabular}
\\\hline
\end{tabular}
\]
where
\[
H^{(H)}=U^{\dagger}H^{(S)}U
\]
It follows that the Schr\"{o}dinger Hamiltonian $H^{(S)}$ and the Heisenberg
Hamiltonian are related as follows:
\[
\frac{\partial H^{(S)}}{\partial t}=U\frac{\partial H^{(H)}}{\partial
t}U^{\dagger}\text{,}%
\]
where terms containing $\frac{\partial U}{\partial t}$ and $\frac{\partial
U^{\dagger}}{\partial t}$ have cancelled out as a result of the Schr\"{o}%
dinger equation.

We should also mention that the Schr\"{o}dinger and Heisenberg pictures can be
transformed into one another via the mappings:
\[%
\begin{tabular}
[c]{|c||c|}\hline\hline
\multicolumn{1}{||c|}{$S\longrightarrow H$} &
\multicolumn{1}{||c||}{$H\longrightarrow S$}\\\hline\hline
$\overset{\mathstrut}{\underset{\mathstrut}{\quad\left|  \psi^{(S)}%
\right\rangle \longmapsto\left|  \psi^{(H)}\right\rangle =U^{\dagger}\left|
\psi^{(S)}\right\rangle \quad}}$ & $\quad\left|  \psi^{(H)}\right\rangle
\longmapsto\left|  \psi^{(S)}\right\rangle =U\left|  \psi^{(H)}\right\rangle
\quad$\\\hline
$\overset{\mathstrut}{\underset{\mathstrut}{\rho^{(S)}\longmapsto\rho
^{(H)}=U^{\dagger}\rho^{(S)}U}}$ & $\rho^{(H)}\longmapsto\rho^{(S)}%
=U\rho^{(H)}U^{\dagger}$\\\hline
$\overset{\mathstrut}{\underset{\mathstrut}{A^{(S)}\longmapsto A^{(H)}%
=U^{\dagger}A^{(S)}U}}$ & $A^{(H)}\longmapsto A^{(S)}=UA^{(H)}U^{\dagger}%
$\\\hline
$\overset{\mathstrut}{\underset{\mathstrut}{A^{(S)}\longmapsto A^{(H)}%
=U^{\dagger}A^{(S)}U}}$ & $A^{(H)}\longmapsto A^{(S)}=UA^{(H)}U^{\dagger}%
$\\\hline
\end{tabular}
\]
\bigskip Obviously, much more could be said on this topic.

\bigskip

For quantum computation from the perspective of the Heisenberg model, please
refer to the work of Deutsch
\index{Deutsch} and Hayden\cite{Deutsch2}, and also to Gottesman's ``study of
the ancient Hittites'' :-) \cite{Gottesman1}.
\index{Gottesman}
\index{Heisenberg, Picture|)}

\section{\textbf{Quantum entanglement}}

\bigskip

\subsection{The juxtaposition of two quantum systems}

\qquad\bigskip

Let $\mathcal{Q}_{1}$ and $\mathcal{Q}_{2}$ be two quantum systems that have
been separately prepared respectively in states $\left|  \psi_{1}\right\rangle
$ and $\left|  \psi_{2}\right\rangle $, and that then have been united without
interacting. \ Because $\mathcal{Q}_{1}$ and $\mathcal{Q}_{2}$ have been
separately prepared without interacting, their states $\left|  \psi
_{1}\right\rangle $ and $\left|  \psi_{2}\right\rangle $ respectively lie in
distinct Hilbert spaces $\mathcal{H}_{1}$ and $\mathcal{H}_{2}$. \ Moreover,
because of the way in which $\mathcal{Q}_{1}$ and $\mathcal{Q}_{2}$ have been
prepared, all physical predictions relating to one of these quantum systems do
not depend in any way whatsoever on the other quantum system.

\index{Entangled, Quantum}
\index{Quantum Entangled}

\bigskip

The global quantum system $\mathcal{Q}$ consisting of the two quantum systems
$\mathcal{Q}_{1}$ and $\mathcal{Q}_{2}$ as prepared above is called a
\textbf{juxtaposition}
\index{Juxtaposition} of the quantum systems $\mathcal{Q}_{1}$ and
$\mathcal{Q}_{2}$. \ The state of the global quantum system $\mathcal{Q}$ is
the tensor product of the states $\left|  \psi_{1}\right\rangle $ and $\left|
\psi_{2}\right\rangle $. \ In other words, the state of $\mathcal{Q}$ is:
\[
\left|  \psi_{1}\right\rangle \otimes\left|  \psi_{2}\right\rangle
\in\mathcal{H}_{1}\otimes\mathcal{H}_{2}%
\]

\bigskip

\subsection{An example: An $n$-qubit register $\mathcal{Q}$ consisting of the
juxtaposition of $n$ qubits.}

\qquad\bigskip

\index{Quantum Register}

Let $\mathcal{H}$ be a two dimensional Hilbert space, and let $\left\{
\left|  0\right\rangle ,\left|  1\right\rangle \right\}  $ denote an
arbitrarily selected orthonormal basis\footnote{We obviously have chosen to
label the basis elements in a suggestive way.}. \ Let $\mathcal{H}_{n-1}$,
$\mathcal{H}_{n-2}$, $\ldots$ ,$\mathcal{H}_{0}$ be \ distinct Hilbert spaces,
each isomorphic to $\mathcal{H}$, with the obvious induced orthonormal bases \ %

\[
\left\{  \ \left|  0_{n-1}\right\rangle ,\ \left|  1_{n-1}\right\rangle
\ \right\}  ,\ \left\{  \ \left|  0_{n-2}\right\rangle ,\ \left|
1_{n-2}\right\rangle \ \right\}  ,\ \ldots\ ,\ \left\{  \ \left|
0_{0}\right\rangle ,\ \left|  1_{0}\right\rangle \ \right\}
\]
respectively.\bigskip

Consider $n$ qubits $\mathcal{Q}_{n-1}$, $\mathcal{Q}_{n-2}$, $\ldots$ ,
$\mathcal{Q}_{0}$ separately prepared in the states
\begin{align*}
&  \frac{1}{\sqrt{2}}\left(  \left|  0_{n-1}\right\rangle +\left|
1_{n-1}\right\rangle \right)  \text{, }\frac{1}{\sqrt{2}}\left(  \left|
0_{n-2}\right\rangle +\left|  1_{n-2}\right\rangle \right)  \text{, }%
\ldots\text{ , }\frac{1}{\sqrt{2}}\left(  \left|  0_{0}\right\rangle +\left|
1_{0}\right\rangle \right)  \text{, }\\
&  \quad\strut
\end{align*}
respectively. \ Let $\mathcal{Q}$ denote the global system consisting of the
separately prepared (without interacting) qubits $\mathcal{Q}_{n-1}$,
$\mathcal{Q}_{n-2}$, $\ldots$ , $\mathcal{Q}_{0}$. \ Then the state $\left|
\psi\right\rangle $ of $\mathcal{Q}$ is:
\begin{align*}
\left|  \psi\right\rangle  &  =\frac{1}{\sqrt{2}}\left(  \left|
0_{n-1}\right\rangle +\left|  1_{n-1}\right\rangle \right)  \otimes\frac
{1}{\sqrt{2}}\left(  \left|  0_{n-2}\right\rangle +\left|  1_{n-2}%
\right\rangle \right)  \otimes\ldots\otimes\frac{1}{\sqrt{2}}\left(  \left|
0_{0}\right\rangle +\left|  1_{0}\right\rangle \right) \\
&  =\left(  \frac{1}{\sqrt{2}}\right)  ^{n}\left(  \left|  0_{n-1}%
0_{n-2}\ldots0_{1}0_{0}\right\rangle +\left|  0_{n-1}0_{n-2}\ldots0_{1}%
1_{0}\right\rangle +\ \ldots\ +\left|  1_{n-1}1_{n-2}\ldots1_{1}%
1_{0}\right\rangle \right) \\
&  \strut
\end{align*}
which lies in the Hilbert space
\[
\mathcal{H}=\mathcal{H}_{n-1}\otimes\mathcal{H}_{n-2}\otimes\ \ldots
\ \otimes\mathcal{H}_{0}.
\]

\bigskip

\begin{description}
\item [Notational Convention]We will usually omit subscripts whenever they can
easily be inferred from context.
\end{description}

\bigskip

Thus, the global system $\mathcal{Q}$ consisting of the $n$ qubits
$\mathcal{Q}_{n-1}$, $\mathcal{Q}_{n-2}$,$\ \ldots\ $, $\mathcal{Q}_{0}$ is in
the state
\[
\left|  \psi\right\rangle =\left(  \frac{1}{\sqrt{2}}\right)  ^{n}\left(
\left|  00\ldots00\right\rangle +\left|  00\ldots01\right\rangle
+\ \ldots\ +\left|  11\ldots11\right\rangle \right)  \in%
{\displaystyle\bigotimes\limits_{0}^{n-1}}
\mathcal{H}%
\]

\bigskip

The reader should note that the $n$-qubit register $\mathcal{Q}$ is a
superposition
\index{Superposition} of kets with labels consisting of all the binary
n-tuples. \ If each binary n-tuple $b_{n-1}b_{n-2}\ldots b_{0}$ is identified
with the integer
\[
b_{n-1}2^{n-1}+b_{n-2}2^{n-2}+\ldots+b_{0}2^{0}\text{ ,}%
\]
i.e., if we interpret each binary n-tuple as the radix 2 representation of an
integer, then we can rewrite the state as
\[
\left|  \psi\right\rangle =\left(  \frac{1}{\sqrt{2}}\right)  ^{n}\left(
\left|  0\right\rangle +\left|  1\right\rangle +\left|  2\right\rangle
+\ \ldots\ +\left|  2^{n}-1\right\rangle \right)  \text{.}%
\]
In other words, this n-qubit register contains all the integers from $0$ to
$2^{n}-1$ in superposition.
\index{Superposition} \ But most importantly, it contains all the integers $0
$ to $2^{n}-1$ \textit{simultaneously}!

\bigskip

This is an example of the massive parallelism that is possible within quantum
computation. \ However, there is a downside. \ If we observe (measure) the
register, then all the massive parallelism disappears. \ On measurement, the
quantum world selects for us one and only one of the $2^{n}$ integers. \ The
\ probability of observing any particular one of the integers is $\left|
\left(  1/\sqrt{2}\right)  ^{n}\right|  ^{2}=(\frac{1}{2})^{n}$. \ The
selection of which integer is observed is unfortunately not made by us, but by
the quantum world.

\index{Quantum Register}

\bigskip

Thus, harnessing the massive parallelism of quantum mechanics is no easy task!
\ As we will see, a more subtle approach is required.\bigskip

\subsection{An example of the dynamic behavior of a 2-qubit register}

\qquad\bigskip

We now consider the previous $n$-qubit register for $n=2$. \ In terms of the
bases described in the previous section, we have:
\[
\left\{
\begin{array}
[c]{ccccccc}%
\left|  0\right\rangle  & = & \left|  00\right\rangle  & = & \left(
\begin{array}
[c]{c}%
1\\
0
\end{array}
\right)  \otimes\left(
\begin{array}
[c]{c}%
1\\
0
\end{array}
\right)  & = & \left(
\begin{array}
[c]{c}%
1\\
0\\
0\\
0
\end{array}
\right) \\
&  &  &  &  &  & \\
\left|  1\right\rangle  & = & \left|  01\right\rangle  & = & \left(
\begin{array}
[c]{c}%
1\\
0
\end{array}
\right)  \otimes\left(
\begin{array}
[c]{c}%
0\\
1
\end{array}
\right)  & = & \left(
\begin{array}
[c]{c}%
0\\
1\\
0\\
0
\end{array}
\right) \\
&  &  & = &  &  & \\
\left|  2\right\rangle  & = & \left|  10\right\rangle  & = & \left(
\begin{array}
[c]{c}%
0\\
1
\end{array}
\right)  \otimes\left(
\begin{array}
[c]{c}%
1\\
0
\end{array}
\right)  & = & \left(
\begin{array}
[c]{c}%
0\\
0\\
1\\
0
\end{array}
\right) \\
&  &  &  &  &  & \\
\left|  3\right\rangle  & = & \left|  11\right\rangle  & = & \left(
\begin{array}
[c]{c}%
0\\
1
\end{array}
\right)  \otimes\left(
\begin{array}
[c]{c}%
0\\
1
\end{array}
\right)  & = & \left(
\begin{array}
[c]{c}%
0\\
0\\
0\\
1
\end{array}
\right)
\end{array}
\right.
\]

Let us assume that the initial state $\left|  \psi\right\rangle _{t=0}$ of our
2-qubit register is
\[
\left|  \psi\right\rangle _{t=0}=\left(  \frac{\left|  0\right\rangle -\left|
1\right\rangle }{\sqrt{2}}\right)  \otimes\left|  0\right\rangle =\frac
{1}{\sqrt{2}}\left(  \left|  00\right\rangle -\left|  10\right\rangle \right)
=\frac{1}{\sqrt{2}}\left(  \left|  0\right\rangle -\left|  2\right\rangle
\right)  =\frac{1}{\sqrt{2}}\left(
\begin{array}
[c]{r}%
1\\
0\\
-1\\
0
\end{array}
\right)
\]

Let us also assume that from time $t=0$ to time $t=1$ the dynamical behavior
of the above 2-qubit register is determined by a constant Hamiltonian $H$,
which when written in terms of the basis $\left\{  \left|  00\right\rangle
,\left|  01\right\rangle ,\left|  10\right\rangle ,\left|  11\right\rangle
\right\}  =\left\{  \left|  0\right\rangle ,\left|  1\right\rangle ,\left|
2\right\rangle ,\left|  3\right\rangle \right\}  $ is given by
\[
H=\frac{\pi\hslash}{2}\left(
\begin{array}
[c]{rrrr}%
0 & 0 & 0 & 0\\
0 & 0 & 0 & 0\\
0 & 0 & 1 & -1\\
0 & 0 & -1 & 1
\end{array}
\right)  \text{ ,}%
\]
where the rows and the columns are listed in the order $\left|
00\right\rangle $, $\left|  01\right\rangle $, $\left|  10\right\rangle $,
$\left|  11\right\rangle $, i.e., in the order $\left|  0\right\rangle $,
$\left|  1\right\rangle $, $\left|  2\right\rangle $, $\left|  3\right\rangle
$.

\bigskip

Then, as a consequence of Schr\"{o}dinger's equation, the Hamiltonian $H$
determines a unitary transformation
\begin{align*}
U_{CNOT}  &  =\ _{\overset{\vspace{5pt}}{t}}%
\raisebox{-4pt}{\includegraphics[
trim=0.000000in -0.042019in -0.046161in 0.000000in,
height=22.875pt,
width=10.25pt
]%
{prodint.ps}%
}%
_{\ 0}e^{-\frac{i}{\hslash}Hdt}=e^{\int_{0}^{1}-\frac{i}{\hslash}%
Hdt}=e^{-\frac{i}{\hslash}H}\\
&  \strut\\
&  =\left(
\begin{array}
[c]{cccc}%
1 & 0 & 0 & 0\\
0 & 1 & 0 & 0\\
0 & 0 & 0 & 1\\
0 & 0 & 1 & 0
\end{array}
\right)  =\left|  0\right\rangle \left\langle 0\right|  +\left|
1\right\rangle \left\langle 1\right|  +\left|  2\right\rangle \left\langle
3\right|  +\left|  3\right\rangle \left\langle 2\right|
\end{align*}
which moves the 2-qubit register from the initial state $\left|
\psi\right\rangle _{t=0}$ at time $t=0$ to $\left|  \psi\right\rangle
_{t=1}=U_{CNOT}\left|  \psi\right\rangle _{t=0}$ at time $t=1$. \ Then
\begin{align*}
\left|  \psi\right\rangle _{t=1}  &  =U_{CNOT}\left|  \psi\right\rangle
_{t=0}=\left(
\begin{array}
[c]{cccc}%
1 & 0 & 0 & 0\\
0 & 1 & 0 & 0\\
0 & 0 & 0 & 1\\
0 & 0 & 1 & 0
\end{array}
\right)  \cdot\frac{1}{\sqrt{2}}\left(
\begin{array}
[c]{r}%
1\\
0\\
-1\\
0
\end{array}
\right) \\
&  =\frac{1}{\sqrt{2}}\left(
\begin{array}
[c]{r}%
1\\
0\\
0\\
-1
\end{array}
\right)  =\frac{1}{\sqrt{2}}\left(  \left|  00\right\rangle -\left|
11\right\rangle \right)  =\frac{1}{\sqrt{2}}\left(  \left|  0\right\rangle
-\left|  3\right\rangle \right)
\end{align*}

\bigskip

The resulting state (called an \textbf{EPR pair}
\index{EPR} of qubits for reasons we shall later explain) can no longer be
written as a tensor product of two states. \ Consequently, we no longer have
the juxtaposition of two qubits.

\bigskip

Somehow, the resulting two qubits have in some sense ``lost their separate
identities.'' \ Measurement of any one of the qubits immediately impacts the other.

\bigskip

For example, if we measure the 0-th qubit (i.e., the right-most qubit), the
EPR state in some sense ``jumps'' to one of two possible states. Each of the
two possibilities occurs with probability $\frac{1}{2}$, as indicated in the
table below:
\[%
\begin{tabular}
[c]{|c||c|}\hline
\multicolumn{2}{|c|}{$\overset{}{\underset{}{\frac{1}{\sqrt{2}}\left(  \left|
0_{1}0_{0}\right\rangle -\left|  1_{1}1_{0}\right\rangle \right)  }}$}\\\hline
\multicolumn{2}{|c|}{$\swarrow\swarrow\swarrow\fbox{$%
\begin{tabular}
[c]{c}%
Meas.\\
0-th\\
Qubit
\end{tabular}
$}\searrow\searrow\searrow$}\\\hline%
\begin{tabular}
[c]{l}%
$\overset{}{Prob=\frac{1}{2}}$\\
\multicolumn{1}{c}{$\underset{}{\left|  0_{1}0_{0}\right\rangle }$}%
\end{tabular}
&
\begin{tabular}
[c]{l}%
$\overset{}{Prob=\frac{1}{2}}$\\
\multicolumn{1}{c}{$\underset{}{\left|  1_{1}1_{0}\right\rangle }$}%
\end{tabular}
\\\hline
\end{tabular}
\]
Thus we see that a measurement of one of the qubits causes a change in the other.

\bigskip

\subsection{Definition of quantum entanglement}

\qquad\bigskip

The above mentioned
\index{Entanglement!Quantum} phenomenon is so unusual and so non-classical
that it warrants a name.

\bigskip

\begin{definition}
Let $\mathcal{Q}_{1}$, $\mathcal{Q}_{2}$, $\ldots$ , $\mathcal{Q}_{n}$ be
quantum systems with underlying Hilbert spaces $\mathcal{H}_{1}$,
$\mathcal{H}_{2}$, $\ldots$ , $\mathcal{H}_{n}$, respectively. \ Then the
global quantum system $\mathcal{Q}$ consisting of the quantum systems
$\mathcal{Q}_{1}$, $\mathcal{Q}_{2}$, $\ldots$ , $\mathcal{Q}_{n}$ is said to
be \textbf{entangled
\index{Quantum Entangled}} if its state $\left|  \psi\right\rangle
\in\mathcal{H}=\bigotimes_{j=1}^{n}\mathcal{H}_{j}$ can not be written in the
form
\[
\left|  \psi\right\rangle =%
{\displaystyle\bigotimes\limits_{j=1}^{n}}
\left|  \psi_{j}\right\rangle \text{ ,}%
\]
where each ket $\left|  \psi_{j}\right\rangle $ lies in the Hilbert space
$\mathcal{H}_{j}$ for, $j=1,2,\ldots,n$. \ We also say that such a state
$\left|  \psi\right\rangle $ is \textbf{entangled}.
\end{definition}

\bigskip

Thus, the state
\[
\left|  \psi\right\rangle _{t=1}=\frac{1}{\sqrt{2}}\left(  \left|
00\right\rangle -\left|  11\right\rangle \right)
\]
of the 2-qubit register of the previous section is entangled.

\bigskip

\begin{remark}
In terms of density operator formalism, a pure ensemble $\rho$ is entangled if
it can not be written in the form
\[
\rho=%
{\displaystyle\bigotimes\limits_{j=1}^{n}}
\rho_{j}\text{ ,}%
\]
where the $\rho_{j}$'s denote density operators.\bigskip
\end{remark}

Please note that we have defined entanglement only for pure ensembles. \ For
mixed ensembles, entanglement is not well understood\footnote{Quantum
entanglement is not even well understood for pure ensembles.}. \ As a result,
the ``right'' definition of entanglement of mixed ensembles is still
unresolved. \ We give one definition below:

\bigskip

\begin{definition}
A density operator $\rho$ on a Hilbert space $\mathcal{H}$ is said to be
entangled with respect to the Hilbert space decomposition
\[
\mathcal{H}=%
{\displaystyle\bigotimes\limits_{j=1}^{n}}
\mathcal{H}_{j}%
\]
if it can not be written in the form
\[
\rho=\sum_{k=1}^{\ell}\lambda_{k}\left(  \bigotimes\limits_{j=1}^{n}%
\rho_{(j,k)}\right)  \text{ ,}%
\]
for some positive integer $\ell$, where the $\lambda_{k}$'s are positive real
numbers such that
\[
\sum_{k=1}^{\ell}\lambda_{k}=1\text{ .}%
\]
and where each $\rho_{(j,k)}$ is a density operator on the Hilbert space
$\mathcal{H}_{j\text{.}}$
\end{definition}

\bigskip

Readers interested in pursuing this topic further should refer to the works of
Bennett
\index{Bennett}, the Horodecki's, Nielsen,
\index{Nielsen} Smolin, Wootters
\index{Wootters}, and others\cite{Bennett2}, \cite{Horodecki1}, \cite{Linden1}%
, \cite{Nielsen1}.

\bigskip

\subsection{Einstein, Podolsky, Rosen's (EPR's) grand challenge to quantum mechanics.}

\qquad\bigskip

Albert Einstein was skeptical of quantum mechanics, so skeptical that he
together with Podolsky and Rosen wrote a joint paper\cite{Einstein1} appearing
in 1935 challenging the very foundations of quantum mechanics. \ Their paper
hit the scientific community like a bombshell. \ For it delivered a direct
frontal attack at the very heart and center of quantum mechanics.

\index{Einstein}
\index{Podolsky}
\index{Rosen}
\index{EPR}

\bigskip

At the core of their objection was quantum entanglement. \ Einstein and his
colleagues had insightfully recognized the central importance of this quantum phenomenon.

\bigskip

Their argument centered around the fact that quantum mechanics violated either
the \textbf{principle of non-locality}\footnote{We will later explain the
principle of non-locality.}
\index{Principle of Non-Locality} or the \textbf{principle of reality}%
\footnote{For an explanation of the principle of reality as well as the
principle of non-localty, please refer, for example, to \cite{Peres1},
\cite{Bub1}.}
\index{Reality!Principle of}. \ They argued that, as a result, quantum
mechanics must be incomplete, and that quantum entanglement could be explained
by missing \textbf{hidden variables}.

\bigskip

For many years, no one was able to conceive of an experiment that could
determine which of the two theories, i.e., quantum mechanics or EPR's hidden
variable theory, was correct. \ In fact, many believed that the two theories
were not distinguishable on physical grounds.

\bigskip

It was not until Bell developed his famous inequalities \cite{Bell1}%
,\cite{Bell2}, \cite{Bub1}, that a physical criterion was found to distinquish
the two theories. \ Bell developed inequalities which, if violated, would
clearly prove that quantum mechanics is correct, and hidden variable theories
are not. \ Many experiments were performed. \ Each emphatically supported
quantum mechanics, and clearly demonstrated the incorrectness of hidden
variable theory. \ Quantum mechanics was the victor!

\bigskip

\subsection{Why did Einstein, Podolsky, Rosen (EPR) object?}

\qquad\bigskip

But why did Einstein and his colleagues object so vehemently to quantum
entanglement? \ 

\bigskip

\index{Non-Locality}

As a preamble to our answer to this question, we note that Einstein and his
colleagues were convinced of the validity of the following two physical orinciples:

\begin{itemize}
\item [1)]The \textbf{principle of local interactions
\index{Local Interaction}}, i.e., that all the known forces of nature are
local interactions,

\item[2)] The \textbf{principle of non-locality}, i.e., that spacelike
separated regions of spacetime are physically independent of one another.
\end{itemize}

\bigskip

Their conviction in regard to principle 1) was based on the fact that all four
known forces of nature, i.e., gravitational, electromagnetic, weak, and strong
forces, are \textbf{local interactions}. \ By this we mean:

\begin{itemize}
\item [i)]They are mediated by another entity, e.g., graviton, photon, etc.

\item[ii)] They propagate no faster than the speed $c$ of light

\item[iii)] Their strength drops off with distance
\end{itemize}

\bigskip

Their conviction in regard to principle 2) was based on the following reasoning:

Two points in spacetime $P_{1}=\left(  x_{1},y_{1},z_{1},t_{1}\right)  $ and
$P_{2}=\left(  x_{2},y_{2},z_{2},t_{2}\right)  $ are separated by a
\textbf{spacelike distance
\index{Spacelike Distance}} provided the distance between $\left(  x_{1}%
,y_{1},z_{1}\right)  $ and $\left(  x_{2},y_{2},z_{2}\right)  $ is greater
than $c\left|  t_{2}-t_{1}\right|  $, i.e.,
\[
Distance\left(  \left(  x_{1},y_{1},z_{1}\right)  ,\left(  x_{2},y_{2}%
,z_{2}\right)  \right)  >c\left|  t_{2}-t_{1}\right|  \text{ ,}%
\]
where $c$ denotes the speed of light. \ In other words, no signal can travel
between points that are said to be separated by a spacelike distance unless
the signal travels faster than the speed of light. \ But because of the basic
principles of relativity, such superluminal communication
\index{Superluminal Communication} is not possible.

\bigskip

Hence we have:

\begin{description}
\item [The principle of non-locality]Spacelike separated regions of spacetime
are physically independent. \ In other words, spacelike separated regions can
not influence one another.
\end{description}

\index{Non-Locality}

\index{Principle of Non-Locality}

\bigskip

\subsubsection{EPR's objection}

\qquad\bigskip

We now are ready to explain why Einstein and his colleagues objected so
vehemently to quantum entanglement. \ We explain Bohm's simplified version of
their argument. \ 

\bigskip

Consider a two qubit quantum system that has been prepared by \textbf{Alice
\index{Alice}}\footnote{Alice is a well known personality in quantum
computation, quantum cryptography, and quantum information theory.} in her
laboratory in the state
\[
\left|  \psi\right\rangle =\frac{1}{\sqrt{2}}\left(  \left|  0_{1}%
0_{0}\right\rangle -\left|  1_{1}1_{0}\right\rangle \right)  \text{ .}%
\]
After the preparation, she decides to keep qubit \#1 in her laboratory, but
enlists Captain James T. Kirk of the Starship Enterprise to transport qubit
\#0 to her friend \textbf{Bob
\index{Bob}}\footnote{Bob is another well known personality in quantum
computation, quantum cryptography, and quantum information theory.} who is at
some far removed distant part of the universe, such as at a Federation outpost
orbiting about the double star Alpha Centauri in the constellation Centaurus.

\bigskip

After Captain Kirk has delivered qubit \#0, Alice's two qubits are now
separated by a spacelike distance. \ Qubit \#1 is located in her Earth based
laboratory. \ Qubits \#0 is located with Bob at a Federation outpost orbiting
Alpha Centauri. \ But the two qubits are still entangled, even in spite of the
fact that they are separated by a spacelike distance.

\bigskip

If Alice now measures qubit \#1 (which is located in her Earth based
laboratory), then the principles of quantum mechanics force her to conclude
that instantly, without any time lapse, both qubits are ``effected.'' \ As a
result of the measurement, both qubits will be either in the state $\left|
0_{1}0_{0}\right\rangle $ or the state $\left|  1_{1}1_{0}\right\rangle $,
each possibility occurring with probability 1/2.

\bigskip

This is a non-local ``interaction.'' \ For,

\begin{itemize}
\item  The ``interaction'' occurred without the presence of any force. \ It
was not mediated by anything.

\item  The measurement produced an instantaneous change, which was certainly
faster than the speed of light.

\item  The strength of the ``effect'' of the measurement did not drop off with distance.
\end{itemize}

\bigskip

No wonder Einstein was highly skeptical of quantum entanglement. \ Yet
puzzlingly enough, since no information is exchanged by the process, the
principles of general relativity are not violated. \ As a result, such an
``effect'' can not be used for superluminal communication.

\bigskip

For a more in-depth discussion of the EPR paradox and the foundations of
quantum mechanics, the reader should refer to \cite{Bub1}.

\index{EPR}

\bigskip

\subsection{\textbf{\bigskip}Quantum entanglement: The Lie group perspective}

\qquad\bigskip

\index{Lie Group}

Many aspects of quantum entanglement can naturally be captured in terms of Lie
groups and their Lie algebras.

\bigskip

Let
\[
\mathcal{H}=\mathcal{H}_{n-1}\otimes\mathcal{H}_{n-2}\otimes\ldots
\otimes\mathcal{H}_{0}=\ _{\overset{}{n-1}}%
{\displaystyle\bigotimes\nolimits_{0}}
\mathcal{H}_{j}%
\]
be a decomposition of a Hilbert space $\mathcal{H}$ into the tensor product of
the Hilbert spaces $\mathcal{H}_{n-1}$, $\mathcal{H}_{n-2}$, $\ldots$
,$\mathcal{H}_{0}$. \ Let $\mathbb{U}=\mathbb{U}(\mathcal{H})$, $\mathbb{U}%
_{n-1}=\mathbb{U}(\mathcal{H}_{n-1})$, $\mathbb{U}_{n-2}=\mathbb{U}%
(\mathcal{H}_{n-2})$, $\ldots\ $,$\mathbb{U}_{0}=\mathbb{U}(\mathcal{H}_{0})$,
denote respectively the Lie groups of all unitary transformations on
$\mathcal{H}$, $\mathcal{H}_{n-1}$, $\mathcal{H}_{n-2}$, $\ldots$
,$\mathcal{H}_{0}$. \ Moreover, let $\mathbf{u}=\mathbf{u}(\mathcal{H})$,
$\mathbf{u}_{n-1}=\mathbf{u}_{n-1}(\mathcal{H}_{n-1})$, $\mathbf{u}%
_{n-2}=\mathbf{u}_{n-2}(\mathcal{H}_{n-2})$, $\ldots\ ,\mathbf{u}%
_{0}=\mathbf{u}_{0}(\mathcal{H}_{0})$ denote the corresponding Lie algebras.

\bigskip

\begin{definition}
The \textbf{local subgroup
\index{Local Subgroup}} $\mathbb{L}=\mathbb{L}(\mathcal{H})$ of $\mathbb{U}%
=\mathbb{U}(\mathcal{H})$ is defined as the subgroup
\[
\mathbb{L}=\mathbb{U}_{n-1}\otimes\mathbb{U}_{n-2}\otimes\ldots\otimes
\mathbb{U}_{0}=\ _{\overset{}{n-1}}%
{\displaystyle\bigotimes\nolimits_{0}}
\mathbb{U}_{j}\text{ .}%
\]
The elements of $\mathbb{L}$ are called \textbf{local unitary transformations
\index{Local Unitary Transformation}}. Unitary transformations which are in
$\mathbb{U}$ but not in $\mathbb{L}$ are called \textbf{global unitary
transformations}.
\index{Global Unitary Transformation} \ The corresponding lie algebra
\[
\mathbf{\ell=u}_{n-1}\boxplus\mathbf{u}_{n-2}\boxplus\ldots\boxplus
\mathbf{u}_{0}%
\]
is called the \textbf{local Lie algebra},
\index{Local Lie Algebra} where `$\boxplus$' denotes the \textbf{Kronecker
sum}\footnote{The Kronecker sum $A\boxplus B$ is defined as
\[
A\boxplus B=A\otimes\mathbf{1}+\mathbf{1}\otimes B\text{ ,}%
\]
where $\mathbf{1}$ denotes the identity transformation.}.
\index{Kronecker Sum}
\end{definition}

\bigskip

Local unitary transformations can not entangle quantum systems with respect to
the above tensor product decomposition. \ However, global unitary
transformations are those unitary transformations which can and often do
produce interactions which entangle quantum systems. This leads to the
following definition:

\bigskip

\begin{definition}
Two states $\left|  \psi_{1}\right\rangle $ and $\left|  \psi_{2}\right\rangle
$ in $\mathcal{H}$ are said to be \textbf{locally equivalent} ( or, of the
\textbf{same entanglement type})
\index{Entanglement!Type}, written
\[
\left|  \psi_{1}\right\rangle \underset{local}{\sim}\left|  \psi
_{2}\right\rangle \text{ ,}%
\]
if there exists a local unitary transformation $U\in\mathbb{L}$ such that
\[
U\left|  \psi_{1}\right\rangle =\left|  \psi_{2}\right\rangle \text{ .}%
\]
The equivalence classes of local equivalence
\index{Local Equivalence} $\underset{local}{\sim}$ are called the
\textbf{entanglement classes of }$\mathcal{H}$. \ Two density operators
$\rho_{1}$ and $\rho_{2}$, (and hence, the corresponding two skew Hermitian
operators $i\rho_{1}$ and $i\rho_{2}$ lying in $\mathbf{u}$) are said to be
\textbf{locally equivalent} ( or, of the \textbf{same entanglement type}),
written
\[
\rho_{1}\underset{local}{\sim}\rho_{2}\text{ ,}%
\]
if there exists a local unitary transformation $U\in\mathbb{L}$ such that
\[
Ad_{U}(\rho_{1})=\rho_{2}\text{ ,}%
\]
where $Ad_{U}$ denotes the big adjoint representation, i.e., $Ad_{U}%
(i\rho)=U(i\rho)U^{\dagger}$. The equivalence classes under this relation are
called \textbf{entanglement classes
\index{Entanglement!Classes}} of the Lie algebra $\mathbf{u}(\mathcal{H})$.
\end{definition}

\bigskip

Thus, the entanglement classes of the Hilbert space $\mathcal{H}$ are just the
\textbf{orbits
\index{Orbits}} of the group action of $\mathbb{L}(\mathcal{H})$ on
$\mathcal{H}$. \ In like manner, the entanglement classes of the Lie algebra
$\mathbf{u}(\mathcal{H})$ are the \textbf{orbits} of the big adjoint action of
$\mathbb{L}(\mathcal{H})$ on $\mathbf{u}(\mathcal{H})$. \ Two states are
entangled in the same way if and only if they lie in the same entanglement
class, i.e., the same orbit.

\bigskip

For example, let us assume that Alice
\index{Alice} and Bob
\index{Bob} collectively possess two qubits $\mathcal{Q}_{AB}$ which are in
the entangled state
\[
\left|  \psi_{1}\right\rangle =\frac{\left|  0_{B}0_{A}\right\rangle +\left|
1_{B}1_{A}\right\rangle }{\sqrt{2}}=\frac{1}{\sqrt{2}}\left(
\begin{array}
[c]{c}%
1\\
0\\
0\\
1
\end{array}
\right)  \text{ ,}%
\]
and moreover that Alice possesses qubit labeled $A$, but not the qubit labeled
$B$, and that Bob holds qubit $B$, but not qubit $A$. \ Let us also assume
that Alice and Bob are also separated by a spacelike distance. \ As a result,
they can only apply local unitary transformations to the qubits that they possess.

\bigskip

Alice could, for example, apply the local unitary transformation
\[
U_{A}=\left(
\begin{array}
[c]{rr}%
0 & 1\\
-1 & 0
\end{array}
\right)  \otimes\left(
\begin{array}
[c]{cc}%
1 & 0\\
0 & 1
\end{array}
\right)  =\left(
\begin{array}
[c]{rrrr}%
0 & 0 & 1 & 0\\
0 & 0 & 0 & 1\\
-1 & 0 & 0 & 0\\
0 & -1 & 0 & 0
\end{array}
\right)
\]
to her qubit to move Alice's and Bob's qubits $A$ and $B$ respectively\ into
the state
\[
\left|  \psi_{2}\right\rangle =\frac{\left|  0_{B}1_{A}\right\rangle -\left|
1_{B}0_{A}\right\rangle }{\sqrt{2}}=\frac{1}{\sqrt{2}}\left(
\begin{array}
[c]{r}%
0\\
1\\
-1\\
0
\end{array}
\right)  \text{ ,}%
\]
Bob also could accomplish the same by applying the local unitary
transformation
\[
U_{B}=\left(
\begin{array}
[c]{cc}%
1 & 0\\
0 & 1
\end{array}
\right)  \otimes\left(
\begin{array}
[c]{rr}%
0 & -1\\
1 & 0
\end{array}
\right)  =\left(
\begin{array}
[c]{rrrr}%
0 & -1 & 0 & 0\\
1 & 0 & 0 & 0\\
0 & 0 & 0 & -1\\
0 & 0 & 1 & 0
\end{array}
\right)
\]
to his qubit. \ 

\bigskip

By local unitary transformations, Alice and Bob can move the state of their
two qubits to any other state within the same entanglement class. \ But with
local unitary transformations, there is no way whatsoever that Alice and Bob
can transform the two qubits into a state lying in a different entanglement
class (i.e., a different orbit), such as
\[
\left|  \psi_{3}\right\rangle =\left|  0_{B}0_{A}\right\rangle .
\]

\bigskip

The only way Alice and Bob could transform the two qubits from state $\left|
\psi_{1}\right\rangle $ to the state $\left|  \psi_{3}\right\rangle $ is for
Alice and Bob to come together, and make the two qubits interact with one
another via a global unitary transformation such as
\[
U_{AB}=\frac{1}{\sqrt{2}}\left(
\begin{array}
[c]{rrrr}%
1 & 0 & 0 & 1\\
0 & 1 & 1 & 0\\
0 & -1 & 1 & 0\\
-1 & 0 & 0 & 1
\end{array}
\right)
\]

\bigskip

The main objective of this approach to quantum entanglement is to determine
when two states lie in the same orbit or in different orbits? \ In other
words, what is needed is a complete set of invariants, i.e., invariants that
completely specify all the orbits ( i.e., all the entanglement classes). \ We
save this topic for another lecture\cite{Lomonaco1}.

\bigskip

\bigskip

At first it would seem that state kets are a much better vehicle than density
operators for the study of quantum entanglement. \ After all, state kets are
much simpler mathematical objects. \ So why should one deal with the
additional mathematical luggage of density operators?

\bigskip

Actually, density operators have a number of advantages over state kets. \ The
most obvious advantage is that density operators certainly have an upper hand
over state kets when dealing with mixed ensembles. \ But their most important
advantage is that the orbits of the adjoint action are actually manifolds,
which have a very rich and pliable mathematical structure. \ Needless to say,
this topic is beyond the scope of this paper.\bigskip

\begin{remark}
It should also be mentioned that the mathematical approach discussed in this
section by no means captures every aspect of the physical phenomenon of
quantum entanglement. \ The use of ancilla
\index{Ancilla} and of classical communication have not been considered. \ For
an in-depth study of the relation between quantum entanglement and classical
communication (including catalysis), please refer to the work of Jonathan,
Nielson, and others\cite{Nielsen1}.
\end{remark}

\bigskip\bigskip

In regard to describing the locality of unitary operations, we will later have
need for a little less precision than that given above in the above
definitions. \ So we give the following (unfortunately rather technical) definitions:

\bigskip

\begin{definition}
Let $\mathcal{H}$, $\mathcal{H}_{n-1}$, $\mathcal{H}_{n-2}$, $\ldots$
,$\mathcal{H}_{0}$ be as stated above. \ Let $\mathcal{P}=\left\{  B_{\alpha
}\right\}  $ be a \textbf{partition} of the set of indices $\left\{
0,1,2,\ldots,n-1\right\}  $, i.e., $\mathcal{P}$ is a collection of disjoint
subsets $B_{\alpha}$ of $\left\{  0,1,2,\ldots,n-1\right\}  $, called
\textbf{blocks}, such that $\bigcup_{\alpha}B_{\alpha}=\left\{  0,1,2,\ldots
,n-1\right\}  $. \ Then the $\mathcal{P}$-\textbf{tensor product
decomposition} of $\mathcal{H}$ is defined as
\[
\mathcal{H}=%
{\displaystyle\bigotimes\limits_{B_{\alpha}\in\mathcal{P}}}
\mathcal{H}_{B_{\alpha}}\text{ ,}%
\]
\index{Partition}where
\[
\mathcal{H}_{B_{\alpha}}=%
{\textstyle\bigotimes\limits_{j\in B_{\alpha}}}
\mathcal{H}_{j}\text{ ,}%
\]
for each block $B_{\alpha}$ in $\mathcal{P}$. \ Also the subgroup of
$\mathcal{P}$-\textbf{local unitary transformations }$\mathbb{L}_{\mathcal{P}%
}(\mathcal{H})$ is defined as the subgroup of local unitary transformations of
$\mathcal{H}$ corresponding to the $\mathcal{P}$-tensor decomposition of
$\mathcal{H}$.

We define the \textbf{fineness of a partition }$\mathcal{P}$, written
$fineness(\mathcal{P})$, as the maximum number of indices in a block of
$\mathcal{P}$. \ We say that a unitary transformation $U$ of $\mathcal{H}$ is
\textbf{sufficiently local
\index{Unitary!Sufficiently Local}
\index{Local!Sufficiently}
\index{Sufficiently Local Unitary Transformation|see{Unitary}}} if there
exists a partition $\mathcal{P}$ with sufficiently small $fineness(\mathcal{P}%
)$ (e.g., $fineness(\mathcal{P})\ \leq3$) such that $U\in\mathbb{L}%
_{\mathcal{P}}(\mathcal{H})$.
\end{definition}

\bigskip

\begin{remark}
The above lack of precision is needed because there is no way to know what
kind (if any) of quantum computing devices will be implemented in the future.
Perhaps we will at some future date be able to construct quantum computing
devices that locally manipulate more than 2 or 3 qubits at a time?
\end{remark}

\index{Entangled, Quantum}

\index{Quantum Entangled}

\section{\textbf{Entropy and quantum mechanics}}

\bigskip

\subsection{Classical entropy, i.e., Shannon Entropy}

\qquad\bigskip

\bigskip Let $\mathcal{S}$ be a probability distribution on a finite set
$\left\{  s_{1},s_{2},\ldots,s_{n}\right\}  $ of elements called
\textbf{symbols} given by
\[
\text{Prob}\left(  s_{j}\right)  =p_{j}\text{ ,}%
\]
where $\sum_{j=1}^{n}p_{j}=1$. \ Let $s$ denote the random variable (i.e.,
\textbf{finite memoryless stochastic source}) that produces the value $s_{j}$
with probability $p_{j}$.

\index{Entropy, Classical}

\bigskip

\textbf{
\index{Entropy, Shannon}}

\begin{definition}
The \textbf{classical entropy} (also called the \textbf{Shannon entropy}%
)\textbf{\ }$H(S)$ of a probability distribution
\index{Probability Distribution}
\index{Stochastic Source} $\mathcal{S}$ (or of the source $s$) is defined as:
\[
H(\mathcal{S})=H(s)=-%
{\displaystyle\sum\limits_{j=1}^{n}}
p_{j}\lg(p_{j})\text{ ,}%
\]
where `$\lg$'
\index{lg@$\lg$} denotes the $\log$ to the base 2 .
\end{definition}

\bigskip

Classical entropy $H(\mathcal{S})$ is a measure of the uncertainty inherent in
the probability distribution $\mathcal{S}$. \ Or in other words, it is the
measure of the uncertainty of an observer before the source $s$ ``outputs'' a
symbol $s_{j}$. \bigskip

One property of such classical \textbf{stochastic sources
\index{Stochastic Source}} we often take for granted is that the output
symbols
\index{Symbols!Output} $s_{j}$ are completely distinguishable from one
another. \ We will see that this is not necessarily the case in the strange
world of the quantum.

\bigskip

\subsection{Quantum entropy, i.e., Von Neumann entropy}

\qquad\bigskip

Let $\mathcal{Q}$ be a quantum system with state given by the desity operator
$\rho$.
\index{Entropy, von Neumann}

\bigskip

Then there are many \textbf{preparations}
\[
\overset{\text{\textbf{Preparation}}}{%
\begin{tabular}
[c]{||c||c||c||c||}\hline\hline
$\left|  \psi_{1}\right\rangle $ & $\left|  \psi_{2}\right\rangle $ &
$\ \ldots\ $ & $\left|  \psi_{n}\right\rangle $\\\hline\hline
$p_{1}$ & $p_{2}$ & $\ldots$ & $p_{n}$\\\hline\hline
\end{tabular}
}%
\]
which will produce the same state $\rho$. \ These preparations are classical
stochastic sources with classical entropy given by
\[
H=-%
{\displaystyle\sum}
p_{j}\lg(p_{j})\text{ .}%
\]
Unfortunately, the classical entropy $H$ of a preparation does not necessarily
reflect the uncertainty in the resulting state $\rho$. \ For two different
preparations $\mathcal{P}_{1}$ and $\mathcal{P}_{2}$, having different
entropies $H\left(  \mathcal{P}_{1}\right)  $ and $H\left(  \mathcal{P}%
_{2}\right)  $, can (and often do) produce the same state $\rho$. \ The
problem is that the states of the preparation my not be completely physically
distinguishable from one another. \ This happens when the states of the
preparation are not orthogonal. \ (Please refer to the Heisenberg uncertainty principle.)

\index{Entropy, Classical}

John von Neumann found that the true measure of quantum entropy can be defined
as follows:\bigskip

\begin{definition}
Let $\mathcal{Q}$ be a quantum system with state given by the density operator
$\rho$. \ Then the \textbf{quantum entropy} (also called the \textbf{von
Neumann entropy}) of $\mathcal{Q}$, written $S(\mathcal{Q})$, is defined as
\[
S(\mathcal{Q})=-Trace\left(  \rho\lg\rho\right)  \text{ ,}%
\]
where `$\lg\rho$' denotes the log to the base 2 of the operator $\rho$.
\end{definition}

\bigskip

\begin{remark}
The operator $\lg\rho$ exists and is an analytic map $\rho\longmapsto\lg\rho$
given by the power series
\[
\lg\rho=\frac{1}{\ln2}%
{\displaystyle\sum\limits_{n=1}^{\infty}}
(-1)^{n+1}\frac{\left(  \rho-I\right)  ^{n}}{n}%
\]
provided that $\rho$ is sufficiently close to the identity operator $I$, i.e.,
provided
\[
\left\|  \rho-I\right\|  <1\text{ ,}%
\]
where
\[
\left\|  A\right\|  =\sup_{v\in\mathcal{H}}\frac{\left\|  Av\right\|
}{\left\|  v\right\|  }\text{ .}%
\]
It can be shown that this is the case for all positive definite Hermitian
operators of trace $1$. \ 

For Hermitian operators $\rho$ of trace $1$ which are not positive definite,
but only positive semi-definite (i.e., which have a zero eigenvalue), the
logarithm $\lg(\rho)$ does not exist. \ However, there exists a sequence
$\rho_{1},\rho_{2},\rho_{3},\ldots$ of positive definite Hermitian operators
of trace $1$ which converges to $\rho$, i.e., such that
\[
\rho=\lim_{k\longrightarrow\infty}\rho_{k}%
\]
It can then be shown that the limit
\[
\lim_{k\longrightarrow\infty}\rho_{k}\lg\rho_{k}%
\]
exists.

Hence, $S(\rho)$ is defined and exists for all density operators $\rho$.
\end{remark}

\bigskip

Quantum entropy is a measure of the uncertainty at the quantum level. \ As we
shall see, it is very different from the classical entropy that arises when a
measurement is made.

\bigskip

One important feature of quantum entropy $S(\rho)$ is that it is invariant
under the adjoint action of unitary transformations, i.e.,
\[
S\left(  \ Ad_{U}(\rho)\ \right)  =S\left(  U\rho U^{\dagger}\right)
=S(\rho)\text{ .}%
\]
It follows that, for closed quantum systems, it is a \textbf{dynamical
invariant.}
\index{Dynamic Invariants} \ As the state $\rho$ moves according to
Schr\"{o}dinger's equation, the quantum entropy $S(\rho)$ of $\rho$ remains
constant. \ It does not change unless \ measurement is made, or, as we shall
see, unless we ignore part of the quantum system.

\bigskip

Because of unitary invariance, the quantum entropy can be most easily computed
by first diagonalizing $\rho$ with a unitary transformation $U$, i.e.,
\[
U\rho U^{\dagger}=\triangle(\overrightarrow{\lambda})\text{ ,}%
\]
where $\triangle(\overrightarrow{\lambda})$ denotes the diagonal matrix with
diagonal $\overrightarrow{\lambda}=\left(  \lambda_{1},\lambda_{2}%
,\ \ldots\ ,\lambda_{n}\right)  $. \ 

\bigskip

Once $\rho$ has been diagonalized
\index{Diagonalization}, we have
\begin{align*}
S(\rho)  &  =-Trace\left(  \triangle(\overrightarrow{\lambda})\lg
\triangle(\overrightarrow{\lambda})\right) \\
&  =-Trace\left(  \ \triangle(\lambda_{1}\lg\lambda_{1},\ \lambda_{2}%
\lg\lambda_{2},\ \ldots\ ,\ \lambda_{n}\lg\lambda_{n})\ \right) \\
&  =-%
{\displaystyle\sum\limits_{j=1}^{n}}
\lambda_{j}\lg\lambda_{j}\text{ ,}%
\end{align*}
where the $\lambda_{j}$'s are the eigenvalues of $\rho$, and where
$0\lg0\equiv0$.

\bigskip

Please note that, because $\rho$ is positive semi-definite Hermitian of trace
$1$, all the eigenvalues of $\rho$ are non-negative real numbers such that
\[%
{\displaystyle\sum\limits_{j=1}^{n}}
\lambda_{j}=1\text{ .}%
\]

\vspace*{0.5in}

As an immediate corollary we have that the quantum entropy of a pure ensemble
\index{Ensemble!Pure} must be zero, i.e.,
\[
\fbox{$\rho$ pure ensemble $\Longrightarrow S(\rho)=0$}%
\]
There is no quantum uncertainty in a pure ensemble. \ However, as expected,
there is quantum uncertainty in mixed ensembles.

\index{Entropy, von Neumann}

\bigskip

\subsection{How is quantum entropy related to classical entropy?}

\qquad\bigskip

But how is classical entropy $H$ related to quantum entropy $S$?

\bigskip

Let $A$ be an observable of the quantum system $\mathcal{Q}$. \ Then a
measurement of $A$ of $\mathcal{Q}$ produces an eigenvalue $a_{i}$ with
probability
\[
p_{i}=Trace\left(  P_{a_{i}}\rho\right)  \text{ ,}%
\]
where $P_{a_{i}}$ denotes the projection operator
\index{Projection Operator} for the eigenspace of the eigenvalue $a_{i}$.
\ For example, if $a_{i}$ is a non-degenerate eigenvalue, then $P_{a_{i}%
}=\left|  a_{i}\right\rangle \left\langle a_{i}\right|  $ .

\bigskip

In other words, measurement of $A$ of the quantum system $\mathcal{Q}$ in
state $\rho$ can be identified with a classical stochastic source with the
eigenvalues $a_{i}$ as output symbols
\index{Symbols!Output} occurring with probability $p_{i}$. \ We denote this
classical stochastic source simply by $(\rho,A)$ .

\bigskip

The two entropies $S(\rho)$ and $H(\rho,A)$ are by no means the same. \ One is
a measure of quantum uncertainty before measurement, the other a measure of
the classical uncertainty that results from measurement. \ The quantum entropy
$S(\rho)$ is usually a lower bound for the classical entropy, i.e.,
\[
S(\rho)\leq H(\rho,A)\text{ .}%
\]
If $A$ is a complete observable (hence, non-degenerate), and if $A$ is
compatible with $\rho$, i.e., $\left[  \rho,A\right]  =0$, then $S(\rho
)=H(\rho,A)$.\bigskip

\subsection{\bigskip When a part is greater than the whole -- Ignorance = uncertainty}

\qquad\bigskip

Let $\mathcal{Q}$ be a multipartite quantum system with constituent parts
$\mathcal{Q}_{n-1}$, $\ldots$ ,$\mathcal{Q}_{1}$, $\mathcal{Q}_{0}$, and let
the density operator $\rho$ denote the state of $\mathcal{Q}$. \ Then from
section 5.6 of this paper we know that the state $\rho_{j}$ of each
constituent ``part'' $\mathcal{Q}_{j}$ is given by the partial trace over all
degrees of freedom except $\mathcal{Q}_{j}$, i.e., by
\[
\rho_{j}=\underset{%
\begin{array}
[c]{c}%
0\leq k\leq n-1\\
k\neq j
\end{array}
}{Trace\left(  \rho\right)  }\text{ .}%
\]

\bigskip

By applying the above partial trace, we are focusing only on the quantum
system $\mathcal{Q}_{j}$, and literally ignoring the remaining constituent
``parts'' of $\mathcal{Q}$. \ By taking the partial trace, we have done
nothing physical to the quantum system. \ We have simply ignored parts of the
quantum system.

\bigskip

What is surprising is that, by intentionally ignoring ``part'' of the quantum
system, we can in some cases create more quantum uncertainty. \ This happens
when the constituent ``parts'' of $\mathcal{Q}$ are quantum entangled.\vspace*{0.5in}

For example, let $\mathcal{Q}$ denote the bipartite quantum system consisting
of two qubits $\mathcal{Q}_{1}$ and $\mathcal{Q}_{0}$ in the entangled state
\[
\left|  \Psi_{\mathcal{Q}}\right\rangle =\frac{\left|  0_{1}0_{0}\right\rangle
-\left|  1_{1}1_{0}\right\rangle }{\sqrt{2}}\text{ .}%
\]
The corresponding density operator $\rho_{\mathcal{Q}}$ is
\begin{align*}
\rho_{\mathcal{Q}}  &  =\frac{1}{2}\left(  \left|  0_{1}0_{0}\right\rangle
\left\langle 0_{1}0_{0}\right|  -\left|  0_{1}0_{0}\right\rangle \left\langle
1_{1}1_{0}\right|  -\left|  1_{1}1_{0}\right\rangle \left\langle 0_{1}%
0_{0}\right|  +\left|  1_{1}1_{0}\right\rangle \left\langle 1_{1}1_{0}\right|
\right) \\
&  =\frac{1}{2}\left(
\begin{array}
[c]{rrrr}%
1 & 0 & 0 & -1\\
0 & 0 & 0 & 0\\
0 & 0 & 0 & 0\\
-1 & 0 & 0 & 1
\end{array}
\right)
\end{align*}

\bigskip

Since $\rho_{\mathcal{Q}}$ is a pure ensemble, there is no quantum
uncertainty, i.e.,
\[
S\left(  \rho_{\mathcal{Q}}\right)  =0\text{ .}%
\]

\bigskip

Let us now focus on qubit \#0 (i.e., $\mathcal{Q}_{0}$). \ The resulting
density operator $\rho_{0}$ for qubit \#0 is obtained by tracing over
$\mathcal{Q}_{1}$, i.e.,
\[
\rho_{0}=Trace_{1}\left(  \rho_{\mathcal{Q}}\right)  =\frac{1}{2}\left(
\ \left|  0\right\rangle \left\langle 0\right|  +\left|  1\right\rangle
\left\langle 1\right|  \ \right)  =\frac{1}{2}\left(
\begin{array}
[c]{cc}%
1 & 0\\
0 & 1
\end{array}
\right)  \text{ .}%
\]
Hence, the quantum uncertainty of qubit \#0 is
\[
S(\rho_{0})=1\text{ .}%
\]

\bigskip

Something most unusual, and non-classical, has happened. \ Simply by ignoring
part of the quantum system, we have increased the quantum uncertainty. \ The
quantum uncertainty of the constituent ``part'' $\mathcal{Q}_{0}$ is greater
than that of he whole quantum system $\mathcal{Q}$. \ This is not possible in
the classical world, i.e., not possible for Shannon entropy. (For more
details, see \cite{Cerf1}.)

\bigskip

\section{\textbf{\bigskip There is much more to quantum mechanics}}

\qquad

\bigskip

There is much more to quantum mechanics. For more in-depth overviews, there
are many outstanding books. Among such books are \cite{Bub1},
\cite{Cohen-Tannoudji1}, \cite{Dirac1}, \cite{Feynman1}, \cite{Haag1},
\cite{Heisenberg1}, \cite{Helstrom1}, \cite{Holevo1}, \cite{Jauch1},
\cite{Mackey1}, \cite{Omnes1}, \cite{Neumann1}, \cite{Peres1}, \cite{Piron1},
\cite{Sakurai1}, and many more.

\bigskip

\part{{\protect\Large Part of a Rosetta Stone for Quantum Computation}\bigskip}

\bigskip

\section{\textbf{The Beginnings of Quantum Computation - Elementary Quantum
Computing Devices}}

\qquad\bigskip

We begin this section with some examples of quantum computing devices. \ By a
\textbf{quantum computing device}\footnote{Unfortunately, Physicists have
``stolen'' the akronym QCD. :-)} we mean a unitary transformation $U$ that is
the composition of finitely many sufficiently local unitary transformations,
i.e.,
\[
U=U_{n-1}U_{n-2}\ldots U_{1}U_{0}\text{, }%
\]
where $U_{n-1}$, $U_{n-2}$, $\ldots$ ,$U_{1}$ ,$U_{0}$ are sufficiently
local\footnote{See Definition 8 in Section 7.7 of this paper for a definition
of the term `sufficiently local'.} unitary transformations. Each $U_{j}$ is
called a \textbf{computational step}
\index{Computational Step} of the device. \ 

\bigskip

Our first examples will be obtained by embedding
\index{Embedding} classical computing devices
\index{Computing Device!Classical} within the realm of quantum mechanics. \ We
will then look at some other quantum computing devices
\index{Computing Device!Quantum} that are not the embeddings of classical
devices. \ 

\subsection{Embedding classical (memoryless) computation in quantum mechanics}

\qquad\bigskip

One objective in this section is to represent\footnote{Double meaning is
intended.} classical computing computing devices as unitary transformations.
\ Since unitary transformations are invertible, i.e., reversible, it follows
that the only classical computing devices that can be represented as such
transformations must of necessity be reversible devices. \ Hence, the keen
interest in reversible computation.

\index{Computation!Reversible}
\index{Reversible Computation}

\bigskip

For a more in depth study of reversible computation, please refer to the work
of Bennett
\index{Bennett} and others.

\subsection{Classical reversible computation without memory}%

\[%
\begin{tabular}
[c]{lll}%
Input$\left\{
\begin{tabular}
[c]{ll}%
$x_{n-1}$ & $\longrightarrow$\\
$x_{n-1}$ & $\longrightarrow$\\
$\vdots$ & $\vdots$\\
$x_{1}$ & $\longrightarrow$\\
$x_{0}$ & $\longrightarrow$%
\end{tabular}
\right.  $ &
\begin{tabular}
[c]{|l|}\hline
\\
\\
\\
CRCD$_{n}$\\
\\
\\
\\\hline
\end{tabular}
& $\left.
\begin{tabular}
[c]{ll}%
$\longrightarrow$ & $y_{n-1}$\\
$\longrightarrow$ & $y_{n-1}$\\
$\vdots$ & $\vdots$\\
$\longrightarrow$ & $y_{1}$\\
$\longrightarrow$ & $y_{0}$%
\end{tabular}
\right\}  $Output
\end{tabular}
\]

Each \textbf{classical} $n$-input/$n$-output (binary memoryless)
\textbf{reversible computing device} (\textbf{CRCD}$_{n}$)
\index{CRCD|see{Computing Device, Classlical Reversible}} can be identified
with a bijection
\[
\pi:\left\{  0,1\right\}  ^{n}\longrightarrow\left\{  0,1\right\}  ^{n}%
\]
on the set $\left\{  0,1\right\}  ^{n}$ of all binary $n$-tuples. \ Thus, we
can in turn identify each CRCD$_{n}$ with an element of the permutation group
\index{Permutation Group} $S_{2^{n}}$ on the $2^{n}$ symbols
\[
\left\{  \ \left\langle \overrightarrow{a}\right|  \quad\mid\quad
\overrightarrow{a}\in\left\{  0,1\right\}  ^{n}\ \right\}  \text{ .}%
\]

Let
\[
\mathcal{B}_{n}=\mathcal{B}\left\langle x_{0},x_{1},\ \ldots\ ,x_{n-1}%
\right\rangle
\]
denote the \textbf{free Boolean ring
\index{Boolean Ring@Boolean Ring $\mathcal{B}_{n}$}} on the symbols
$x_{0},x_{1},\ \ldots\ ,x_{n-1}$ . \ Then the binary $n$-tuples
$\overrightarrow{a}\in\left\{  0,1\right\}  ^{n}$ are in one-to-one
correspondence with the \textbf{minterms} of $\mathcal{B}_{n}$, i.e.,
\[
\overrightarrow{a}\longleftrightarrow x^{\overrightarrow{a}}=\prod_{j=0}%
^{n-1}x_{j}^{a_{j}}\text{ ,}%
\]
where
\[
\left\{
\begin{array}
[c]{ccc}%
x_{j}^{0} & = & \overline{x}_{j}\\
&  & \\
x_{j}^{1} & = & x_{j}%
\end{array}
\right.
\]
Since there is a one-to-one correspondence between the automorphisms of
$\mathcal{B}_{n}$ and the permutations on the set of minterms, it follows that
CRCD$_{n}$'s can also be identified with the \textbf{automorphism group}
\index{Automorphism Group} $Aut\left(  \mathcal{B}_{n}\right)  $ of \ the free
Boolean ring $\mathcal{B}_{n}$.

\bigskip

Moreover, since the set of binary $n$-tuples $\left\{  0,1\right\}  ^{n}$ is
in one-to-one correspondence with the set of integers $\left\{  0,1,2,\ \ldots
\ ,2^{n}-1\right\}  $ via the radix 2 representation of integers, i.e.,
\[
\left(  b_{n-1},b_{n-2},\ \ldots\ ,b_{1},b_{0}\right)  \longleftrightarrow
\sum_{j=0}^{n-1}b_{j}2^{j}\text{ ,}%
\]
we can, and frequently do, identify binary $n$-tuples with integers.

\bigskip

For example, consider the Controlled-NOT
\index{Controlled-NOT} gate, called $\mathbf{CNOT}$
\index{CNOT|see{Controlled-NOT}}, which is defined by the following
\textbf{wiring diagram}:
\index{Wiring Diagram|(}
\[
\mathbf{CNOT}=\fbox{$%
\begin{array}
[c]{ccc}%
c & \longrightarrow\oplus\longrightarrow &  b+c\\
& \mid & \\
b & \longrightarrow\bullet\longrightarrow &  b\\
&  & \\
a & \longrightarrow\longrightarrow\longrightarrow &  a
\end{array}
$}\text{ ,}%
\]
where `$\bullet$' and `$\oplus$' denote respectively a \textbf{control bit}
and a \textbf{target bit}, and where `$a+b$' denotes the exclusive `or' of
bits $a$ and $b$. \ This corresponds to the permutation $\pi=(26)(37)$, i.e.,
\[
\left\{
\begin{array}
[c]{ccccc}%
\left|  0\right\rangle = & \left|  000\right\rangle  & \longmapsto & \left|
000\right\rangle  & =\left|  0\right\rangle \\
\left|  1\right\rangle = & \left|  001\right\rangle  & \longmapsto & \left|
001\right\rangle  & =\left|  1\right\rangle \\
\left|  2\right\rangle = & \left|  010\right\rangle  & \longmapsto & \left|
110\right\rangle  & =\left|  6\right\rangle \\
\left|  3\right\rangle = & \left|  011\right\rangle  & \longmapsto & \left|
111\right\rangle  & =\left|  7\right\rangle \\
&  &  &  & \\
\left|  4\right\rangle = & \left|  100\right\rangle  & \longmapsto & \left|
100\right\rangle  & =\left|  4\right\rangle \\
\left|  5\right\rangle = & \left|  101\right\rangle  & \longmapsto & \left|
101\right\rangle  & =\left|  5\right\rangle \\
\left|  6\right\rangle = & \left|  110\right\rangle  & \longmapsto & \left|
010\right\rangle  & =\left|  2\right\rangle \\
\left|  7\right\rangle = & \left|  111\right\rangle  & \longmapsto & \left|
011\right\rangle  & =\left|  3\right\rangle
\end{array}
\right.  ,
\]
where we have used the following indexing conventions:
\[
\left\{
\begin{tabular}
[c]{l}%
First=Right=Bottom\\
\\
Last=Left=Top
\end{tabular}
\right.
\]

As another example, consider the \textbf{Toffoli} gate
\index{Toffoli Gate}, which is defined by the following wiring diagram:
\[
\mathbf{Toffoli}=\fbox{$%
\begin{array}
[c]{ccc}%
c & \longrightarrow\oplus\longrightarrow &  c+ab\\
& \mid & \\
b & \longrightarrow\bullet\longrightarrow &  b\\
& \mid & \\
a & \longrightarrow\bullet\longrightarrow &  a
\end{array}
$}\text{ ,}%
\]
where `$ab$' denotes the logical `and' of $a$ and $b$. \ As before, `$+$'
denotes exclusive `or'. \ This gate corresponds to the permutation $\pi=(67) $.

\bigskip

In summary, we have:
\[
\fbox{$\overset{}{\underset{}{\left\{  \ CRCD_n\ \right\}  }}=S_2^{n}%
=Aut\left(  \mathcal{B}_n\right)  $}%
\]

\bigskip

\subsection{Embedding classical irreversible
\index{Computation!Irreversible} computation within classical reversible computation}

\qquad\bigskip

A
\index{Embedding} classical 1-input/n-output (binary memoryless) irreversible
computing device can be thought of as a Boolean function $f=f(x_{n-2}%
,\ldots,x_{1},x_{0})$ in $\mathcal{B}_{n-1}=\mathcal{B}\left\langle
x_{0},x_{1},\ldots,x_{n-2}\right\rangle $. Such irreversible computing devices
can be transformed into reversible computing devices via the monomorphism
\[
\iota:\mathcal{B}_{n-1}\longrightarrow Aut(\mathcal{B}_{n}),
\]
where $\iota(f)$ is the automorphism in $Aut(\mathcal{B}_{n})$ defined by
\[
\left(  x_{n-1},x_{n-2},\ldots,x_{1},x_{0}\right)  \longmapsto\left(
x_{n-1}\oplus f,x_{n-2},\ldots,x_{1},x_{0}\right)  ,
\]
and where `$\oplus$' denotes exclusive `or'. Thus, the image of each Boolean
function $f$ is a product of disjoint transpositions in $S_{2^{n}}$.

As an additive group (ignoring ring structure), $\mathcal{B}_{n-1}$ is the
abelian group $\bigoplus_{j=0}^{2^{(n-1)}-1}\mathbb{Z}_{2}$, where
$\mathbb{Z}_{2}$ denotes the cyclic group of order two.

\bigskip

Classical Binary Memoryless Computation is summarized in the table below:
\[%
\begin{tabular}
[c]{||c||}\hline\hline
$\overset{}{\text{Summary}}$\\
$\underset{}{\text{Classical Binary Memoryless Computation}}$\\\hline\hline
$\overset{}{\underset{}{\mathcal{B}_{n-1}=\bigoplus_{j=0}^{2^{(n-1)}%
-1}\mathbb{Z}_{2}\overset{\iota}{\longrightarrow}S_{2^{n}}=Aut(\mathcal{B}%
_{n})}}$\\\hline\hline
\end{tabular}
\]

\bigskip

\subsection{The unitary representation of reversible computing devices}

\quad\bigskip

It
\index{Embedding} is now a straight forward task to represent CRCD$_{n}$'s as
unitary transformations. \ We simply use the \textbf{standard unitary
representation}
\index{Standard Unitary Representation}
\[
\fbox{\fbox{$\nu:S_2^{n}\longrightarrow\mathbb{U}(2^{n};\mathbb{C})$}}%
\]
of the symmetric group
\index{Symmetric Group} $S_{2^{n}}$ into the group of $2^{n}\times2^{n}$
unitary matrices $\mathbb{U}(2^{n};\mathbb{C})$. \ This is the representation
defined by
\[
\pi\longmapsto\left(  \delta_{k,\pi k}\right)  _{2^{n}\times2^{n}}\text{ ,}%
\]
where $\delta_{k\ell}$ denotes the Kronecker delta, i.e.,
\[
\delta_{k\ell}=\left\{
\begin{array}
[c]{cl}%
1 & \text{if }k=\ell\\
& \\
0 & \text{otherwise}%
\end{array}
\right.
\]
We think of such unitary transformations as quantum computing devices.\bigskip

For example, consider the controlled-NOT
\index{Controlled-NOT} gate $\mathbf{CNOT}^{\prime}=(45)(67)\in S_{8}$ given
by the wiring diagram
\[
\mathbf{CNOT}^{\prime}=\fbox{$%
\begin{array}
[c]{ccc}%
c & \longrightarrow\bullet\longrightarrow &  c\\
& \mid & \\
b & \longrightarrow\longrightarrow\longrightarrow &  b\\
& \mid & \\
a & \longrightarrow\oplus\longrightarrow &  a+c
\end{array}
$}%
\]
This corresponds to the unitary transformation
\[
U_{\mathbf{CNOT}^{\prime}}=\nu(\mathbf{CNOT}^{\prime})=\left(
\begin{array}
[c]{cccccccc}%
1 & 0 & 0 & 0 & 0 & 0 & 0 & 0\\
0 & 1 & 0 & 0 & 0 & 0 & 0 & 0\\
0 & 0 & 1 & 0 & 0 & 0 & 0 & 0\\
0 & 0 & 0 & 1 & 0 & 0 & 0 & 0\\
0 & 0 & 0 & 0 & 0 & 1 & 0 & 0\\
0 & 0 & 0 & 0 & 1 & 0 & 0 & 0\\
0 & 0 & 0 & 0 & 0 & 0 & 0 & 1\\
0 & 0 & 0 & 0 & 0 & 0 & 1 & 0
\end{array}
\right)
\]

\bigskip

Moreover, consider the Toffoli gate
\index{Toffoli Gate} $\mathbf{Toffoli}^{\prime}=(57)\in S_{8}$ given by the
wiring diagram
\[
\mathbf{Toffoli}^{\prime}=\fbox{$%
\begin{array}
[c]{ccc}%
c & \longrightarrow\bullet\longrightarrow &  c\\
& \mid & \\
b & \longrightarrow\oplus\longrightarrow &  b+ac\\
& \mid & \\
a & \longrightarrow\bullet\longrightarrow &  a
\end{array}
$}\text{ }%
\]
This corresponds to the unitary transformation
\[
U_{\mathbf{Toffoli}^{\prime}}=\nu(\mathbf{Toffoli}^{\prime})=\left(
\begin{array}
[c]{cccccccc}%
1 & 0 & 0 & 0 & 0 & 0 & 0 & 0\\
0 & 1 & 0 & 0 & 0 & 0 & 0 & 0\\
0 & 0 & 1 & 0 & 0 & 0 & 0 & 0\\
0 & 0 & 0 & 1 & 0 & 0 & 0 & 0\\
0 & 0 & 0 & 0 & 1 & 0 & 0 & 0\\
0 & 0 & 0 & 0 & 0 & 0 & 0 & 1\\
0 & 0 & 0 & 0 & 0 & 0 & 1 & 0\\
0 & 0 & 0 & 0 & 0 & 1 & 0 & 0
\end{array}
\right)
\]

\begin{description}
\item [Abuse of Notation and a Caveat]Whenever it is clear from context, we
will use the name of a CRCD$_{n}$ to also refer to the unitary transformation
corresponding to the CRCD$_{n}$. For example, we will denote $\nu(CNOT)$ and
$\nu(Toffoli)$ simply by $CNOT$ and $Toffoli$. Moreover we will also use the
wiring diagram of a CRCD$_{n}$ to refer to the unitary transformation
corresponding to the CRCD$_{n}$. For quantum computation beginners, this can
lead to some confusion. Be careful!
\end{description}

\bigskip

\subsection{Some other simple quantum computing devices}

\qquad\bigskip

After CRCD$_{n}$'s are embedded as quantum computing devices, they are no
longer classical computing devices. \ After the embedding, they suddenly have
acquired much more computing power. \ Their inputs and outputs can be a
superposition of many states. \ They can entangle their outputs. It is
misleading to think of their input qubits as separate, for they could be entangled.

\bigskip

As an illustration of this fact, please note that the quantum computing device
$\mathbf{CNOT}^{\prime\prime}$
\index{Controlled-NOT} given by the wiring diagram
\[
\mathbf{CNOT}^{\prime\prime}=\fbox{$%
\begin{array}
[c]{ccc}%
b & \longrightarrow\bullet\longrightarrow &  a+b\\
& \mid & \\
a & \longrightarrow\oplus\longrightarrow &  a
\end{array}
$}=\left(
\begin{array}
[c]{cccc}%
1 & 0 & 0 & 0\\
0 & 1 & 0 & 0\\
0 & 0 & 0 & 1\\
0 & 0 & 1 & 0
\end{array}
\right)
\]
is far from classical. \ It is more than a permutation. \ It is a linear
operator that respects quantum superposition. \ 

\bigskip

For example, $\mathbf{CNOT}^{\prime\prime}$ can take two non-entangled qubits
as input, and then produce two entangled qubits as output. \ This is something
no classical computing device can do. \ For example,
\[
\frac{\left|  0\right\rangle -\left|  1\right\rangle }{\sqrt{2}}\otimes\left|
0\right\rangle =\frac{1}{\sqrt{2}}\left(  \left|  00\right\rangle -\left|
10\right\rangle \right)  \longmapsto\frac{1}{\sqrt{2}}\left(  \left|
00\right\rangle -\left|  11\right\rangle \right)
\]

\vspace{0.5in}

For completeness, we list two other quantum computing devices that are
embeddings of CRCD$_{n}$'s, $\mathbf{NOT}$ and
\index{NOT gate} $\mathbf{SWAP}$:
\index{SWAP Gate}
\[
\mathbf{NOT}=\fbox{$%
\begin{array}
[c]{ccc}%
a & \longrightarrow\fbox{$\textbf{NOT}$}\longrightarrow &  a+1
\end{array}
$}=\left(
\begin{array}
[c]{cc}%
0 & 1\\
1 & 0
\end{array}
\right)  =\sigma_{1}%
\]
and
\[
\mathbf{SWAP}=\fbox{$%
\begin{array}
[c]{ccccc}%
b & \longrightarrow\bullet\longrightarrow & \longrightarrow\oplus
\longrightarrow & \longrightarrow\bullet\longrightarrow &  a\\
& \mid & \mid & \mid & \\
a & \longrightarrow\oplus\longrightarrow & \longrightarrow\bullet
\longrightarrow & \longrightarrow\oplus\longrightarrow &  b
\end{array}
$}=\left(
\begin{array}
[c]{cccc}%
1 & 0 & 0 & 0\\
0 & 0 & 1 & 0\\
0 & 1 & 0 & 0\\
0 & 0 & 0 & 1
\end{array}
\right)
\]

\vspace{0.5in}

\subsection{Quantum computing devices that are not embeddings}

\qquad\bigskip

We now consider quantum computing devices that are not embeddings of
CRCD$_{n}$'s.

\vspace{0.5in}

The \textbf{Hadamard} gate $\mathbf{H}$ is defined as:
\[
\mathbf{H}=\fbox{$%
\begin{array}
[c]{c}%
\longrightarrow\mathbf{H}\longrightarrow
\end{array}
$}=\frac{1}{\sqrt{2}}\left(
\begin{array}
[c]{rr}%
1 & 1\\
1 & -1
\end{array}
\right)  \text{ .}%
\]

\bigskip

Another quantum gate is the \textbf{square root of NOT},
\index{Square Root of!NOT} i.e., $\sqrt{\mathbf{NOT}}$, which is given by
\[
\sqrt{\mathbf{NOT}}=\fbox{$%
\begin{array}
[c]{c}%
\longrightarrow\sqrt{\mathbf{NOT}}\longrightarrow
\end{array}
$}=\frac{1-i}{2}\left(
\begin{array}
[c]{rr}%
i & 1\\
1 & i
\end{array}
\right)  =\frac{1+i}{2}\left(
\begin{array}
[c]{rr}%
1 & -i\\
-i & 1
\end{array}
\right)  \text{ .}%
\]

\bigskip

There is also the \textbf{square root of swap}
\index{Square Root of!SWAP} $\sqrt{\mathbf{SWAP}}$ which is defined as:
\[
\sqrt{\mathbf{SWAP}}=\fbox{$%
\begin{array}
[c]{c}%
\longrightarrow\sqrt{\mathbf{SWAP}}\longrightarrow
\end{array}
$}=\left(
\begin{array}
[c]{cccc}%
1 & 0 & 0 & 0\\
0 & \frac{1+i}{2} & \frac{1-i}{2} & 0\\
0 & \frac{1-i}{2} & \frac{1+i}{2} & 0\\
0 & 0 & 0 & 1
\end{array}
\right)  \text{ .}%
\]

\bigskip

Three frequently used unary quantum gates are the rotations:
\index{Rotation}
\[
\fbox{$%
\begin{array}
[c]{c}%
\longrightarrow\fbox{$e^{i\theta\sigma_1}$}\longrightarrow
\end{array}
$}=\left(
\begin{array}
[c]{rr}%
\cos\theta &  i\sin\theta\\
i\sin\theta & \cos\theta
\end{array}
\right)  =e^{i\theta\sigma_{1}}%
\]%
\[
\fbox{$%
\begin{array}
[c]{c}%
\longrightarrow\fbox{$e^{i\theta\sigma_2}$}\longrightarrow
\end{array}
$}=\left(
\begin{array}
[c]{rr}%
\cos\theta & \sin\theta\\
-\sin\theta & \cos\theta
\end{array}
\right)  =e^{i\theta\sigma2}%
\]%
\[
\fbox{$%
\begin{array}
[c]{c}%
\longrightarrow\fbox{$e^{i\theta\sigma_3}$}\longrightarrow
\end{array}
$}=\left(
\begin{array}
[c]{cc}%
e^{i\theta} & 0\\
0 & e^{-i\theta}%
\end{array}
\right)  =e^{i\theta\sigma_{3}}%
\]

\bigskip

\subsection{\textbf{\bigskip}The implicit frame of a wiring diagram}

\qquad\bigskip

\index{Wiring Diagrams!Implicit Frame}

Wiring diagrams have the advantage of being a simple means of describing some
rather complicated unitary transformations. \ However, they do have their
drawbacks, and they can, if we are not careful, be even misleading.

\bigskip

One problem with wiring diagrams is that they are not frame (i.e., basis)
independent descriptions of unitary transformations. \ Each wiring diagram
describes a unitary transformation using an implicitly understood basis.

\vspace{0.5in}

For example, consider $\mathbf{CNOT}^{\prime\prime}$ given by the wiring
diagram:
\[
\mathbf{CNOT}^{\prime\prime}=\fbox{$%
\begin{array}
[c]{ccc}%
b & \longrightarrow\bullet\longrightarrow &  a+b\\
& \mid & \\
a & \longrightarrow\oplus\longrightarrow &  a
\end{array}
$}\text{ .}%
\]
The above wiring diagram defines $\mathbf{CNOT}^{\prime\prime}$ in terms of
the implicitly understood basis
\[
\left\{  \left|  0\right\rangle =\left(
\begin{array}
[c]{c}%
1\\
0
\end{array}
\right)  ,\ \left|  1\right\rangle =\left(
\begin{array}
[c]{c}%
0\\
1
\end{array}
\right)  \right\}  \text{ .}%
\]
This wiring diagram suggests that qubit \#1 controls qubit \#0, and that qubit
\#1 is not effected by qubit \#0. \ But this is far from the truth. \ For,
$\mathbf{CNOT}^{\prime\prime}$ transforms
\[
\frac{\left|  0\right\rangle +\left|  1\right\rangle }{\sqrt{2}}\otimes
\frac{\left|  0\right\rangle -\left|  1\right\rangle }{\sqrt{2}}%
\]
into
\[
\frac{\left|  0\right\rangle -\left|  1\right\rangle }{\sqrt{2}}\otimes
\frac{\left|  0\right\rangle -\left|  1\right\rangle }{\sqrt{2}}\text{ ,}%
\]
where we have used our indexing conventions
\[
\left\{
\begin{tabular}
[c]{l}%
First=Right=Bottom\\
\\
Last=Left=Top
\end{tabular}
\right.  \text{ .}%
\]

\bigskip

In fact, in the basis
\[
\left\{  \left|  0^{\prime}\right\rangle =\frac{\left|  0\right\rangle
+\left|  1\right\rangle }{\sqrt{2}},\ \left|  1^{\prime}\right\rangle
=\frac{\left|  0\right\rangle -\left|  1\right\rangle }{\sqrt{2}}\right\}
\]
the wiring diagram of the same unitary transformation $\mathbf{CNOT}%
^{\prime\prime}$ is:
\[
\fbox{$%
\begin{array}
[c]{ccc}%
b & \longrightarrow\oplus\longrightarrow &  a+b\\
& \mid & \\
a & \longrightarrow\bullet\longrightarrow &  a
\end{array}
$}%
\]
The roles of the target and control qubits appeared to have switched!

\index{Wiring Diagram}

\section{\textbf{\bigskip The No-Cloning Theorem}}

\qquad\bigskip

\index{No-Cloning Theorem}

In this section, we prove the no-cloning theorem of Wootters
\index{Wootters} and Zurek
\index{Zurek}\cite{Wootters1}. \ The theorem states that there can be no
device that produces exact replicas or copies of a quantum state. \ (See also
\cite{Wootters1} for an elegant proof using the creation operators of quantum electrodynamics.)

\bigskip

The proof is an amazingly simple application of the linearity of quantum
mechanics. \ The key idea is that copying is an inherently quadratic
transformation, while the unitary transformations of quantum mechanics are
inherently linear. \ Ergo, copying can not be a unitary transformation. \bigskip

But what do we mean by a quantum replicator?\bigskip

\begin{definition}
Let $\mathcal{H}$ be a Hilbert space. \ Then a \textbf{quantum replicator
\index{Quantum!Replicator}
\index{Replicator!Quantum}} consists of an auxiliary Hilbert space
$\mathcal{H}_{A}$, a fixed state $\left|  \psi_{0}\right\rangle \in
\mathcal{H}_{A}$ (called the \textbf{initial state of replicator}), and a
unitary transformation
\[
U:\mathcal{H}_{A}\otimes\mathcal{H}\oplus\mathcal{H}\longrightarrow
\mathcal{H}_{A}\otimes\mathcal{H}\oplus\mathcal{H}%
\]
such that, for some fixed state $\left|  blank\right\rangle \in\mathcal{H}$,
\[
U\left|  \psi_{0}\right\rangle \left|  a\right\rangle \left|
blank\right\rangle =\left|  \psi_{a}\right\rangle \left|  a\right\rangle
\left|  a\right\rangle \text{ ,}%
\]
for all states $\left|  a\right\rangle \in\mathcal{H}$, where $\left|
\psi_{a}\right\rangle \in\mathcal{H}_{A}$ (called the \textbf{replicator state
after replication} of $\left|  a\right\rangle $) depends on $\left|
a\right\rangle $.
\end{definition}

\bigskip

Since a quantum state is determined by a ket up to a multiplicative non-zero
complex number, we can without loss of generality assume that $\left|
\psi_{0}\right\rangle $, $\left|  a\right\rangle $, $\left|
blank\right\rangle $ are all of unit length. \ From unitarity, it follows that
$\left|  \psi_{a}\right\rangle $ is also of unit length.

\bigskip

Let $\left|  a\right\rangle $, $\left|  b\right\rangle $ be two kets of unit
length in $\mathcal{H}$ such that
\[
0<\left|  \ \left\langle \ a\mid b\ \right\rangle \ \right|  <1\text{ .}%
\]
Then
\[
\left\{
\begin{array}
[c]{ccc}%
U\left|  \psi_{0}\right\rangle \left|  a\right\rangle \left|
blank\right\rangle  & = & \left|  \psi_{a}\right\rangle \left|  a\right\rangle
\left|  a\right\rangle \\
&  & \\
U\left|  \psi_{0}\right\rangle \left|  b\right\rangle \left|
blank\right\rangle  & = & \left|  \psi_{b}\right\rangle \left|  b\right\rangle
\left|  b\right\rangle
\end{array}
\right.
\]

\bigskip

Hence,
\begin{align*}
\left\langle blank\right|  \left\langle a\right|  \left\langle \psi
_{0}\right|  U^{\dagger}U\left|  \psi_{0}\right\rangle \left|  b\right\rangle
\left|  blank\right\rangle  &  =\left\langle blank\right|  \left\langle
a\right|  \left\langle \ \psi_{0}\mid\psi_{0}\ \right\rangle \left|
b\right\rangle \left|  blank\right\rangle \\
&  =\left\langle \ a\mid b\ \right\rangle
\end{align*}

\bigskip

On the other hand,\bigskip%

\begin{align*}
\left\langle blank\right|  \left\langle a\right|  \left\langle \psi
_{0}\right|  U^{\dagger}U\left|  \psi_{0}\right\rangle \left|  b\right\rangle
\left|  blank\right\rangle  &  =\left\langle a\right|  \left\langle a\right|
\left\langle \ \psi_{a}\mid\psi_{b}\ \right\rangle \left|  b\right\rangle
\left|  b\right\rangle \\
&  =\left\langle \ a\mid b\ \right\rangle ^{2}\left\langle \ \psi_{a}\mid
\psi_{b}\ \right\rangle
\end{align*}

\bigskip

Thus,
\[
\left\langle \ a\mid b\ \right\rangle ^{2}\left\langle \ \psi_{a}\mid\psi
_{b}\ \right\rangle =\left\langle \ a\mid b\ \right\rangle \text{ .}%
\]
And \ so,
\[
\left\langle \ a\mid b\ \right\rangle \left\langle \ \psi_{a}\mid\psi
_{b}\ \right\rangle =1\text{ .}%
\]

\bigskip

But this equation can not be satisfied since
\[
\left|  \left\langle \ a\mid b\ \right\rangle \right|  <1
\]
and
\[
\left|  \left\langle \ \psi_{a}\mid\psi_{b}\ \right\rangle \right|
\leq\left\|  \ \left|  \psi_{a}\right\rangle \ \right\|  \left\|  \ \left|
\psi_{b}\right\rangle \ \right\|  =1
\]

\bigskip

Hence, a quantum replicator cannot exist.

\index{No-Cloning Theorem}

\bigskip

\section{\textbf{\bigskip Quantum teleportation}}

\qquad\bigskip

\bigskip

We now give a brief description of quantum teleportation, a means possibly to
be used by future quantum computers to bus qubits from one location to
another. \
\index{Teleportation, Quantum}

\bigskip

As stated earlier, qubits can not be copied as a result of the no-cloning
theorem. \ (Please refer to the previous section.) \ However, they can be
teleported, as has been demonstrated in laboratory settings. \ Such a
mechanism could be used to bus qubits from one computer location to another.
\ It could be used to create devices called \textbf{quantum repeaters}.
\index{Quantum!Repeater}
\index{Repeater!Quantum}

\vspace{0.5in}

But what do we mean by teleportation?

\bigskip

\index{Teleportation}

\textbf{Teleportation} is the transferring of an object from one location to
another by a process that:

\begin{itemize}
\item [1)]Firstly dissociates (i.e., destroys) the object to obtain
information. -- The object to be teleported is first scanned to extract
sufficient information to reassemble the original object.

\item[2)] Secondly transmits the acquired information from one location to another.

\item[3)] Lastly reconstructs the object at the new location from the received
information. -- An exact replicas re-assembled at the destination out of
locally available materials.
\end{itemize}

\bigskip

Two key effects of teleportation should be noted:

\begin{itemize}
\item [1)]The original object is destroyed during the process of
teleportation. \ Hence, the no-cloning theorem is not violated.

\item[2)] An exact replica of the original object is created at the intended destination.
\end{itemize}

\vspace{0.5in}

Scotty of the Starship Enterprise was gracious enough to loan me the following
teleportation manual. \ So I am passing it on to you.

\bigskip

\begin{center}
{\Huge Quantum Teleportation Manual}
\end{center}

\bigskip

\begin{description}
\item [Step. 1 .(Location A): Preparation]At location A, construct an EPR pair
of qubits (qubits \#2 and \#3) in $\mathcal{H}_{2}\otimes\mathcal{H}_{3}$.
\[%
\begin{tabular}
[c]{ccc}%
$\left|  00\right\rangle \longmapsto$ & $\overset{}{\underset{}{\fbox{%
\begin{tabular}
[c]{c}%
Unitary\\
Matrix
\end{tabular}
}}}$ & $\longmapsto\frac{\left|  01\right\rangle -\left|  10\right\rangle
}{\sqrt{2}}$\\
$\mathcal{H}_{2}\otimes\mathcal{H}_{3}$ & $\longrightarrow$ & $\mathcal{H}%
_{2}\otimes\mathcal{H}_{3}$%
\end{tabular}
\]

\item[Step 2. Transport] Physically transport entangled qubit \#3 from
location A to location B.\bigskip

\item[Step 3. ] The qubit to be teleported, i.e., qubit \#1, is delivered to
location A in an unknown state
\[
a\left|  0\right\rangle +b\left|  1\right\rangle
\]
\bigskip
\end{description}

As a result of Steps 1 - 3, we have:

\begin{itemize}
\item  Locations A and B share an EPR pair, i.e.

\begin{itemize}
\item  The qubit which is to be teleported, i.e., qubit \#1, is at Location A

\item  Qubit \#2 is at Location A

\item  Qubit \#3 is at Location B

\item  Qubits \#2 \& \#3 are entangled
\end{itemize}

\item  The current state $\left|  \Phi\right\rangle $ of all three qubits is:
\[
\left|  \Phi\right\rangle =\left(  a\left|  0\right\rangle +b\left|
1\right\rangle \right)  \left(  \frac{\left|  01\right\rangle -\left|
10\right\rangle }{\sqrt{2}}\right)  \in\mathcal{H}_{1}\otimes\mathcal{H}%
_{2}\otimes\mathcal{H}_{3}%
\]
\end{itemize}

\bigskip

To better understand what is about to happen, we re-express the state $\left|
\Phi\right\rangle $ of the three qubits in terms of the following basis
(called the \textbf{Bell basis}) of $\mathcal{H}_{1}\otimes\mathcal{H}_{2}$ :
\[
\left\{
\begin{array}
[c]{ccc}%
\left|  \Psi_{A}\right\rangle  & = & \frac{\left|  10\right\rangle -\left|
01\right\rangle }{\sqrt{2}}\\
&  & \\
\left|  \Psi_{B}\right\rangle  & = & \frac{\left|  10\right\rangle +\left|
01\right\rangle }{\sqrt{2}}\\
&  & \\
\left|  \Psi_{C}\right\rangle  & = & \frac{\left|  00\right\rangle -\left|
11\right\rangle }{\sqrt{2}}\\
&  & \\
\left|  \Psi_{D}\right\rangle  & = & \frac{\left|  00\right\rangle +\left|
11\right\rangle }{\sqrt{2}}%
\end{array}
\right.
\]
The result is:
\[%
\begin{array}
[c]{ccc}%
\left|  \Phi\right\rangle  & = &
\begin{array}
[c]{c}%
\frac{1}{2}[\quad\left|  \Psi_{A}\right\rangle \left(  -a\left|
0\right\rangle -b\left|  1\right\rangle \right) \\
\ +\left|  \Psi_{B}\right\rangle \left(  -a\left|  0\right\rangle +b\left|
1\right\rangle \right) \\
+\left|  \Psi_{C}\right\rangle \left(  a\left|  1\right\rangle +b\left|
0\right\rangle \right) \\
\quad+\left|  \Psi_{D}\right\rangle \left(  a\left|  1\right\rangle -b\left|
0\right\rangle \right)  \quad]
\end{array}
\end{array}
\text{ ,}%
\]
where, as you might have noticed, we have written the expression in a
suggestive way.

\bigskip

\begin{remark}
Please note that since the completion of Step 3, we have done nothing
physical. \ We have simply performed some algebraic manipulation of the
expression representing the state $\left|  \Phi\right\rangle $ of the three qubits.
\end{remark}

\bigskip

Let $U:\mathcal{H}_{1}\otimes\mathcal{H}_{2}\longrightarrow\mathcal{H}%
_{1}\otimes\mathcal{H}_{2}$ be the unitary transformation defined by
\[
\left\{
\begin{array}
[c]{ccc}%
\overset{}{\underset{}{\left|  \Psi_{A}\right\rangle }} & \longmapsto &
\left|  00\right\rangle \\
\overset{}{\underset{}{\left|  \Psi_{B}\right\rangle }} & \longmapsto &
\left|  01\right\rangle \\
\overset{}{\underset{}{\left|  \Psi_{C}\right\rangle }} & \longmapsto &
\left|  10\right\rangle \\
\overset{}{\underset{}{\left|  \Psi_{D}\right\rangle }} & \longmapsto &
\left|  11\right\rangle
\end{array}
\right.
\]

\bigskip

\begin{description}
\item [Step 4. (Location A)]\footnote{Actually, there is no need to apply the
unitary transformation $U$. \ We could have instead made a complete Bell state
measurement, i.e., a measurement with respect to the compatible observables
$\left|  \Psi_{A}\right\rangle \left\langle \Psi_{A}\right|  $, $\left|
\Psi_{B}\right\rangle \left\langle \Psi_{B}\right|  $, $\left|  \Psi
_{C}\right\rangle \left\langle \Psi_{C}\right|  $, $\left|  \Psi
_{D}\right\rangle \left\langle \Psi_{D}\right|  $. \ We have added the
additional step 4 to make quantum teleportation easier to understand for
quantum computation beginners. \
\par
Please note that a complete Bell state measurement has, of this writing, yet
to be achieved in a laboratoy setting.}Apply the local unitary transformation
$U\otimes I:\mathcal{H}_{1}\otimes\mathcal{H}_{2}\otimes\mathcal{H}%
_{3}\longrightarrow\mathcal{H}_{1}\otimes\mathcal{H}_{2}\otimes\mathcal{H}%
_{3}$ to the three qubits (actually more precisely, to qubits \#1 and \#2).
\ Thus, under $U\otimes I$ the state $\left|  \Phi\right\rangle $ of all three
qubits becomes
\[%
\begin{array}
[c]{ccl}%
\left|  \Phi^{\prime}\right\rangle  & = &
\begin{array}
[c]{c}%
\frac{1}{2}[\quad\left|  00\right\rangle \left(  -a\left|  0\right\rangle
-b\left|  1\right\rangle \right) \\
\quad+\left|  01\right\rangle \left(  -a\left|  0\right\rangle +b\left|
1\right\rangle \right) \\
\quad+\left|  10\right\rangle \left(  a\left|  1\right\rangle +b\left|
0\right\rangle \right) \\
\quad\quad+\left|  11\right\rangle \left(  a\left|  1\right\rangle -b\left|
0\right\rangle \right)  \quad]
\end{array}
\end{array}
\]
\bigskip

\item[Step 5. (Location A)] Measure qubits \#1 and \#2 to obtain two bits of
classical information. \ The result of this measurement will be one of the bit
pairs $\left\{  00,01,10,11\right\}  $. \bigskip

\item[Step 6.] Send from location A to location B (via a classical
communication channel) the two classical bits obtained in Step 6.
\end{description}

\vspace{0.5in}

As an intermediate summary, we have:

\begin{itemize}
\item [1)]Qubit \#1 has been disassembled, and

\item[2)] The information obtained during disassembly (two classical bits) has
been sent to location B.
\end{itemize}

\vspace{0.5in}

\begin{description}
\item [Step 7. (Location B)]The two bits $(i,j)$ received from location A are
used to select from the following table a unitary transformation $U^{(i,j)}$
of $\mathcal{H}_{3}$, (i.e., a local unitary transformation $I_{4}\otimes
U^{(i,j)}$ on $\mathcal{H}_{1}\otimes\mathcal{H}_{2}\otimes\mathcal{H}_{3}$)
\[%
\begin{tabular}
[c]{||c||c||c||}%
Rec. Bits & $U^{(i,j)}$ & Future effect on qubit \#3\\\hline\hline
$00$ & $U^{(00)}=\left(
\begin{array}
[c]{rr}%
-1 & 0\\
0 & -1
\end{array}
\right)  $ & $-a\left|  0\right\rangle -b\left|  1\right\rangle \longmapsto
a\left|  0\right\rangle +b\left|  1\right\rangle $\\\hline\hline
$01$ & $U^{(01)}=\left(
\begin{array}
[c]{cc}%
-1 & 0\\
0 & 1
\end{array}
\right)  $ & $-a\left|  0\right\rangle +b\left|  1\right\rangle \longmapsto
a\left|  0\right\rangle +b\left|  1\right\rangle $\\\hline\hline
$10$ & $U^{(10)}=\left(
\begin{array}
[c]{cc}%
0 & 1\\
1 & 0
\end{array}
\right)  $ & $a\left|  1\right\rangle +b\left|  0\right\rangle \longmapsto
a\left|  0\right\rangle +b\left|  1\right\rangle $\\\hline\hline
$11$ & $U^{(11)}=\left(
\begin{array}
[c]{rr}%
0 & 1\\
-1 & 0
\end{array}
\right)  $ & $a\left|  1\right\rangle -b\left|  0\right\rangle \longmapsto
a\left|  0\right\rangle +b\left|  1\right\rangle $\\\hline\hline
\end{tabular}
\]
\end{description}

\bigskip

\begin{description}
\item [Step 8. (Location B)]The unitary transformation $U^{(i,j)}$ selected in
Step 7 is applied to qubit \#3.
\end{description}

\bigskip

As a result, qubit \#3 is at location B and has the original state of qubit
\#1 when qubit \#1 was first delivered to location A, i.e., the state
\[
a\left|  0\right\rangle +b\left|  1\right\rangle
\]

\vspace{0.5in}

It is indeed amazing that no one knows the state of the quantum teleported
qubit except possibly the individual that prepared the qubit. \ Knowledge of
the actual state of the qubit is not required for teleportaton. \ If its state
is unknown before the teleportation, it remains unknown after the
teleportation. \ All that we know is that the states before and after the
teleportation are the same.

\index{Teleportation, Quantum}

\bigskip

\section{\textbf{Shor's algorithm}}

\bigskip

The following description of Shor's algorithm is based on \cite{Ekert1},
\cite{Hoyer1}, \cite{Jozsa2}, \cite{Kitaev1}, and \cite{Shor1}
\index{Shor}.

\subsection{Preamble to Shor's algorithm}

\quad\bigskip

\index{Shor's Algorithm}

There are cryptographic systems (such as RSA\footnote{RSA is a public key
cryptographic system invented by Rivest, Shamir, Adleman. \ Hence the name.
\ For more information, please refer to \cite{Stinson1}.}) that are
extensively used today (e.g., in the banking industry) which are based on the
following questionable assumption, i.e., conjecture: \ 

\bigskip

\noindent\textbf{Conjecture(Assumption). \ }\textit{Integer factoring is
computationally much harder than integer multiplication. \ In other words,
while there are obviously many polynomial time algorithms for integer
multiplication, there are no polynomial time algorithms for integer factoring.
\ I.e., integer factoring computationally requires super-polynomial time.}

\bigskip

This assumption is based on the fact that, in spite of the intensive efforts
over many centuries of the best minds to find a polynomial time factoring
algorithm, no one has succeeded so far. \ As of this writing, the most
asymptotically efficient \emph{classical} algorithm is the number theoretic
sieve \cite{Lenstra1}, \cite{Lenstra2}, which factors an integer $N$ in time
$O\left(  \exp\left[  \left(  \lg N\right)  ^{1/3}\left(  \lg\lg N\right)
^{2/3}\right]  \right)  $. \ Thus, this is a super-polynomial time algorithm
in the number $O\left(  \lg N\right)  $ of digits in $N$. \ 

\vspace{0.5in}

However, ... Peter Shor suddenly changed the rules of the game. \ \bigskip

Hidden in the above conjecture is the unstated, but implicitly understood,
assumption that all algorithms run on computers based on the principles of
classical mechanics, i.e., on \textbf{classical computers}. \ But what if a
computer could be built that is based not only on classical mechanics, but on
quantum mechanics as well? \ I.e., what if we could build a \textbf{quantum
computer}?

\bigskip

Shor, starting from the works of Benioff, Bennett, Deutsch
\index{Deutsch}, Feynman, Simon, and others, created an algorithm to be run on
a quantum computer, i.e., a \textbf{quantum algorithm}, that factors integers
in polynomial time! \ Shor's algorithm takes asymptotically $O\left(  \left(
\lg N\right)  ^{2}\left(  \lg\lg N\right)  \left(  \lg\lg\lg N\right)
\right)  $ steps on a quantum computer, which is polynomial time in the number
of digits $O\left(  \lg N\right)  $ of $N$. \ 

\vspace{0.5in}

\subsection{Number theoretic preliminaries}

\quad\bigskip

Since the time of Euclid, it has been known that every positive integer $N$
can be uniquely (up to order) factored into the product of primes. \ Moreover,
it is a computationally easy (polynomial time) task to determine whether or
not $N$ is a prime or composite number. \ For the primality testing algorithm
\index{Primality Testing} of Miller-Rabin\cite{Miller1} makes such a
determination at the cost of $O\left(  s\lg N\right)  $ arithmetic operations
[$O\left(  s\lg^{3}N\right)  $ bit operations] \ with probability of error
$Prob_{Error}\leq2^{-s}$. \ 

\bigskip

However, once an odd positive integer $N$ is known to be composite, it does
not appear to be an easy (polynomial time) task on a classical computer to
determine its prime factors. \ As mentioned earlier, so far the most
asymptotically efficient \emph{classical} algorithm known is the number
theoretic sieve \cite{Lenstra1}, \cite{Lenstra2}, which factors an integer $N
$ in time $O\left(  \exp\left[  \left(  \lg N\right)  ^{1/3}\left(  \lg\lg
N\right)  ^{2/3}\right]  \right)  $.

\bigskip

\noindent\textbf{Prime Factorization Problem.} \ \textit{Given a composite odd
positive integer }$N$\textit{, find its prime factors.}

\bigskip

It is well known\cite{Miller1} that factoring $N$ can be reduced to the task
of choosing at random an integer $m$ relatively prime to $N$, and then
determining its modulo $N$ multiplicative order $P$, i.e., to finding the
smallest positive integer $P$ such that
\[
m^{P}=1\operatorname{mod}N\text{ .}%
\]
It was precisely this approach to factoring that enabled Shor to construct his
factoring algorithm.

\vspace{0.5in}

\subsection{Overview of Shor's algorithm}

\quad\bigskip

But what is Shor's quantum factoring algorithm?

\vspace{0.5in}

Let $\mathbb{N}=\left\{  0,1,2,3,\ldots\right\}  $ denote the set of natural numbers.

\vspace{0.5in}

Shor's algorithm provides a solution to the above problem. \ His algorithm
consists of the five steps (\textbf{steps} \textbf{1} through \textbf{5}),
with only $\mathbb{STEP}$\textbf{\ 2} requiring the use of a quantum computer.
\ The remaining four other steps of the algorithm are to be performed on a
classical computer.

\bigskip

We begin by briefly describing all five steps. \ After that, we will then
focus in on the quantum part of the algorithm, i.e., $\mathbb{STEP}%
$\textbf{\ 2}.

\vspace{0.5in}

\begin{itemize}
\item [\fbox{\textbf{Step 1.}}]Choose a random positive even integer $m$.
\ Use the polynomial time Euclidean algorithm\footnote{The Euclidean algorithm
is $O\left(  \lg^{2}N\right)  $. \ For a description of the Euclidean
algorithm, see for example \cite{Cox1} or \cite{Cormen1}.} to compute the
greatest common divisor $\gcd\left(  m,N\right)  $ of $m$ and $N$. \ If the
greatest common divisor $\gcd\left(  m,N\right)  \neq1$, then we have found a
non-trivial factor of $N$, and we are done. \ If, on the other hand,
$\gcd\left(  m,N\right)  =1$, then proceed to $\mathbb{STEP}$\textbf{\ 2}.
\end{itemize}

\bigskip

\begin{itemize}
\item [\fbox{$\mathbb{STEP}$ \textbf{2.}}]Use a \textsc{quantum computer} to
determine the unknown period $P$ of the function
\[%
\begin{array}
[c]{ccc}%
\mathbb{N} & \overset{f_{N}}{\longrightarrow} & \mathbb{N}\\
a & \longmapsto &  m^{a}\operatorname{mod}N
\end{array}
\]
\end{itemize}

\bigskip

\begin{itemize}
\item [\fbox{\textbf{Step 3.}}]If $P$ is an odd integer, then goto
\textbf{Step 1}. \ [The probability of $P$ being odd is $(%
\frac12
)^{k}$, where $k$ is the number of distinct prime factors of $N$.] \ If $P$ is
even, then proceed to \textbf{Step 4}.
\end{itemize}

\bigskip\ 

\begin{itemize}
\item [\fbox{\textbf{Step 4.}}]Since $P$ is even,
\[
\left(  m^{P/2}-1\right)  \left(  m^{P/2}+1\right)  =m^{P}%
-1=0\operatorname{mod}N\text{ .}%
\]
If $m^{P/2}+1=0\operatorname{mod}N$, then goto \textbf{Step 1}. \ If
$m^{P/2}+1\neq0\operatorname{mod}N$, then proceed to \textbf{Step 5}. \ It can
be shown that the probability that $m^{P/2}+1=0\operatorname{mod}N$ is less
than $(%
\frac12
)^{k-1}$, where $k$ denotes the number of distinct prime factors of $N$.
\end{itemize}

\bigskip

\begin{itemize}
\item [\fbox{\textbf{Step 5.}}]Use the Euclidean algorithm to compute
$d=\gcd\left(  m^{P/2}-1,N\right)  $. \ Since $m^{P/2}+1\neq
0\operatorname{mod}N$, it can easily be shown that $d$ is a non-trivial factor
of $N$. \ Exit with the answer $d$.
\end{itemize}

\vspace{0.5in}

Thus, the task of factoring an odd positive integer $N$ reduces to the
following problem:

\bigskip

\noindent\textbf{Problem.} \textit{Given a periodic function }
\[
f:\mathbb{N}\longrightarrow\mathbb{N}\text{ ,}%
\]
\textit{find the period }$P$\textit{\ of }$f$\textit{.}

\vspace{0.5in}

\subsection{Preparations for the quantum part of Shor's algorithm}

\quad\bigskip

Choose a power of 2
\[
Q=2^{L}%
\]
such that
\[
N^{2}\leq Q=2^{L}<2N^{2}\text{ ,}%
\]
and consider $f$ restricted to the set
\[
S_{Q}=\left\{  0,1,\ldots,Q-1\right\}
\]
which we also denote by $f$, i.e.,
\[
f:S_{Q}\longrightarrow S_{Q}\text{ .}%
\]

\bigskip

In preparation for a discussion of $\mathbb{STEP}$ 2 of Shor's algorithm, we
construct two $L$-qubit quantum registers, \textsc{Register1} and
\textsc{Register2} to hold respectively the arguments and the values of the
function $f$, i.e.,
\[
\left|  \text{\textsc{Reg1}}\right\rangle \left|  \text{\textsc{Reg2}%
}\right\rangle =\left|  a\right\rangle \left|  f(a)\right\rangle =\left|
a\right\rangle \left|  b\right\rangle =\left|  a_{0}a_{1}\cdots a_{L-1}%
\right\rangle \left|  b_{0}b_{1}\cdots b_{L-1}\right\rangle
\]

In doing so, we have adopted the following convention for representing
integers in these registers:

\bigskip

\noindent\textbf{Notation Convention.} \ In a quantum computer, \textit{we
represent an integer }$a$\textit{\ with radix }$2$\textit{\ representation
\index{Radix 2 Representation} }
\[
a=\sum_{j=0}^{L-1}a_{j}2^{j}\text{ , }%
\]
\textit{as a quantum register consisting of the }$2^{n}$\textit{\ qubits }
\[
\left|  a\right\rangle =\left|  a_{0}a_{1}\cdots a_{L-1}\right\rangle =%
{\displaystyle\bigotimes\limits_{j=0}^{L-1}}
\left|  a_{j}\right\rangle
\]

\bigskip

For example, the integer $23$ is represented in our quantum computer as $n$
qubits in the state:
\[
\left|  23\right\rangle =\left|  10111000\cdots0\right\rangle
\]

\vspace{0.5in}

Before continuing, we remind the reader of the classical definition of the $Q
$-point Fourier transform.

\begin{definition}
Let $\omega$ be a primitive $Q$-th root of unity, e.g., $\omega=e^{2\pi i/Q}$.
Then the $Q$-point Fourier transform
\index{Fourier Transform} is the map
\begin{gather*}
Map(S_{Q},\mathbb{C})\overset{\mathcal{F}}{\longrightarrow}Map(S_{Q}%
,\mathbb{C})\\
\left[  f:S_{Q}\longrightarrow\mathbb{C}\right]  \longmapsto\left[
\widehat{f}:S_{Q}\longrightarrow\mathbb{C}\right]
\end{gather*}
where
\[
\widehat{f}\left(  y\right)  =\frac{1}{\sqrt{Q}}\sum_{x\in S_{Q}}%
f(x)\omega^{xy}%
\]
\end{definition}

\bigskip

We implement the Fourier transform $\mathcal{F}$\ as a unitary transformation,
which in the standard basis
\[
\left|  0\right\rangle ,\left|  1\right\rangle ,\ldots,\left|
Q-1\right\rangle
\]
is given by the $Q\times Q$ unitary matrix
\[
\mathcal{F}=\frac{1}{\sqrt{Q}}\left(  \omega^{xy}\right)  \text{ .}%
\]
This unitary transformation can be factored into the product of $O\left(
\lg^{2}Q\right)  =O\left(  \lg^{2}N\right)  $ sufficiently local unitary
transformations. (See \cite{Shor1}, \cite{Hoyer1}.)

\vspace{0.5in}

\subsection{The quantum part of Shor's algorithm}

\quad\bigskip

The quantum part of Shor's algorithm, i.e., $\mathbb{STEP}$ \textbf{2}, is the following:

\bigskip

\begin{itemize}
\item [\fbox{$\mathbb{STEP}$ \textbf{2.0}}]Initialize registers 1 and 2,
i.e.,
\[
\left|  \psi_{0}\right\rangle =\left|  \text{\textsc{Reg1}}\right\rangle
\left|  \text{\textsc{Reg2}}\right\rangle =\left|  0\right\rangle \left|
0\right\rangle =\left|  00\cdots0\right\rangle \left|  0\cdots0\right\rangle
\]

\item[\fbox{$\mathbb{STEP}$ \textbf{2.1}}] \footnote{In this step we could
have instead applied the Hadamard transform to \textsc{Register1} with the
same result, but at the computational cost of $O\left(  \lg N\right)  $
sufficiently local unitary transformations.}Apply the $Q$-point Fourier
transform $\mathcal{F}$ to \textsc{Register1}.
\[
\left|  \psi_{0}\right\rangle =\left|  0\right\rangle \left|  0\right\rangle
\overset{\mathcal{F}\otimes I}{\longmapsto}\left|  \psi_{1}\right\rangle
=\frac{1}{\sqrt{Q}}\sum_{x=0}^{Q-1}\omega^{0\cdot x}\left|  x\right\rangle
\left|  0\right\rangle =\frac{1}{\sqrt{Q}}\sum_{x=0}^{Q-1}\left|
x\right\rangle \left|  0\right\rangle
\]
\end{itemize}

\begin{remark}
Hence, \textsc{Register1} now holds all the integers
\[
0,1,2,\ldots,Q-1
\]
in superposition.
\end{remark}

\bigskip

\begin{itemize}
\item [\fbox{$\mathbb{STEP}$ \textbf{2.2}}]Let $U_{f}$ be the unitary
transformation that takes $\left|  x\right\rangle \left|  0\right\rangle $ to
$\left|  x\right\rangle \left|  f(x)\right\rangle $. \ Apply the linear
transformation $U_{f}$ to the two registers. \ The result is:
\[
\left|  \psi_{1}\right\rangle =\frac{1}{\sqrt{Q}}\sum_{x=0}^{Q-1}\left|
x\right\rangle \left|  0\right\rangle \overset{U_{f}}{\longmapsto}\left|
\psi_{2}\right\rangle =\frac{1}{\sqrt{Q}}\sum_{x=0}^{Q-1}\left|
x\right\rangle \left|  f(x)\right\rangle
\]
\end{itemize}

\bigskip

\begin{remark}
The state of the two registers is now more than a superposition of states.
\ In this step, we have quantum entangled the two registers.
\end{remark}

\bigskip

\bigskip

\begin{itemize}
\item [\fbox{$\mathbb{STEP}$ \textbf{2.3.}}]Apply the $Q$-point Fourier
transform $\mathcal{F}$ to \textsc{Reg1}. \ The resulting state is:
\[%
\begin{array}
[c]{ccl}%
\left|  \psi_{2}\right\rangle =\frac{1}{\sqrt{Q}}%
{\displaystyle\sum\limits_{x=0}^{Q-1}}
\left|  x\right\rangle \left|  f(x)\right\rangle  & \overset{\mathcal{F}%
\otimes I}{\longmapsto} & \left|  \psi_{3}\right\rangle =\frac{1}{Q}%
{\displaystyle\sum\limits_{x=0}^{Q-1}}
{\displaystyle\sum\limits_{y=0}^{Q-1}}
\omega^{xy}\left|  y\right\rangle \left|  f(x)\right\rangle \\
&  & \\
&  & \qquad=\frac{1}{Q}%
{\displaystyle\sum\limits_{y=0}^{Q-1}}
\left\|  \left|  \Upsilon(y)\right\rangle \right\|  \cdot\left|
y\right\rangle \frac{\left|  \Upsilon(y)\right\rangle }{\left\|  \left|
\Upsilon(y)\right\rangle \right\|  }\text{ ,}%
\end{array}
\]
where
\[
\left|  \Upsilon(y)\right\rangle =%
{\displaystyle\sum\limits_{x=0}^{Q-1}}
\omega^{xy}\left|  f(x)\right\rangle \text{. }%
\]
\end{itemize}

\bigskip

\begin{itemize}
\item [\fbox{$\mathbb{STEP}$ \textbf{2.4.}}]Measure \textsc{Reg1}, i.e.,
perform a measurement with respect to the orthogonal projections
\[
\left|  0\right\rangle \left\langle 0\right|  \otimes I,\ \left|
1\right\rangle \left\langle 1\right|  \otimes I,\ \left|  2\right\rangle
\left\langle 2\right|  \otimes I,\ \ldots\ ,\ \left|  Q-1\right\rangle
\left\langle Q-1\right|  \otimes I\text{ ,}%
\]
where $I$ denotes the identity operator on the Hilbert space of the second
register \textsc{Reg2}. \ 
\end{itemize}

\bigskip

As a result of this measurement, we have, with probability
\[
Prob\left(  y_{0}\right)  =\frac{\left\|  \left|  \Upsilon(y_{0})\right\rangle
\right\|  ^{2}}{Q^{2}}\text{ ,}%
\]
moved to the state
\[
\left|  y_{0}\right\rangle \frac{\left|  \Upsilon(y_{0})\right\rangle
}{\left\|  \left|  \Upsilon(y_{0})\right\rangle \right\|  }%
\]
and measured the value
\[
y_{0}\in\left\{  0,1,2,\ldots,Q-1\right\}  \text{ . }%
\]

\bigskip

If after this computation, we ignore the two registers \textsc{Reg1} and
\textsc{Reg2}, we see that what we have created is nothing more than a
classical probability distribution $\mathcal{S}$ on the sample space
\[
\left\{  0,1,2,\ldots,Q-1\right\}  \text{ .}%
\]
In other words, the sole purpose of executing STEPS 2.1 to 2.4 is to create a
classical finite memoryless stochastic source $\mathcal{S}$ which outputs a
symbol $y_{0}\in\left\{  0,1,2,\ldots,Q-1\right\}  $ with the probability
\[
Prob(y_{0})=\frac{\left\|  \left|  \Upsilon(y_{0})\right\rangle \right\|
^{2}}{Q^{2}}\text{ .}%
\]
(For more details, please refer to section 8.1 of this paper.)

\bigskip

As we shall see, the objective of the remander of Shor's algorithm is to glean
information about the period $P$ of $f$ from the just created stochastic
source $\mathcal{S}$. The stochastic source was created exactly for that reason.

\bigskip

\subsection{Peter Shor's stochastic source $\mathcal{S}$}

\qquad\bigskip

Before continuing to the final part of Shor's algorithm, we need to analyze
the probability distribution $Prob\left(  y\right)  $ a little more carefully.

\bigskip

\begin{proposition}
Let $q$ and $r$ be the unique non-negative integers such that $Q=Pq+r$ , where
$0\leq r<P$ ; and let $Q_{0}=Pq$. \ Then
\[
Prob\left(  y\right)  =\left\{
\begin{array}
[c]{lrl}%
\frac{r\sin^{2}\left(  \frac{\pi Py}{Q}\cdot\left(  \frac{Q_{0}}{P}+1\right)
\right)  +\left(  P-r\right)  \sin^{2}\left(  \frac{\pi Py}{Q}\cdot\frac
{Q_{0}}{P}\right)  }{Q^{2}\sin^{2}\left(  \frac{\pi Py}{Q}\right)  } &
\text{if} & Py\neq0\operatorname{mod}Q\\
&  & \\
\frac{r\left(  Q_{0}+P\right)  ^{2}+\left(  P-r\right)  Q_{0}^{2}}{Q^{2}P^{2}}%
& \text{if} & Py=0\operatorname{mod}Q
\end{array}
\right.
\]
\end{proposition}

\begin{proof}
We begin by deriving a more usable expression for $\left|  \Upsilon
(y)\right\rangle $.
\[%
\begin{array}
[c]{rrl}%
\left|  \Upsilon(y)\right\rangle  & = &
{\displaystyle\sum\limits_{x=0}^{Q-1}}
\omega^{xy}\left|  f(x)\right\rangle =%
{\displaystyle\sum\limits_{x=0}^{Q_{0}-1}}
\omega^{xy}\left|  f(x)\right\rangle +%
{\displaystyle\sum\limits_{x=Q_{0}}^{Q-1}}
\omega^{xy}\left|  f(x)\right\rangle \\
&  & \\
& = &
{\displaystyle\sum\limits_{x_{0}=0}^{P-1}}
{\displaystyle\sum\limits_{x_{1}=0}^{\frac{Q_{0}}{P}-1}}
\omega^{\left(  Px_{1}+x_{0}\right)  y}\left|  f(Px_{1}+x_{0})\right\rangle +%
{\displaystyle\sum\limits_{x_{0}=0}^{r-1}}
\omega^{\left[  P\left(  \frac{Q_{0}}{P}\right)  +x_{0}\right]  y}\left|
f(Px_{1}+x_{0})\right\rangle \\
&  & \\
& = &
{\displaystyle\sum\limits_{x_{0}=0}^{P-1}}
\omega^{x_{0}y}\cdot\left(
{\displaystyle\sum\limits_{x_{1}=0}^{\frac{Q_{0}}{P}-1}}
\omega^{Pyx_{1}}\right)  \left|  f(x_{0})\right\rangle +%
{\displaystyle\sum\limits_{x_{0}=0}^{r-1}}
\omega^{x_{0}y}\cdot\omega^{Py\left(  \frac{Q_{0}}{P}\right)  }\left|
f(x_{0})\right\rangle \\
&  & \\
& = &
{\displaystyle\sum\limits_{x_{0}=0}^{r-1}}
\omega^{x_{0}y}\cdot\left(
{\displaystyle\sum\limits_{x_{1}=0}^{\frac{Q_{0}}{P}}}
\omega^{Pyx_{1}}\right)  \left|  f(x_{0})\right\rangle +%
{\displaystyle\sum\limits_{x_{0}=r}^{P-1}}
\omega^{x_{0}y}\cdot\left(
{\displaystyle\sum\limits_{x_{1}=0}^{\frac{Q_{0}}{P}-1}}
\omega^{Pyx_{1}}\right)  \left|  f(x_{0})\right\rangle
\end{array}
\]
where we have used the fact that $f$ is periodic of period $P$.

\bigskip

Since $f$ is one-to-one when restricted to its period $0,1,2,\ldots,P-1$, all
the kets
\[
\left|  f(0)\right\rangle ,\ \left|  f(1)\right\rangle ,\ \left|
f(2)\right\rangle ,\ \ldots\ ,\ \left|  f(P-1)\right\rangle ,\
\]
are mutually orthogonal. \ Hence,
\[
\left\langle \Upsilon(y)\mid\Upsilon(y)\right\rangle =r\left|
{\displaystyle\sum\limits_{x_{1}=0}^{\frac{Q_{0}}{P}}}
\omega^{Pyx_{1}}\right|  ^{2}+(P-r)\left|
{\displaystyle\sum\limits_{x_{1}=0}^{\frac{Q_{0}}{P}-1}}
\omega^{Pyx_{1}}\right|  ^{2}\text{ .}%
\]

\bigskip

If $Py=0\operatorname{mod}Q$, then since $\omega$ is a $Q$-th root of unity,
we have
\[
\left\langle \Upsilon(y)\mid\Upsilon(y)\right\rangle =r\left(  \frac{Q_{0}}%
{P}+1\right)  ^{2}+\left(  P-r\right)  \left(  \frac{Q_{0}}{P}\right)
^{2}\text{ .}%
\]

\bigskip

On the other hand, if $Py\neq0\operatorname{mod}Q$, then we can sum the
geometric series to obtain
\begin{align*}
\left\langle \Upsilon(y)\mid\Upsilon(y)\right\rangle  &  =\left|  \frac
{\omega^{Py\cdot\left(  \frac{Q_{0}}{P}+1\right)  }-1}{\omega^{Py}-1}\right|
^{2}+\left(  P-r)\right)  \left|  \frac{\omega^{Py\cdot\left(  \frac{Q_{0}}%
{P}\right)  }-1}{\omega^{Py}-1}\right|  ^{2}\\
& \\
&  =\left|  \frac{e^{\frac{2\pi i}{Q}\cdot Py\cdot\left(  \frac{Q_{0}}%
{P}+1\right)  }-1}{e^{\frac{2\pi i}{Q}\cdot Py}-1}\right|  ^{2}+\left(
P-r)\right)  \left|  \frac{e^{\frac{2\pi i}{Q}\cdot Py\cdot\left(  \frac
{Q_{0}}{P}\right)  }-1}{e^{\frac{2\pi i}{Q}\cdot Py}-1}\right|  ^{2}%
\end{align*}
where we have used the fact that $\omega$ is the primitive $Q$-th root of
unity given by
\[
\omega=e^{2\pi i/Q}\text{ .}%
\]

\bigskip

The remaining part of the proposition is a consequence of the trigonometric
identity
\[
\left|  e^{i\theta}-1\right|  ^{2}=4\sin^{2}\left(  \frac{\theta}{2}\right)
\text{ .}%
\]
\end{proof}

\bigskip

As a corollary, we have

\bigskip

\begin{corollary}
If $P$ is an exact divisor of $Q$, then
\[
Prob\left(  y\right)  =\left\{
\begin{array}
[c]{lrl}%
0 & \text{if} & Py\neq0\operatorname{mod}Q\\
&  & \\
\frac{1}{P} & \text{if} & Py=0\operatorname{mod}Q
\end{array}
\right.
\]
\end{corollary}

\bigskip

\subsection{A momentary digression: Continued fractions}

\qquad\bigskip

We digress for a moment to review the theory of continued fractions. (For a
more in-depth explanation of the theory of continued fractions, please refer
to \cite{Hardy1} and \cite{LeVeque1}.)

\bigskip

Every positive rational number $\xi$ can be written as an expression in the
form
\[
\xi=a_{0}+\frac{1}{a_{1}+\frac{\overset{}{\underset{}{1}}}{a_{2}%
+\frac{\overset{}{\underset{}{1}}}{a_{3}+\frac{\overset{}{\underset{}{1}}%
}{\cdots+\frac{\overset{}{\underset{}{1}}}{\overset{}{a_{N}}}}}}}\text{ ,}%
\]
where $a_{0}$ is a non-negative integer, and where $a_{1},\ldots,a_{N}$ are
positive integers. \ Such an expression is called a (finite, simple)
\textbf{continued fraction}
\index{Continued Fraction}, and is uniquely determined by $\xi$ provided we
impose the condition $a_{N}>1$. \ For typographical simplicity, we denote the
above continued fraction by
\[
\left[  a_{0},a_{1},\ldots,a_{N}\right]  \text{ .}%
\]

The continued fraction expansion of $\xi$ can be computed with the following
recurrence relation, which always terminates if $\xi$ is rational:
\[
\fbox{$\overset{}{\underset{}{%
\begin{array}
[c]{lll}%
\left\{
\begin{array}
[c]{r}%
a_{0}=\left\lfloor \xi\right\rfloor \\
\\
\xi_{0}=\xi-a_{0}%
\end{array}
\right.  \text{ ,} & \text{and if }\xi_{n}\neq0\text{, then} & \left\{
\begin{array}
[c]{l}%
a_{n+1}=\left\lfloor 1/\xi_{n}\right\rfloor \\
\\
\xi_{n+1}=\frac{1}{\xi_{n}}-a_{n+1}%
\end{array}
\right.
\end{array}
}}$}%
\]

\bigskip

The $n$-th \textbf{convergent}
\index{Continued Fraction!Convergent of}
\index{Convergent|see{Continued Fraction}} ($0\leq n\leq N$) of the above
continued fraction is defined as the rational number $\xi_{n}$ given by
\[
\xi_{n}=\left[  a_{0},a_{1},\ldots,a_{n}\right]  \text{ .}%
\]
Each convergent $\xi_{n}$ can be written in the for, $\xi_{n}=\frac{p_{n}%
}{q_{n}}$, where $p_{n}$ and $q_{n}$ are relatively prime integers (
$\gcd\left(  p_{n},q_{n}\right)  =1$). The integers $p_{n}$ and $q_{n}$ are
determined by the recurrence relation
\[
\fbox{$%
\begin{array}
[c]{lll}%
p_{0}=a_{0}, & p_{1}=a_{1}a_{0}+1, & p_{n}=a_{n}p_{n-1}+p_{n-2},\\
&  & \\
q_{0}=1, & q_{1}=a_{1}, & q_{n}=a_{n}q_{n-1}+q_{n-2}\text{ \ .}%
\end{array}
$}%
\]

\bigskip

\bigskip

\subsection{Preparation for the final part of Shor's algorithm}

\qquad\bigskip

\begin{definition}
\footnote{$\left\{  a\right\}  _{Q}=a-Q\cdot round\left(  \frac{a}{Q}\right)
=a-Q\cdot\left\lfloor \frac{a}{Q}+\frac{1}{2}\right\rfloor $.}For each integer
$\ \ a$, let $\left\{  a\right\}  _{Q}$ denote the \textbf{residue} of
$\ \ a\ \ $ modulo $Q$ \textbf{of smallest magnitude}. \ In other words,
$\left\{  a\right\}  _{Q}$ is the unique integer such that
\[
\left\{
\begin{array}
[c]{l}%
a=\left\{  a\right\}  _{Q}\operatorname{mod}Q\\
\\
-Q/2<\left\{  a\right\}  _{Q}\leq Q/2
\end{array}
\right.  \text{ .}%
\]
\end{definition}

\bigskip

\begin{proposition}
Let $y$ be an integer lying in $S_{Q}$. \ Then
\[
Prob\left(  y\right)  \geq\left\{
\begin{array}
[c]{lrl}%
\frac{4}{\pi^{2}}\cdot\frac{1}{P}\cdot\left(  1-\frac{1}{N}\right)  ^{2} &
\text{if} & 0<\left|  \left\{  Py\right\}  _{Q}\right|  \leq\frac{P}{2}%
\cdot\left(  1-\frac{1}{N}\right) \\
&  & \\
\frac{1}{P}\cdot\left(  1-\frac{1}{N}\right)  ^{2} & \text{if} & \left\{
Py\right\}  _{Q}=0
\end{array}
\right.
\]
\end{proposition}

\begin{proof}
We begin by noting that
\[%
\begin{array}
[c]{ll}%
\left|  \frac{\pi\left\{  Py\right\}  _{Q}}{Q}\cdot\left(  \frac{Q_{0}}%
{P}+1\right)  \right|  & \leq\frac{\pi}{Q}\cdot\frac{P}{2}\cdot\left(
1-\frac{1}{N}\right)  \cdot\left(  \frac{Q_{0}+P}{P}\right)  \leq\frac{\pi}%
{2}\cdot\left(  1-\frac{1}{N}\right)  \cdot\left(  \frac{Q+P}{Q}\right) \\
& \\
& \leq\frac{\pi}{2}\cdot\left(  1-\frac{1}{N}\right)  \cdot\left(  1+\frac
{P}{Q}\right)  \leq\frac{\pi}{2}\cdot\left(  1-\frac{1}{N}\right)
\cdot\left(  1+\frac{N}{N^{2}}\right)  <\frac{\pi}{2}\text{ ,}%
\end{array}
\]
where we have made use of the inequalities
\[
N^{2}\leq Q<2N^{2}\text{ \ and \ }0<P\leq N\text{ \ .}%
\]
It immediately follows that
\[
\left|  \frac{\pi\left\{  Py\right\}  _{Q}}{Q}\cdot\frac{Q_{0}}{P}\right|
<\frac{\pi}{2}\text{ \ .}%
\]

\bigskip

As a result, we can legitimately use the inequality
\[
\frac{4}{^{\pi^{2}}}\theta^{2}\leq\sin^{2}\theta\leq\theta^{2}\text{, for
}\left|  \theta\right|  <\frac{\pi}{2}%
\]
to simplify the expression for $Prob\left(  y\right)  $.

\bigskip

Thus,
\[%
\begin{array}
[c]{lll}%
Prob\left(  y\right)  & = & \frac{r\sin^{2}\left(  \frac{\pi\left\{
Py\right\}  _{Q}}{Q}\cdot\left(  \frac{Q_{0}}{P}+1\right)  \right)  +\left(
P-r\right)  \sin^{2}\left(  \frac{\pi\left\{  Py\right\}  _{Q}}{Q}\cdot
\frac{Q_{0}}{P}\right)  }{Q^{2}\sin^{2}\left(  \frac{\pi Py}{Q}\right)  }\\
&  & \\
& \geq & \frac{r\cdot\frac{4}{\pi^{2}}\cdot\left(  \frac{\pi\left\{
Py\right\}  _{Q}}{Q}\cdot\left(  \frac{Q_{0}}{P}+1\right)  \right)
^{2}+\left(  P-r\right)  \cdot\frac{4}{\pi^{2}}\cdot\left(  \frac{\pi\left\{
Py\right\}  _{Q}}{Q}\cdot\frac{Q_{0}}{P}\right)  ^{2}}{Q^{2}\left(  \frac
{\pi\left\{  Py\right\}  _{Q}}{Q}\right)  ^{2}}\\
&  & \\
& \geq & \frac{4}{\pi^{2}}\cdot\frac{P\cdot\left(  \frac{Q_{0}}{P}\right)
^{2}}{Q^{2}}=\frac{4}{\pi^{2}}\cdot\frac{1}{P}\cdot\left(  \frac{Q-r}%
{Q}\right)  ^{2}\\
&  & \\
& = & \frac{4}{\pi^{2}}\cdot\frac{1}{P}\cdot\left(  1-\frac{r}{Q}\right)
^{2}\geq\frac{4}{\pi^{2}}\cdot\frac{1}{P}\cdot\left(  1-\frac{1}{N}\right)
^{2}%
\end{array}
\]

\bigskip

The remaining case, $\left\{  Py\right\}  _{Q}=0$ is left to the reader.
\end{proof}

\bigskip

\begin{lemma}
Let
\[
Y=\left\{  y\in S_{Q}\mid\left|  \left\{  Py\right\}  _{Q}\right|  \leq
\frac{P}{2}\right\}  \text{ \ \ and \ \ }S_{P}=\left\{  d\in S_{Q}\mid0\leq
d<P\right\}  \text{ .}%
\]
Then the map
\[%
\begin{array}
[c]{lll}%
Y & \longrightarrow &  S_{P}\\
y & \longmapsto &  d=d(y)=round\left(  \frac{P}{Q}\cdot y\right)
\end{array}
\]
is a bijection with inverse
\[
y=y(d)=round\left(  \frac{Q}{P}\cdot d\right)  \text{ .}%
\]
Hence, $Y$ and $S_{P}$ are in one-to-one correspondence. \ Moreover,
\[
\left\{  Py\right\}  _{Q}=P\cdot y-Q\cdot d(y)\text{ .}%
\]
\end{lemma}

\bigskip

\begin{remark}
Moreover, the following two sets of rationals are in one-to-one
correspondence
\[
\left\{  \frac{y}{Q}\mid y\in Y\right\}  \longleftrightarrow\left\{  \frac
{d}{P}\mid0\leq d<P\right\}
\]
\end{remark}

\bigskip

As a result of the measurement performed in $\mathbb{STEP}$ 2.4, we have in
our possession an integer $y\in Y$. \ We now show how $y$ \ can be use to
determine the unknown period $P$. \ 

\bigskip

We now need the following theorem\footnote{See \cite[Theorem 184, Section
10.15]{Hardy1}.} from the theory of continued fractions:

\bigskip

\begin{theorem}
Let $\xi$ be a real number, and let $a$ and $b$ be integers with $b>0$. If
\[
\left|  \xi-\frac{a}{b}\right|  \leq\frac{1}{2b^{2}}\text{ ,}%
\]
then the rational number $a/b$ is a convergent of the continued fraction
expansion of $\xi$. \ 
\end{theorem}

\bigskip

As a corollary, we have:

\bigskip

\begin{corollary}
If $\left|  \left\{  Py\right\}  _{Q}\right|  \leq\frac{P}{2}$, then the
rational number $\frac{d(y)}{P}$ is a convergent of the continued fraction
expansion of $\frac{y}{Q}$. \ 
\end{corollary}

\begin{proof}
Since
\[
Py-Qd(y)=\left\{  Py\right\}  _{Q}\text{ ,}%
\]
we know that
\[
\left|  Py-Qd(y)\right|  \leq\frac{P}{2}\text{, }%
\]
which can be rewritten as
\[
\left|  \frac{y}{Q}-\frac{d(y)}{P}\right|  \leq\frac{1}{2Q}\text{ .}%
\]
But, since $Q\geq N^{2}$, it follows that
\[
\left|  \frac{y}{Q}-\frac{d(y)}{P}\right|  \leq\frac{1}{2N^{2}}\text{ .}%
\]
Finally, since $P\leq N$ (and hence $\frac{1}{2N}\leq\frac{1}{2P^{2}})$, the
above theorem can be applied. \ Thus, $\frac{d(y)}{P}$ is a convergent of the
continued fraction expansion of $\xi=\frac{y}{Q}$. \ 
\end{proof}

\bigskip

Since $\frac{d(y)}{P}$ is a convergent of the continued fraction expansion of
$\frac{y}{Q}$, it follows that, for some $n$,
\[
\frac{d(y)}{P}=\frac{p_{n}}{q_{n}}\text{ ,}%
\]
where $p_{n}$ and $q_{n}$ are relatively prime positive integers given by a
recurrence relation found in the previous subsection. \ So it would seem that
we have found a way of deducing the period $P$ from the output $y$ of
$\mathbb{STEP}$ 2.4, and so we are done. \ 

\bigskip

Not quite! \ 

\bigskip

We can determine $P$ from the measured $y$ produced by $\mathbb{STEP}$ 2.4,
only if
\[
\left\{
\begin{array}
[c]{l}%
p_{n}=d(y)\\
\\
q_{n}=P
\end{array}
\right.  \text{ ,}%
\]
which is true only when $d(y)$ and $P$ are relatively prime.

\bigskip

So what is the probability that the $y\in Y$ produced by $\mathbb{STEP}$ 2.4
satisfies the additional condition that
\[
\gcd\left(  P,d(y)\right)  =1\text{ ?}%
\]

\bigskip

\begin{proposition}
The probability that the random $y$ produced by $\mathbb{STEP}$ 2.4 is such
that $d(y)$ and $P$ are relatively prime is bounded below by the following
expression
\[
Prob\left\{  y\in Y\mid\gcd(d(y),P)=1\right\}  \geq\frac{4}{\pi^{2}}\cdot
\frac{\phi(P)}{P}\cdot\left(  1-\frac{1}{N}\right)  ^{2}\text{ ,}%
\]
where $\phi(P)$ denotes Euler's totient function, i.e., $\phi(P)$ is the
number of positive integers less than $P$ which are relatively prime to $P$. \ 
\end{proposition}

\bigskip

The following theorem can be found in \cite[Theorem 328, Section 18.4]{Hardy1}:

\bigskip

\begin{theorem}%
\[
\lim\inf\frac{\phi(N)}{N/\ln\ln N}=e^{-\gamma}\text{,}%
\]
where $\gamma$ denotes Euler's constant
\index{Euler's constant}
\index{Gamma@$\gamma$|see{Euler's Constant}} $\gamma
=0.57721566490153286061\ldots$ , and where $e^{-\gamma}=0.5614594836\ldots$ . \ 
\end{theorem}

\bigskip

As a corollary, we have:

\bigskip

\begin{corollary}%
\[
Prob\left\{  y\in Y\mid\gcd(d(y),P)=1\right\}  \geq\frac{4}{\pi^{2}\ln2}%
\cdot\frac{e^{-\gamma}-\epsilon\left(  P\right)  }{\lg\lg N}\cdot\left(
1-\frac{1}{N}\right)  ^{2}\text{ ,}%
\]
where $\epsilon\left(  P\right)  $ is a monotone decreasing sequence
converging to zero. \ In terms of asymptotic notation,
\[
Prob\left\{  y\in Y\mid\gcd(d(y),P)=1\right\}  =\Omega\left(  \frac{1}{\lg\lg
N}\right)  \text{ .}%
\]
Thus
\index{W@$\Omega(-)$, asymptotic lower bound}, if $\ \mathbb{STEP}$ 2.4 is
repeated $O(\lg\lg N)$ times, then the probability of success is
$\Omega\left(  1\right)  $.
\end{corollary}

\begin{proof}
From the above theorem, we know that
\[
\frac{\phi(P)}{P/\ln\ln P}\geq e^{-\gamma}-\epsilon\left(  P\right)  \text{ .}%
\]
where $\epsilon\left(  P\right)  $ is a monotone decreasing sequence of
\ positive reals converging to zero. \ Thus,
\[
\frac{\phi(P)}{P}\geq\frac{e^{-\gamma}-\epsilon\left(  P\right)  }{\ln\ln
P}\geq\frac{e^{-\gamma}-\epsilon\left(  P\right)  }{\ln\ln N}=\frac
{e^{-\gamma}-\epsilon\left(  P\right)  }{\ln\ln2+\ln\lg N}\geq\frac
{e^{-\gamma}-\epsilon\left(  P\right)  }{\ln2}\cdot\frac{1}{\lg\lg N}%
\]
\end{proof}

\bigskip

\begin{remark}
$\Omega(\frac{1}{\lg\lg N})$ denotes an asymptotic lower bound. \ Readers not
familiar with the big-oh $O(\ast)$ and big-omega $\Omega\left(  \ast\right)  $
notation should refer to \cite[Chapter 2]{Cormen1} or \cite[Chapter
2]{Brassard1}.
\end{remark}

\bigskip

\begin{remark}
For the curious reader, lower bounds $LB(P)$ of $e^{-\gamma}-\epsilon\left(
P\right)  $ for $3\leq P\leq841$ are given in the following table:
\[%
\begin{tabular}
[c]{|l||l|}\hline\hline
\multicolumn{1}{||l||}{$P$} & \multicolumn{1}{||l||}{$LB(P)$}\\\hline\hline
3 & 0.062\\\hline
4 & 0.163\\\hline
5 & 0.194\\\hline
7 & 0.303\\\hline
13 & 0.326\\\hline
31 & 0.375\\\hline
61 & 0.383\\\hline
211 & 0.411\\\hline
421 & 0.425\\\hline
631 & 0.435\\\hline
841 & 0.468\\\hline
\end{tabular}
\]
Thus, if one wants a reasonable bound on the $Prob\left\{  y\in Y\mid
\gcd(d(y),P)=1\right\}  $ before continuing with Shor's algorithm, it would
pay to first use a classical algorithm to verify that the period $P$ of the
randomly chosen integer $m$ is not too small.
\end{remark}

\bigskip

\subsection{The final part of Shor's algorithm}

\qquad\bigskip

We are now prepared to give the last step in Shor's algorithm. \ This step can
be performed on a classical computer.

\bigskip

\begin{itemize}
\item [\fbox{\textbf{Step 2.5}}]Compute the period $P$ from the integer $y$
produced by $\mathbb{STEP}$ 2.4.
\end{itemize}

\bigskip

\begin{itemize}
\item
\begin{itemize}
\item [\qquad]\textsc{Loop} \textsc{for each} $n$ \textsc{from} $n=1$
\textsc{Until} $\xi_{n}=0$.
\end{itemize}
\end{itemize}

\bigskip

\begin{itemize}
\item
\begin{itemize}
\item
\begin{itemize}
\item [\qquad]Use the recurrence relations given in subsection 13.7, to
compute the $p_{n}$ and $q_{n}$ of the $n$-th convergent $\frac{p_{n}}{q_{n}}
$ of $\frac{y}{Q}$.
\end{itemize}
\end{itemize}
\end{itemize}

\bigskip

\begin{itemize}
\item
\begin{itemize}
\item
\begin{itemize}
\item [\qquad]Test to see if $q_{n}=P$ by computing\footnote{The indicated
algorithm for computing $m^{q_{n}}\operatorname{mod}N$ requires $O(\lg q_{n})$
arithmetic operations.}
\[
m^{q_{n}}=%
{\displaystyle\prod\limits_{i}}
\left(  m^{2^{i}}\right)  ^{q_{n,i}}\operatorname{mod}N\text{ ,}%
\]
where $q_{n}=\sum_{i}q_{n,i}2^{i}$ is the binary expansion of $q_{n}$.

\item[\qquad] If $m^{q_{n}}=1\operatorname{mod}N$, then exit with the answer
$P=q_{n}$, and proceed to \textbf{Step 3}. \ If not, then continue the loop.
\end{itemize}
\end{itemize}
\end{itemize}

\bigskip

\begin{itemize}
\item
\begin{itemize}
\item [\qquad]\textsc{End of Loop}
\end{itemize}
\end{itemize}

\bigskip

\begin{itemize}
\item
\begin{itemize}
\item [\qquad]If you happen to reach this point, you are a very unlucky
quantum computer scientist. \ You must start over by returning to
$\mathbb{STEP}$ 2.0. \ But don't give up hope! \ The probability that the
integer $y$ produced by $\mathbb{STEP}$ 2.4 will lead to a successful
completion of Step 2.5 is bounded below by
\[
\frac{4}{\pi^{2}\ln2}\cdot\frac{e^{-\gamma}-\epsilon\left(  P\right)  }{\lg\lg
N}\cdot\left(  1-\frac{1}{N}\right)  ^{2}>\frac{0.232}{\lg\lg N}\cdot\left(
1-\frac{1}{N}\right)  ^{2}\text{ ,}%
\]
provided the period $P$ is greater than $3$. \ [ $\gamma$ denotes Euler's constant.]
\end{itemize}
\end{itemize}

\bigskip

\subsection{\textbf{\bigskip}An example of Shor's algorithm}

\quad\bigskip

Let us now show how $N=91\ (=7\cdot13)$ can be factored using Shor's algorithm.

\bigskip

We choose $Q=2^{14}=16384$ so that $N^{2}\leq Q<2N^{2}$.

\bigskip

\begin{itemize}
\item [\fbox{\textbf{Step 1}}]Choose a random positive integer $m$, say $m=3$.
\ Since $\gcd(91,3)=1$, we proceed to $\mathbb{STEP}$\textbf{\ 2} to find the
period of the function $f$ given by \
\[
f(a)=3^{a}\operatorname{mod}91
\]
\end{itemize}

\begin{remark}
Unknown to us, $f$ has period $P=6$. For,
\[%
\begin{tabular}
[c]{||l||}\hline\hline
$%
\begin{array}
[c]{ccccccccccc}%
a &  & 0 & 1 & 2 & 3 & 4 & 5 & 6 & 7 & \cdots\\
&  &  &  &  &  &  &  &  &  & \\
f(a) &  & 1 & 3 & 9 & 27 & 81 & 61 & 1 & 3 & \cdots
\end{array}
$\\\hline\hline
\multicolumn{1}{||c||}{$\therefore\text{ Unknown period }P=6$}\\\hline\hline
\end{tabular}
\]
\end{remark}

\vspace{0.5in}

\begin{itemize}
\item [\fbox{$\mathbb{STEP}$ \textbf{2.0}}]Initialize registers 1 and 2. Thus,
the state of the two registers becomes:
\[
\left|  \psi_{0}\right\rangle =\left|  0\right\rangle \left|  0\right\rangle
\]
\end{itemize}

\vspace{0.5in}

\begin{itemize}
\item [\fbox{$\mathbb{STEP}$ \textbf{2.1}}]Apply the $Q$-point Fourier
transform $\mathcal{F}$ to register \#1, where
\[
\mathcal{F}\left|  k\right\rangle =\frac{1}{\sqrt{16384}}\sum_{x=0}%
^{16383}\omega^{kj}\left|  x\right\rangle \text{ ,}%
\]
and where $\omega$ is a primitive $Q$-th root of unity, e.g., $\omega
=e^{\frac{2\pi i}{16384}}$. Thus the state of the two registers becomes:
\[
\left|  \psi_{1}\right\rangle =\frac{1}{\sqrt{16384}}\sum_{x=0}^{16383}\left|
x\right\rangle \left|  0\right\rangle
\]
\end{itemize}

\vspace{0.5in}

\begin{itemize}
\item [\fbox{$\mathbb{STEP}$ \textbf{2.2}}]Apply the unitary transformation
$U_{f}$ to registers \#1 and \#2, where
\[
U_{f}\left|  x\right\rangle \left|  \ell\right\rangle =\left|  x\right\rangle
\left|  \ f(x)-\ell\ \operatorname{mod}91\right\rangle \text{ .}%
\]
(Please note that $U_{f}^{2}=I$.) Thus, the state of the two registers
becomes:
\[%
\begin{array}
[c]{rrrl}%
\left|  \psi_{2}\right\rangle  & = & \frac{1}{\sqrt{16384}} & \sum
_{x=0}^{16383}\left|  x\right\rangle \left|  3^{x}\operatorname{mod}%
91\right\rangle \\
&  &  & \\
& = & \frac{1}{\sqrt{16384}}( & \quad\left|  \ 0\right\rangle \left|
1\right\rangle \ +\left|  \ 1\right\rangle \left|  3\right\rangle +\left|
\ 2\right\rangle \left|  9\right\rangle \ +\left|  \ 3\right\rangle \left|
27\right\rangle +\left|  \ 4\right\rangle \left|  81\right\rangle +\left|
\ 5\right\rangle \left|  61\right\rangle \\
&  &  & \\
&  &  & +\ \left|  \ 6\right\rangle \left|  1\right\rangle \ +\left|
\ 7\right\rangle \left|  3\right\rangle +\left|  \ 8\right\rangle \left|
9\right\rangle \ +\left|  \ 9\right\rangle \left|  27\right\rangle +\left|
10\right\rangle \left|  81\right\rangle +\left|  11\right\rangle \left|
61\right\rangle \\
&  &  & \\
&  &  & +\ \left|  12\right\rangle \left|  1\right\rangle \ +\left|
13\right\rangle \left|  3\right\rangle \ +\left|  14\right\rangle \left|
9\right\rangle \ +\left|  15\right\rangle \left|  27\right\rangle +\left|
16\right\rangle \left|  81\right\rangle +\left|  17\right\rangle \left|
61\right\rangle \\
&  &  & \\
&  &  & +\ \ldots\\
&  &  & \\
&  &  & +\ \left|  16380\right\rangle \left|  1\right\rangle +\left|
16381\right\rangle \left|  3\right\rangle +\left|  16382\right\rangle \left|
9\right\rangle +\left|  16383\right\rangle \left|  27\right\rangle \\
&  & ) &
\end{array}
\]
\end{itemize}

\bigskip

\begin{remark}
The state of the two registers is now more than a superposition of states.
\ We have in the above step quantum entangled the two registers.
\end{remark}

\vspace{0.5in}

\begin{itemize}
\item [\fbox{$\mathbb{STEP}$ \textbf{2.3}}]Apply the $Q$-point $\mathcal{F} $
again to register \#1. Thus, the state of the system becomes:
\[%
\begin{array}
[c]{rrl}%
\left|  \psi_{3}\right\rangle  & = & \frac{1}{\sqrt{16384}}\sum_{x=0}%
^{16383}\frac{1}{\sqrt{16384}}\sum_{y=0}^{16383}\omega^{xy}\left|
y\right\rangle \left|  3^{x}\operatorname{mod}91\right\rangle \\
&  & \\
& = & \frac{1}{16384}\sum_{x=0}^{16383}\left|  y\right\rangle \sum
_{x=0}^{16383}\omega^{xy}\left|  3^{x}\operatorname{mod}91\right\rangle \\
&  & \\
& = & \frac{1}{16384}\sum_{x=0}^{16383}\left|  y\right\rangle \left|
\Upsilon\left(  y\right)  \right\rangle \text{ ,}%
\end{array}
\]
where
\[
\left|  \Upsilon\left(  y\right)  \right\rangle =\sum_{x=0}^{16383}\omega
^{xy}\left|  3^{x}\operatorname{mod}91\right\rangle
\]
Thus,
\[%
\begin{array}
[c]{rl}%
\left|  \Upsilon\left(  y\right)  \right\rangle = & \quad\quad\ \ \ \left|
1\right\rangle \ +\ \ \omega^{y}\left|  3\right\rangle +\ \omega^{2y}\left|
9\right\rangle \ +\ \omega^{3y}\left|  27\right\rangle +\ \ \omega^{4y}\left|
81\right\rangle +\ \ \omega^{5y}\left|  61\right\rangle \\
& \\
& +\ \ \omega^{6y}\left|  1\right\rangle \ +\ \omega^{7y}\left|
3\right\rangle +\ \omega^{8y}\left|  9\right\rangle \ +\ \omega^{9y}\left|
27\right\rangle +\ \omega^{10y}\left|  81\right\rangle +\omega^{11y}\left|
61\right\rangle \\
& \\
& +\ \omega^{12y}\left|  1\right\rangle \ +\omega^{13y}\left|  3\right\rangle
+\omega^{14y}\left|  9\right\rangle \ +\omega^{15y}\left|  27\right\rangle
+\omega^{16y}\left|  81\right\rangle +\omega^{17y}\left|  61\right\rangle \\
& \\
& +\ \ldots\\
& \\
& +\ \omega^{16380y}\left|  1\right\rangle +\omega^{16381y}\left|
3\right\rangle +\omega^{16382y}\left|  9\right\rangle +\omega^{16383y}\left|
27\right\rangle
\end{array}
\]
\end{itemize}

\vspace{0.5in}

\begin{itemize}
\item [\fbox{$\mathbb{STEP}$ \textbf{2.4}}]Measure \textsc{Reg1}. \ The result
of our measurement just happens to turn out to be
\[
y=13453
\]
\end{itemize}

\bigskip

Unknown to us, the probability of obtaining this particular $y$ is: \ \
\[
0.3189335551\times10^{-6}\text{ . }%
\]
Moreover, unknown to us, we're lucky! \ The corresponding $d$ is relatively
prime to $P$, i.e.,
\[
d=d(y)=round(\frac{P}{Q}\cdot y)=5
\]

\bigskip

However, we do know that the probability of $d(y)$ being relatively prime to
$P$ is greater than
\[
\frac{0.232}{\lg\lg N}\cdot\left(  1-\frac{1}{N}\right)  ^{2}\thickapprox
8.4\%\text{ \ (provided }P>3\text{),}%
\]
and we also know that
\[
\frac{d(y)}{P}%
\]
is a convergent of the continued fraction expansion of
\[
\xi=\frac{y}{Q}=\frac{13453}{16384}%
\]

So with a reasonable amount of confidence, we proceed to \textbf{Step 2.5}.

\bigskip

\begin{itemize}
\item [\fbox{\textbf{Step 2.5}}]Using the recurrence relations found in
subsection 13.7 of this paper, we successively compute (beginning with $n=0$)
the $a_{n}$'s and $q_{n}$'s for the continued fraction expansion of
\[
\xi=\frac{y}{Q}=\frac{13453}{16384}\text{ .}%
\]
For each non-trivial $n$ in succession, we check to see if
\[
3^{q_{n}}=1\operatorname{mod}91\text{. }%
\]
If this is the case, then we know $q_{n}=P$, and we immediately exit from
\textbf{Step 2.5} and proceed to \textbf{Step 3}.
\end{itemize}

\bigskip

\begin{itemize}
\item  In this example, $n=0$ and $n=1$ are trivial cases. \ 
\end{itemize}

\bigskip

\begin{itemize}
\item  For $n=2$, $a_{2}=4$ and $q_{2}=5$ . \ We test $q_{2}$ by computing
\[
3^{q_{2}}=3^{5}=\left(  3^{2^{0}}\right)  ^{1}\cdot\left(  3^{2^{1}}\right)
^{0}\cdot\left(  3^{2^{0}}\right)  ^{1}=61\neq1\operatorname{mod}91\text{ .}%
\]
Hence, $q_{2}\neq P$.
\end{itemize}

\bigskip

\begin{itemize}
\item  We proceed to $n=3$, and compute
\[
a_{3}=1\text{ and }q_{3}=6\text{. }%
\]
We then test $q_{3}$ by computing
\[
3^{q_{3}}=3^{6}=\left(  3^{2^{0}}\right)  ^{0}\cdot\left(  3^{2^{1}}\right)
^{1}\cdot\left(  3^{2^{0}}\right)  ^{1}=1\operatorname{mod}91\text{ .}%
\]
Hence, $q_{3}=P$. \ Since we now know the period $P$, there is no need to
continue to compute the remaining $a_{n}$'s and $q_{n}$'s. \ We proceed
immediately to \textbf{Step 3}.
\end{itemize}

\bigskip

To satisfy the reader's curiosity we have listed in the table below all the
values of $a_{n}$, $p_{n}$, and $q_{n}$ for $n=0,1,\ldots,14$. \ But it should
be mentioned again that we need only to compute $a_{n}$ and $q_{n}$ for
$n=0,1,2,3$, as indicated above. \
\[%
\begin{tabular}
[c]{|c||r|r|r|r|r|r|r|r|r|r|r|r|r|r|r|}\hline
$n$ & 0 & 1 & 2 & \textbf{3} & 4 & 5 & 6 & 7 & 8 & 9 & 10 & 11 & 12 & 13 &
14\\\hline\hline
$a_{n}$ & 0 & 1 & 4 & \textbf{1} & 1 & 2 & 3 & 1 & 1 & 3 & 1 & 1 & 1 & 1 &
3\\\hline
$p_{n}$ & 0 & 1 & 4 & \textbf{5} & 9 & 23 & 78 & 101 & 179 & 638 & 817 &
1455 & 2272 & 3727 & 13453\\\hline
$q_{n}$ & 1 & 1 & 5 & \textbf{6} & 11 & 28 & 95 & 123 & 218 & 777 & 995 &
1772 & 2767 & 4539 & 16384\\\hline
\end{tabular}
\]

\bigskip

\begin{itemize}
\item [\fbox{\textbf{Step 3.}}]Since $P=6$ is even, we proceed to \textbf{Step
4}.
\end{itemize}

\bigskip

\begin{itemize}
\item [\fbox{\textbf{Step 4.}}]Since
\[
3^{P/2}=3^{3}=27\neq-1\operatorname{mod}91\text{, }%
\]
we goto \textbf{Step 5}. \ 
\end{itemize}

\bigskip

\begin{itemize}
\item [\fbox{\textbf{Step 5.}}]With the Euclidean algorithm, we compute
\[
\gcd\left(  3^{P/2}-1,91\right)  =\gcd\left(  3^{3}-1,91\right)  =\gcd\left(
26,91\right)  =13\text{ .}%
\]
We have succeeded in finding a non-trivial factor of $N=91$, namely $13$. \ We
exit Shor's algorithm, and proceed to celebrate!
\end{itemize}

\bigskip

\index{Shor's Algorithm}

\bigskip

\section{\textbf{Grover's Algorithm}}

\bigskip

\index{Grover's Algorithm}

The the following description of Grover's algorithm is based on \cite{Grover1}%
, \cite{Grover2}, and \cite{Jozsa1}. \ 

\bigskip

\subsection{Problem definition}

\qquad\bigskip

We consider the problem of searching an unstructured database of $N=2^{n}$
records for exactly one record which has been specifically marked. \ This can
be rephrased in mathematical terms as an oracle problem as follows:

\bigskip

Label the records of the database with the integers
\[
0,1,2,\ \ldots\ ,N-1\text{ ,}%
\]
and denote the label of the unknown marked record by $x_{0}$. \ We are given
\ an oracle which computes the $n$ bit binary function
\[
f:\left\{  0,1\right\}  ^{n}\longrightarrow\left\{  0,1\right\}
\]
defined by
\[
f(x)=\left\{
\begin{array}
[c]{cl}%
1 & \text{if }x=x_{0}\\
& \\
0 & \text{otherwise}%
\end{array}
\right.
\]

\bigskip

We remind the readers that, as a standard oracle idealization, we have no
access to the internal workings of the function $f$. \ It operates simply as a
blackbox
\index{Blackbox} function, which we can query as many times as we like. \ But
with each such a query comes an associated computational cost.

\bigskip

\noindent\textbf{Search Problem for an Unstructured Database.} \ \textit{Find
the record labeled as }$x_{0}$\textit{\ with the minimum amount of
computational work, i.e., with the minimum number of queries of the oracle
}$f$\textit{.}

\bigskip

From probability theory, we know that if we examine $k$ records, i.e., if we
compute the oracle $f$ for $k$ randomly chosen records, then the probability
of finding the record labeled as $x_{0}$ is $k/N$. \ Hence, on a classical
computer it takes $O(N)=O(2^{n})$ queries to find the record labeled $x_{0}$.

\bigskip

\subsection{The quantum mechanical perspective}

\qquad\bigskip

However, as Luv Grover so astutely observed, on a quantum computer the search
of an unstructured database can be accomplished in $O(\sqrt{N})$ steps, or
more precisely, with the application of $O(\sqrt{N}\lg N)$ sufficiently local
unitary transformations. \ Although this is not exponentially faster, it is a
significant speedup.

\vspace{0.5in}

Let $\mathcal{H}_{2}$ be a 2 dimensional Hilbert space with orthonormal basis
\[
\left\{  \left|  0\right\rangle ,\left|  1\right\rangle \right\}  \text{ ;}%
\]
and let
\[
\left\{  \left|  0\right\rangle ,\left|  1\right\rangle ,\ \ldots\ ,\left|
N-1\right\rangle \right\}
\]
denote the induced orthonormal basis of the Hilbert space
\[
\mathcal{H}=%
{\displaystyle\bigotimes\limits_{0}^{N-1}}
\mathcal{H}_{2}\text{ .}%
\]

\vspace{0.5in}

From the quantum mechanical perspective, the oracle function $f$ is given as a
blackbox unitary transformation $U_{f}$, i.e., by
\[%
\begin{array}
[c]{ccc}%
\mathcal{H}\otimes\mathcal{H}_{2} & \overset{U_{f}}{\longrightarrow} &
\mathcal{H}\otimes\mathcal{H}_{2}\\
&  & \\
\left|  x\right\rangle \otimes\left|  y\right\rangle  & \longmapsto & \left|
x\right\rangle \otimes\left|  f(x)\oplus y\right\rangle
\end{array}
\]
where `$\oplus$' denotes exclusive `OR', i.e., addition modulo
2.\footnote{Please note that $U_{f}=\left(  \nu\circ\iota\right)  (f)$, as
defined in sections 10.3 and 10.4 of this paper.}

\vspace{0.5in}

Instead of $U_{f}$, we will use the computationally equivalent unitary
transformation
\[
I_{\left|  x_{0}\right\rangle }\left(  \left|  x\right\rangle \right)
=(-1)^{f(x)}\left|  x\right\rangle =\left\{
\begin{array}
[c]{cl}%
-\left|  x_{0}\right\rangle  & \text{if \ }x=x_{0}\\
& \\
\left|  x\right\rangle  & \text{otherwise}%
\end{array}
\right.
\]
That $I_{\left|  x_{0}\right\rangle }$ is computationally equivalent to $U_{f}
$ follows from the easily verifiable fact that
\[
U_{f}\left(  \left|  x\right\rangle \otimes\frac{\left|  0\right\rangle
-\left|  1\right\rangle }{\sqrt{2}}\right)  =\left(  I_{\left|  x_{0}%
\right\rangle }\left(  \left|  x\right\rangle \right)  \right)  \otimes
\frac{\left|  0\right\rangle -\left|  1\right\rangle }{\sqrt{2}}\text{ ,}%
\]
and also from the fact that $U_{f}$ can be constructed from a
controlled-$I_{\left|  x_{0}\right\rangle }$ and two one qubit Hadamard
transforms. \ (For details, please refer to \cite{Jozsa3}, \cite{Kitaev1}.)

\vspace{0.5in}

The unitary transformation $I_{\left|  x_{0}\right\rangle }$ is actually an
\textbf{inversion}
\index{Inversion}
\index{Mobius Inversion} \cite{Beardon1} in $\mathcal{H}$ about the hyperplane
perpendicular to $\left|  x_{0}\right\rangle $. \ This becomes evident when
$I_{\left|  x_{0}\right\rangle }$ is rewritten in the form
\[
I_{\left|  x_{0}\right\rangle }=I-2\left|  x_{0}\right\rangle \left\langle
x_{0}\right|  \text{ ,}%
\]
where `$I$' denotes the identity transformation. \ More generally, for any
unit length ket $\left|  \psi\right\rangle $, the unitary transformation
\[
I_{\left|  \psi\right\rangle }=I-2\left|  \psi\right\rangle \left\langle
\psi\right|  \text{ }%
\]
is an inversion in $\mathcal{H}$ about the hyperplane orthogonal to $\left|
\psi\right\rangle $.

\vspace{0.5in}

\subsection{Properties of the inversion $I_{\left|  \psi\right\rangle }$}

\qquad\bigskip

We digress for a moment to discuss the properties of the unitary
transformation $I_{\left|  \psi\right\rangle }$. \ To do so, we need the
following definition.

\bigskip

\begin{definition}
Let $\left|  \psi\right\rangle $ and $\left|  \chi\right\rangle $ be two kets
in $\mathcal{H}$ for which the bracket product $\left\langle \psi\mid
\chi\right\rangle $ is a real number. \ We define
\[
\mathcal{S}_{\mathbb{C}}=Span_{\mathbb{C}}\left(  \left|  \psi\right\rangle
,\left|  \chi\right\rangle \right)  =\left\{  \alpha\left|  \psi\right\rangle
+\beta\left|  \chi\right\rangle \in\mathcal{H}\mid\alpha,\beta\in
\mathbb{C}\right\}
\]
as the sub-Hilbert space of $\mathcal{H}$ spanned by $\left|  \psi
\right\rangle $ and $\left|  \chi\right\rangle $. \ We associate with the
Hilbert space $\mathcal{S}_{\mathbb{C}}$ a real inner product space lying in
$\mathcal{S}_{\mathbb{C}}$ defined by
\[
\mathcal{S}_{\mathbb{R}}=Span_{\mathbb{R}}\left(  \left|  \psi\right\rangle
,\left|  \chi\right\rangle \right)  =\left\{  a\left|  \psi\right\rangle
+b\left|  \chi\right\rangle \in\mathcal{H}\mid a,b\in\mathbb{R}\right\}
\text{ ,}%
\]
where the inner product on $\mathcal{S}_{\mathbb{R}}$ is that induced by the
bracket product on $\mathcal{H}$. \ If $\left|  \psi\right\rangle $ and
$\left|  \chi\right\rangle $ are also linearly independent, then
$\mathcal{S}_{\mathbb{R}}$ is a 2 dimensional real inner product space (i.e.,
the 2 dimensional Euclidean plane) lying inside of the complex 2 dimensional
space $\mathcal{S}_{\mathbb{C}}$.
\end{definition}

\bigskip

\begin{proposition}
Let $\left|  \psi\right\rangle $ and $\left|  \chi\right\rangle $ be two
\ linearly independent unit length kets in $\mathcal{H}$ with real bracket
product; and let $\mathcal{S}_{\mathbb{C}}=Span_{\mathbb{C}}\left(  \left|
\psi\right\rangle ,\left|  \chi\right\rangle \right)  $ and $\mathcal{S}%
_{\mathbb{R}}=Span_{\mathbb{R}}\left(  \left|  \psi\right\rangle ,\left|
\chi\right\rangle \right)  $. \ Then

\begin{itemize}
\item [1)]Both $\mathcal{S}_{\mathbb{C}}$ and $\mathcal{S}_{\mathbb{R}}$ are
invariant under the transformations $I_{\left|  \psi\right\rangle }$,
$I_{\left|  \chi\right\rangle }$, and hence $I_{\left|  \psi\right\rangle
}\circ I_{\left|  \chi\right\rangle }$, i.e.,
\[
\fbox{$%
\begin{array}
[c]{lrl}%
I_{\left|  \psi\right\rangle }\left(  \mathcal{S}_{\mathbb{C}}\right)
=\mathcal{S}_{\mathbb{C}} & \ \text{and\ } & I_{\left|  \psi\right\rangle
}\left(  \mathcal{S}_{\mathbb{R}}\right)  =\mathcal{S}_{\mathbb{R}}\\
&  & \\
I_{\left|  \chi\right\rangle }\left(  \mathcal{S}_{\mathbb{C}}\right)
=\mathcal{S}_{\mathbb{C}} & \ \text{and\ } & I_{\left|  \chi\right\rangle
}\left(  \mathcal{S}_{\mathbb{R}}\right)  =\mathcal{S}_{\mathbb{R}}\\
&  & \\
I_{\left|  \psi\right\rangle }I_{\left|  \chi\right\rangle }\left(
\mathcal{S}_{\mathbb{C}}\right)  =\mathcal{S}_{\mathbb{C}} & \ \text{and\ } &
I_{\left|  \psi\right\rangle }I_{\left|  \chi\right\rangle }\left(
\mathcal{S}_{\mathbb{R}}\right)  =\mathcal{S}_{\mathbb{R}}%
\end{array}
$}%
\]
\bigskip

\item[2)] If $L_{\left|  \psi^{\perp}\right\rangle }$ is the line in the plane
$\mathcal{S}_{\mathbb{R}}$ which passes through the origin and which is
perpendicular to $\left|  \psi\right\rangle $, then $I_{\left|  \psi
\right\rangle }$ restricted to $\mathcal{S}_{\mathbb{R}}$ is a reflection in
(i.e., a M\"{o}bius inversion \cite{Beardon1} about) the line $L_{\left|
\psi^{\perp}\right\rangle }$. \ A similar statement can be made in regard to
$\left|  \chi\right\rangle $.

\item[3)] If $\left|  \psi^{\perp}\right\rangle $ is a unit length vector in
$\mathcal{S}_{\mathbb{R}}$ perpendicular to $\left|  \psi\right\rangle $,
then
\[
-I_{\left|  \psi\right\rangle }=I_{\left|  \psi^{\perp}\right\rangle }\text{
.}%
\]
(Hence, $\left\langle \psi^{\perp}\mid\chi\right\rangle $ is real.)
\end{itemize}
\end{proposition}

\vspace{0.5in}

Finally we note that, since $I_{\left|  \psi\right\rangle }=I-2\left|
\psi\right\rangle \left\langle \psi\right|  $, it follows that

\bigskip

\begin{proposition}
If $\left|  \psi\right\rangle $ is a unit length ket in $\mathcal{H}$, and if
$U$ is a unitary transformation on $\mathcal{H}$, then
\[
UI_{\left|  \psi\right\rangle }U^{-1}=I_{U\left|  \psi\right\rangle }\text{ .}%
\]
\end{proposition}

\vspace{0.5in}

\subsection{The method in Luv's ``madness''}

\qquad\bigskip

Let $H:\mathcal{H}\longrightarrow\mathcal{H}$ be the Hadamard transform,
i.e.,
\[
H=%
{\displaystyle\bigotimes\limits_{0}^{n-1}}
H^{(2)}\text{ , }%
\]
where
\[
H^{(2)}=\left(
\begin{array}
[c]{rr}%
1 & 1\\
1 & -1
\end{array}
\right)
\]
with respect to the basis $\left|  0\right\rangle $, $\left|  1\right\rangle $.

\bigskip

We begin by using the Hadamard transform $H$ to construct a state $\left|
\psi_{0}\right\rangle $ which is an equal superposition of all the standard
basis states $\left|  0\right\rangle $, $\left|  1\right\rangle $,$\ldots
$,$\left|  N-1\right\rangle $ (including the unknown state $\left|
x_{0}\right\rangle $), i.e.,
\[
\left|  \psi_{0}\right\rangle =H\left|  0\right\rangle =\frac{1}{\sqrt{N}}%
\sum_{k=0}^{N-1}\left|  k\right\rangle \text{ .}%
\]

\bigskip

Both $\left|  \psi_{0}\right\rangle $ and the unknown state $\left|
x_{0}\right\rangle $ lie in the Euclidean plane $\mathcal{S}_{\mathbb{R}%
}=Span_{\mathbb{R}}\left(  \left|  \psi_{0}\right\rangle ,\left|
x_{0}\right\rangle \right)  $. \ Our strategy is to rotate within the plane
$\mathcal{S}_{\mathbb{R}}$ the state $\left|  \psi_{0}\right\rangle $ about
the origin until it is as close as possible to $\left|  x_{0}\right\rangle $.
\ Then a measurement with respect to the standard basis of the state resulting
from rotating $\left|  \psi_{0}\right\rangle $, will produce $\left|
x_{0}\right\rangle $ with high probability.

\bigskip

To achieve this objective, we use the oracle $I_{\left|  x_{0}\right\rangle }$
to construct the unitary transformation
\[
Q=-HI_{\left|  0\right\rangle }H^{-1}I_{\left|  x_{0}\right\rangle }\text{ ,}%
\]

\bigskip

which by proposition 2 above, can be reexpressed as
\[
Q=-I_{\left|  \psi_{0}\right\rangle }I_{\left|  x_{0}\right\rangle }\text{ .}%
\]

\bigskip

Let $\left|  x_{0}^{\perp}\right\rangle $ and $\left|  \psi_{0}^{\perp
}\right\rangle $ denote unit length vectors in $\mathcal{S}_{\mathbb{R}}$
perpendicular to $\left|  x_{0}\right\rangle $ and $\left|  \psi
_{0}\right\rangle $, respectively. \ There are two possible choices for each
of $\left|  x_{0}^{\perp}\right\rangle $ and $\left|  \psi_{0}^{\perp
}\right\rangle $ respectively. \ To remove this minor, but nonetheless
annoying, ambiguity, we select $\left|  x_{0}^{\perp}\right\rangle $ and
$\left|  \psi_{0}^{\perp}\right\rangle $ so that the orientation of the plane
$\mathcal{S}_{\mathbb{R}}$ induced by the ordered spanning vectors $\left|
\psi_{0}\right\rangle $, $\left|  x_{0}\right\rangle $ is the same orientation
as that induced by each of the ordered bases $\left|  x_{0}^{\perp
}\right\rangle $, $\left|  x_{0}\right\rangle $ and $\left|  \psi
_{0}\right\rangle $, $\left|  \psi_{0}^{\perp}\right\rangle $. \ (Please refer
to Figure 2.)

\bigskip

\begin{remark}
The removal of the above ambiguities is really not essential. \ However, it
does simplify the exposition given below.
\end{remark}

\bigskip%

\begin{center}
\fbox{\includegraphics[
height=1.8248in,
width=2.8323in
]%
{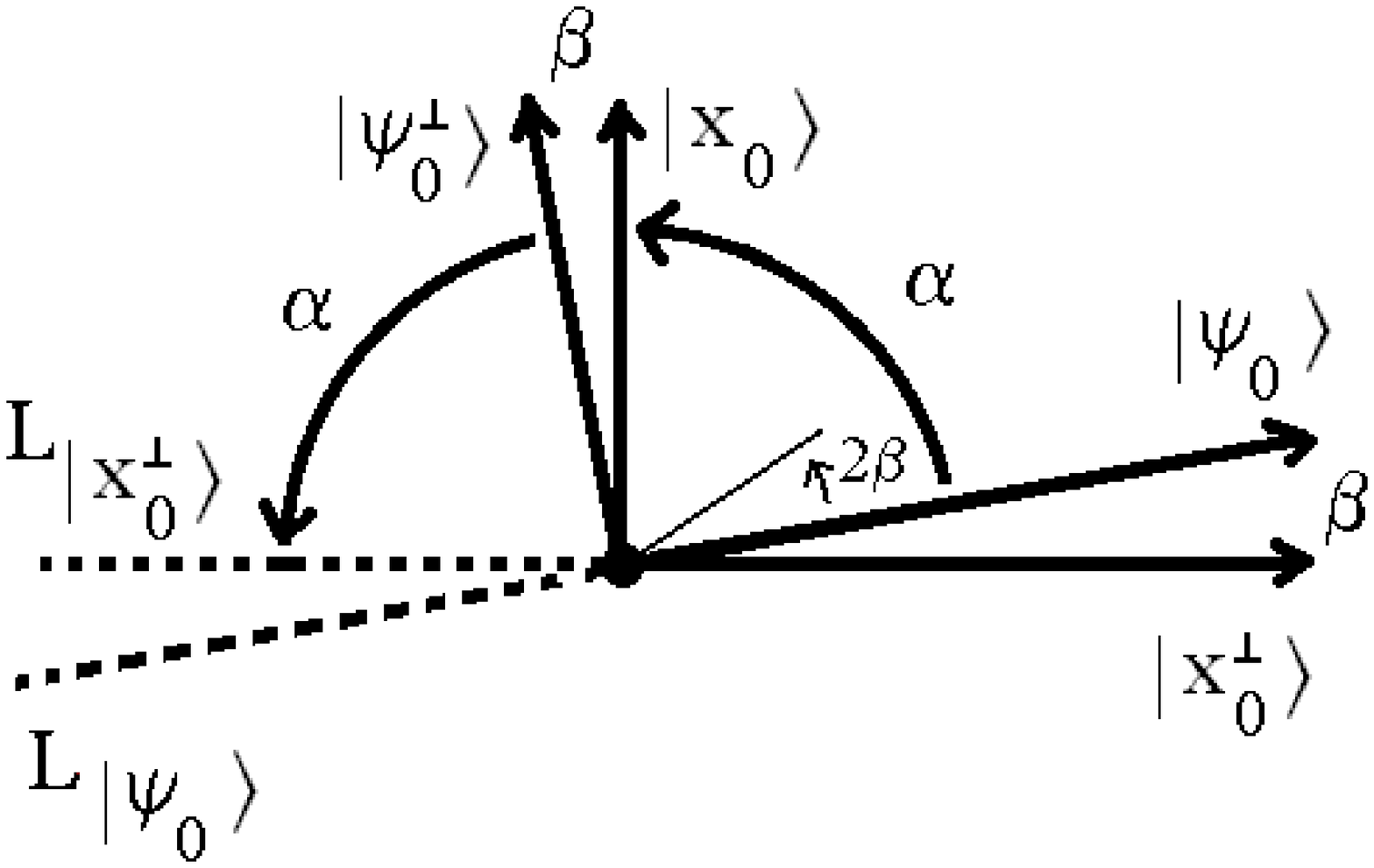}%
}\\
Figure 2. \ The linear transformation $\left.  Q\right|  _{\mathcal{S}%
_{\mathbb{R}}}$ is reflection in the line $L_{\left|  x_{0}^{\perp
}\right\rangle }$ followed by reflection in the line $L_{\left|  \psi
_{0}\right\rangle }$ which is the same as rotation by the angle $2\beta$.
\ Thus, $\left.  Q\right|  _{\mathcal{S}_{\mathbb{R}}}$ rotates $\left|
\psi_{0}\right\rangle $ by the angle $2\beta$ toward $\left|  x_{0}%
\right\rangle $.
\end{center}

\vspace{0.5in}

We proceed by noting that, by the above proposition 1, the plane
$\mathcal{S}_{\mathbb{R}}$ lying in $\mathcal{H}$ is invariant under the
linear transformation $Q$, and that, when $Q$ is restricted to the plane
$\mathcal{S}_{\mathbb{R}}$, it can be written as the composition of two
inversions, i.e.,
\[
\left.  Q\right|  _{\mathcal{S}_{\mathbb{R}}}=I_{\left|  \psi_{0}^{\perp
}\right\rangle }I_{\left|  x_{0}\right\rangle }\text{ .}%
\]

In particular, $\left.  Q\right|  _{\mathcal{S}_{\mathbb{R}}}$ is the
composition of two inversions in $\mathcal{S}_{\mathbb{R}}$, the first in the
line $L_{\left|  x_{0}^{\perp}\right\rangle }$ in $\mathcal{S}_{\mathbb{R}}$
passing through the origin having $\left|  x_{0}\right\rangle $ as normal, the
second in the line $L_{\left|  \psi_{0}\right\rangle }$ through the origin
having $\left|  \psi_{0}^{\perp}\right\rangle $ as normal.\footnote{The line
$L_{\left|  x_{0}^{\perp}\right\rangle }$ is the intersection of the plane
$\mathcal{S}_{\mathbb{R}}$ with the hyperplane in $\mathcal{H}$ orthogonal to
$\left|  x_{0}\right\rangle $. \ A similar statement can be made in regard to
$L_{\left|  \psi_{0}\right\rangle }$.} \ 

\vspace{0.5in}

We can now apply the following theorem from plane geometry:

\bigskip

\begin{theorem}
If $L_{1}$ and $L_{2}$ are lines in the Euclidean plane $\mathbb{R}^{2}$
intersecting at a point $O$; and if $\beta$ is the angle in the plane from
$L_{1}$ to $L_{2}$, then the operation of reflection in $L_{1}$ followed by
reflection in $L_{2}$ is just rotation by angle $2\beta$ about the point $O$.
\end{theorem}

\bigskip

Let $\beta$ denote the angle in $S_{\mathbb{R}}$ from $L_{\left|  x_{0}%
^{\perp}\right\rangle }$ to $L_{\left|  \psi_{0}\right\rangle }$, which by
plane geometry is the same as the angle from $\left|  x_{0}^{\perp
}\right\rangle $ to $\left|  \psi_{0}\right\rangle $, which in turn is the
same as the angle from $\left|  x_{0}\right\rangle $ to $\left|  \psi
_{0}^{\perp}\right\rangle $. \ Then by the above theorem $\left.  Q\right|
_{\mathcal{S}_{\mathbb{R}}}=I_{\left|  \psi_{0}^{\perp}\right\rangle
}I_{\left|  x_{0}\right\rangle }$ is a rotation about the origin by the angle
$2\beta$. \ 

\vspace{0.5in}

The key idea in Grover's algorithm is to move $\left|  \psi_{0}\right\rangle $
toward the unknown state $\left|  x_{0}\right\rangle $ by successively
applying the rotation $Q$ to $\left|  \psi_{0}\right\rangle $ to rotate it
around to $\left|  x_{0}\right\rangle $. \ This process is called
\textbf{amplitude amplification}. \ \ Once this process is completed, the
measurement of the resulting state (with respect to the standard basis) will,
with high probability, yield the unknown state $\left|  x_{0}\right\rangle $.
\ This is the essence of Grover's algorithm. \ 

\vspace{0.5in}

But how many times $K$ should we apply the rotation $Q$ to $\left|  \psi
_{0}\right\rangle $? \ If we applied $Q$ too many or too few times, we would
over- or undershoot our target state $\left|  x_{0}\right\rangle $. \ 

\vspace{0.5in}

We determine the integer $K$ as follows:

\bigskip

Since
\[
\left|  \psi_{0}\right\rangle =\sin\beta\left|  x_{0}\right\rangle +\cos
\beta\left|  x_{0}^{\perp}\right\rangle \text{ ,}%
\]
the state resulting after $k$ applications of $Q$ is
\[
\left|  \psi_{k}\right\rangle =Q^{k}\left|  \psi_{0}\right\rangle =\sin\left[
\left(  2k+1\right)  \beta\right]  \left|  x_{0}\right\rangle +\cos\left[
\left(  2k+1\right)  \beta\right]  \left|  x_{0}^{\perp}\right\rangle \text{
.}%
\]
Thus, we seek to find the smallest positive integer $K=k$ such that
\[
\sin\left[  \left(  2k+1\right)  \beta\right]
\]
is as close as possible to $1$. \ In other words, we seek to find the smallest
positive integer $K=k$ such that
\[
\left(  2k+1\right)  \beta
\]
is as close as possible to $\pi/2$. \ It follows that\footnote{The reader may
prefer to use the $floor$ function instead of the $round$ function.}
\[
K=k=round\left(  \frac{\pi}{4\beta}-\frac{1}{2}\right)  \text{ ,}%
\]
where ``$round$''
\index{round function@$round$ function} is the function that rounds to the
nearest integer.

\vspace{0.5in}

We can determine the angle $\beta$ by noting that the angle $\alpha$ from
$\left|  \psi_{0}\right\rangle $ and $\left|  x_{0}\right\rangle $ is
complementary to $\beta$, i.e.,
\[
\alpha+\beta=\pi/2\text{ ,}%
\]
and hence,
\[
\frac{1}{\sqrt{N}}=\left\langle x_{0}\mid\psi_{0}\right\rangle =\cos
\alpha=\cos(\frac{\pi}{2}-\beta)=\sin\beta\text{ .}%
\]
Thus, the angle $\beta$ is given by
\[
\beta=\sin^{-1}\left(  \frac{1}{\sqrt{N}}\right)  \approx\frac{1}{\sqrt{N}%
}\text{ \ (for large }N\text{) ,}%
\]
and hence,
\[
K=k=round\left(  \frac{\pi}{4\sin^{-1}\left(  \frac{1}{\sqrt{N}}\right)
}-\frac{1}{2}\right)  \approx round\left(  \frac{\pi}{4}\sqrt{N}-\frac{1}%
{2}\right)  \text{ (for large }N\text{).}%
\]

\vspace{0.5in}

\subsection{Summary of Grover's algorithm}

\qquad\bigskip

In summary, we provide the following outline of Grover's algorithm:

\bigskip

\fbox{%
\begin{tabular}
[c]{ll}\hline\hline
& \hspace{0.75in}\textbf{Grover's Algorithm}\\\hline\hline
& \\
$%
\begin{array}
[c]{r}%
\fbox{$\mathbb{STEP}$ 0.}\\
\bigskip\\
\bigskip
\end{array}
$ & $%
\begin{array}
[c]{l}%
\text{(Initialization)}\\
\qquad\left|  \psi\right\rangle \longleftarrow H\left|  0\right\rangle
=\frac{1}{\sqrt{N}}%
{\displaystyle\sum\limits_{j=0}^{N-1}}
\left|  j\right\rangle \\
\qquad k\quad\longleftarrow0
\end{array}
$\\
& \\
$%
\begin{array}
[c]{r}%
\fbox{$\mathbb{STEP}$ 1.}\\
\bigskip\\
\bigskip
\end{array}
$ & $%
\begin{array}
[c]{r}%
\text{Loop until }k=\underset{}{round\left(  \frac{\pi}{4\sin^{-1}\left(
1/\sqrt{N}\right)  }-\frac{1}{2}\right)  }\approx round\left(  \frac{\pi}%
{4}\sqrt{N}-\frac{1}{2}\right) \\
\multicolumn{1}{l}{\qquad\left|  \psi\right\rangle \longleftarrow\underset
{}{Q}\left|  \psi\right\rangle =-HI_{\left|  0\right\rangle }HI_{\left|
x_{0}\right\rangle }\left|  \psi\right\rangle }\\
\multicolumn{1}{l}{\qquad k\quad\longleftarrow k+1}%
\end{array}
$\\
& \\
$%
\begin{array}
[c]{r}%
\fbox{$\mathbb{STEP}$ 2.}\\
\bigskip
\end{array}
$ & $%
\begin{array}
[c]{l}%
\text{Measure }\left|  \psi\right\rangle \text{ with respect to the standard
basis}\\
\left|  0\right\rangle ,\left|  1\right\rangle ,\ \ldots\ ,\left|
N-1\right\rangle \text{ to obtain the marked unknown }\\
\text{state }\left|  x_{0}\right\rangle \text{ with probability }\geq
1-\frac{1}{N}\text{.}%
\end{array}
$%
\end{tabular}
}

\vspace{0.5in}

We complete our summary with the following theorem:

\bigskip

\begin{theorem}
With a probability of error\footnote{If the reader prefers to use the $floor$
function rather than the $round$ function, then probability of error becomes
$Prob_{E}\leq\frac{4}{N}-\frac{4}{N^{2}}$.}
\[
Prob_{E}\leq\frac{1}{N}\text{, }%
\]
Grover's algorithm finds the unknown state $\left|  x_{0}\right\rangle $ at a
computational cost of
\[
O\left(  \sqrt{N}\lg N\right)
\]
\end{theorem}

\begin{proof}
\qquad\bigskip

\begin{itemize}
\item [Part 1.]The probability of error $Prob_{E}$ of finding the hidden state
$\left|  x_{0}\right\rangle $ is given by
\[
Prob_{E}=\cos^{2}\left[  \left(  2K+1\right)  \beta\right]  \text{ ,}%
\]
where
\[
\left\{
\begin{array}
[c]{rrl}%
\beta & = & \sin^{-1}\left(  \frac{1}{\sqrt{N}}\right) \\
&  & \\
K & = & round\left(  \frac{\pi}{4\beta}-\frac{1}{2}\right)
\end{array}
\right.  \text{,}%
\]
where ``$round$'' is the function that rounds to the nearest integer. Hence,
\[%
\begin{array}
[c]{rrl}%
\frac{\pi}{4\beta}-1\leq K\leq\frac{\pi}{4\beta} & \Longrightarrow & \frac
{\pi}{2}-\beta\leq\left(  2K+1\right)  \beta\leq\frac{\pi}{2}+\beta\\
&  & \\
& \Longrightarrow & \sin\beta=\cos\left(  \frac{\pi}{2}-\beta\right)  \geq
\cos\left[  \left(  2K+1\right)  \beta\right]  \geq\cos\left(  \frac{\pi}%
{2}+\beta\right)  =-\sin\beta
\end{array}
\]
Thus,
\[
Prob_{E}=\cos^{2}\left[  \left(  2K+1\right)  \beta\right]  \leq\sin^{2}%
\beta=\sin^{2}\left(  \sin^{-1}\left(  \frac{1}{\sqrt{N}}\right)  \right)
=\frac{1}{N}%
\]
\end{itemize}

\bigskip

\begin{itemize}
\item [Part 2.]The computational cost of the Hadamard transform $H=\bigotimes
_{0}^{n-1}H^{(2)}$ is $O(n)=O(\lg N)$ single qubit operations. \ The
transformations $-I_{\left|  0\right\rangle }$ and $I_{\left|  x_{0}%
\right\rangle }$ each carry a computational cost of $O(1)$.

$\mathbb{STEP}$ 1 is the computationally dominant step. \ In $\mathbb{STEP}$ 1
there are $O\left(  \sqrt{N}\right)  $ iterations. \ In each iteration, the
Hadamard transform is applied twice. \ The transformations $-I_{\left|
0\right\rangle }$ and $I_{\left|  x_{0}\right\rangle }$ are each applied once.
Hence, each iteration comes with a computational cost of $O\left(  \lg
N\right)  $, and so the total cost of $\mathbb{STEP}$ 1 is $O(\sqrt{N}\lg N)$.
\end{itemize}
\end{proof}

\bigskip

\subsection{\textbf{\bigskip}An example of Grover's algorithm}

\qquad\bigskip

As an example, we search a database consisting of $N=2^{n}=8$ records for an
unknown record with the unknown label $x_{0}=5$. \ The calculations for this
example were made with OpenQuacks
\index{OpenQuacks Public Domain Software}, which is an open source quantum
simulator Maple package developed at UMBC and publically available.

\vspace{0.5in}

We are given a blackbox
\index{Blackbox} computing device
\[
\text{In}\rightarrow\fbox{\fbox{%
\begin{tabular}
[c]{l}%
$I_{\left|  ?\right\rangle }$%
\end{tabular}
}}\rightarrow\text{Out}%
\]
that implements as an oracle the unknown unitary transformation
\[
I_{\left|  x_{0}\right\rangle }=I_{\left|  5\right\rangle }=\left(
\begin{array}
[c]{rrrrrrrrr}%
1 & 0 & 0 & 0 &  & 0 & 0 & 0 & 0\\
0 & 1 & 0 & 0 &  & 0 & 0 & 0 & 0\\
0 & 0 & 1 & 0 &  & 0 & 0 & 0 & 0\\
0 & 0 & 0 & 1 &  & 0 & 0 & 0 & 0\\
&  &  &  &  &  &  &  & \\
0 & 0 & 0 & 0 &  & -1 & 0 & 0 & 0\\
0 & 0 & 0 & 0 &  & 0 & 1 & 0 & 0\\
0 & 0 & 0 & 0 &  & 0 & 0 & 1 & 0\\
0 & 0 & 0 & 0 &  & 0 & 0 & 0 & 1
\end{array}
\right)
\]

\bigskip

We cannot open up the blackbox $\rightarrow\fbox{$\fbox{%
\begin{tabular}
[c]{l}%
$I_{\left|  ?\right\rangle }$%
\end{tabular}
}$}\rightarrow$ to see what is inside. \ So we do not know what $I_{\left|
x_{0}\right\rangle }$ and $x_{0}$ are. \ \ The only way that we can glean some
information about $x_{0}$ is to apply some chosen state $\left|
\psi\right\rangle $ as input, and then make use of the resulting output.

\vspace{0.5in}

Using of the blackbox $\rightarrow\fbox{\fbox{%
\begin{tabular}
[c]{l}%
$I_{\left|  ?\right\rangle }$%
\end{tabular}
}}\rightarrow$ as a component device, we construct a computing device
$\rightarrow\fbox{\fbox{%
\begin{tabular}
[c]{l}%
$-HI_{\left|  0\right\rangle }HI_{\left|  ?\right\rangle }$%
\end{tabular}
}}\rightarrow$ which implements the unitary operator
\[
Q=-HI_{\left|  0\right\rangle }HI_{\left|  x_{0}\right\rangle }=\frac{1}%
{4}\left(
\begin{array}
[c]{rrrrrrrrr}%
-3 & 1 & 1 & 1 &  & -1 & 1 & 1 & 1\\
1 & -3 & 1 & 1 &  & -1 & 1 & 1 & 1\\
1 & 1 & -3 & 1 &  & -1 & 1 & 1 & 1\\
1 & 1 & 1 & -3 &  & -1 & 1 & 1 & 1\\
&  &  &  &  &  &  &  & \\
1 & 1 & 1 & 1 &  & 3 & 1 & 1 & 1\\
1 & 1 & 1 & 1 &  & -1 & -3 & 1 & 1\\
1 & 1 & 1 & 1 &  & -1 & 1 & -3 & 1\\
1 & 1 & 1 & 1 &  & -1 & 1 & 1 & -3
\end{array}
\right)
\]

\vspace{0.5in}

We do not know what unitary transformation $Q$ \ is implemented by the device
$\rightarrow\fbox{\fbox{%
\begin{tabular}
[c]{l}%
$-HI_{\left|  0\right\rangle }HI_{\left|  ?\right\rangle }$%
\end{tabular}
}}\rightarrow$ because the blackbox $\rightarrow\fbox{\fbox{%
\begin{tabular}
[c]{l}%
$I_{\left|  ?\right\rangle }$%
\end{tabular}
}}\rightarrow$ is one of its essential components.

\bigskip

\begin{itemize}
\item [\fbox{$\mathbb{STEP}$ 0.}]We begin by preparing the known state
\[
\fbox{$\left|  \psi_0\right\rangle =H\left|  0\right\rangle =\frac{1}{\sqrt
{8}}\left(  1,1,1,1,1,1,1,1\right)  ^{transpose}$}%
\]
\end{itemize}

\bigskip

\begin{itemize}
\item [\fbox{$\mathbb{STEP}$ 1.}]We proceed to loop
\[
K=round\left(  \frac{\pi}{4\sin^{-1}\left(  1/\sqrt{8}\right)  }-\frac{1}%
{2}\right)  =2
\]
times in $\mathbb{STEP}$ 1.

\begin{itemize}
\item [\textsc{Iteration} 1.]On the first iteration, we obtain the unknown
state
\[
\fbox{$\left|  \psi_1\right\rangle =Q\left|  \psi_0\right\rangle =\frac
{1}{4\sqrt{2}}\left(  1,1,1,1,5,1,1,1\right)  ^{transpose}$}%
\]

\item[\textsc{Iteration} 2.] On the second iteration, we obtain the unknown
state
\[
\fbox{$\left|  \psi_2\right\rangle =Q\left|  \psi_1\right\rangle =\frac
{1}{8\sqrt{2}}\left(  -1,-1,-1,-1,11,-1,-1,-1\right)  ^{transpose}$}%
\]
and branch to $\mathbb{STEP}$ 2.
\end{itemize}
\end{itemize}

\bigskip

\begin{itemize}
\item [\fbox{$\mathbb{STEP}$ 2.}]We measure the unknown state $\left|
\psi_{2}\right\rangle $ to obtain either
\[
\left|  5\right\rangle
\]
with probability
\[
Prob_{Success}=\sin^{2}\left(  \left(  2K+1\right)  \beta\right)  =\frac
{121}{128}=0.9453
\]
or some other state with probability
\[
Prob_{Failure}=\cos^{2}\left(  \left(  2K+1\right)  \beta\right)  =\frac
{7}{128}=0.0547
\]
and then exit.
\end{itemize}

\index{Grover's Algorithm}

\bigskip

\section{\textbf{\bigskip There is much more to quantum computation}}

\qquad\bigskip

Needles to say, there is much more to quantum computation. \ I hope that you
found this introductory paper useful.

\vspace{0.5in}


\quad

\bigskip


\printindex
\end{document}